\documentclass[aps,preprint,nofootinbib,preprintnumbers,eqsecnum,superscriptaddress]{revtex4-1}

\usepackage[usenames, dvipsnames]{color}

\newcommand{\nb}[1]{\color{blue}}

\newcommand{\hl}[1]{\color{magenta}}

\usepackage[
	  pagebackref=false,
	  colorlinks=true,
      linkcolor=blue,
      urlcolor=blue,
      filecolor=black,
      citecolor=red,
      pdfstartview=FitV,
      pdftitle={},
        pdfauthor={},
        pdfsubject={},
        pdfkeywords={},
        pdfpagemode=None,
        bookmarksopen=true
      ]{hyperref}

\usepackage{graphicx}
\usepackage{xcolor}
\usepackage{rotating}
\usepackage{bm,amsmath,amssymb}
\usepackage[utf8]{inputenc}
\usepackage{footmisc}

\usepackage[normalem]{ulem}
\usepackage{enumerate}
\usepackage{amsfonts}
\usepackage{mathbbol}

\def\Tr{\mathop{\rm Tr}}

\def\Re{\mathop{\rm Re} }

\newcommand\half{{\ensuremath{\frac{1}{2}}}}
\newcommand\p{\ensuremath{\partial}}

\newcommand\field[1]{{\ensuremath{\mathbb{{#1}}}}}

\newcommand\vev[1]{{\ensuremath{\left\langle{#1}\right\rangle}}}

\newcommand\ket[1]{\ensuremath{\lvert{#1}\rangle}}
\newcommand\bra[1]{\ensuremath{\langle{#1}\rvert}}

\newcommand{\RR}{\field{R}}

\newcommand{\ZZ}{\field{Z}}

\newcommand{\be}{\begin{equation}}
\newcommand{\ee}{\end{equation}}
\newcommand{\bea}{\begin{eqnarray}}
\newcommand{\eea}{\end{eqnarray}}
\newcommand{\bega}{\begin{gather}}
\newcommand{\eega}{\end{gather}}

\newcommand{\bi}{\begin{itemize}}
\newcommand{\ei}{\end{itemize}}
\newcommand{\ben}{\begin{enumerate}}
\newcommand{\een}{\end{enumerate}}
\newcommand{\bca}{\begin{cases}}
\newcommand{\eca}{\end{cases}}
\newcommand{\bln}{\begin{align}}
\newcommand{\eln}{\end{align}}
\newcommand{\bst}{\begin{split}}
\newcommand{\est}{\end{split}}
\def\ie{\begin{equation}\begin{aligned}}
\def\fe{\end{aligned}\end{equation}}
\newcommand{\bma}{\le(\begin{matrix}}
\newcommand{\ema}{\end{matrix}\ri)}

\newcommand\al{{\alpha}}
\def\b{{\beta}}
\newcommand\ep{\epsilon}

\newcommand\Sig{\Sigma}

\newcommand\om{\omega}
\newcommand\Om{\Omega}

\newcommand\ga{{\ensuremath{{\gamma}}}}
\newcommand\Ga{{\ensuremath{{\Gamma}}}}
\newcommand\de{{\ensuremath{{\delta}}}}
\newcommand\De{{\ensuremath{{\Delta}}}}

\newcommand\ka{\kappa}

\newcommand\da{{\dagger}}

\def\th{{\theta}}

\newcommand\Lra{{\Longrightarrow}}

\newcommand\ov{\over}
\newcommand\ha{{\half}}

\def\le{\left}
\def\ri{\right}

\newcommand\sA{{\ensuremath{{\mathcal A}}}}
\newcommand\sB{{\ensuremath{{\mathcal B}}}}

\newcommand\sH{{\ensuremath{{\mathcal H}}}}

\newcommand\sM{{\ensuremath{{\mathcal M}}}}
\newcommand\sN{{\ensuremath{{\mathcal N}}}}
\newcommand\sO{{\ensuremath{{\mathcal O}}}}
\newcommand\sP{{\ensuremath{{\mathcal P}}}}

\newcommand\sJ{{\mathcal J}}

\newcommand\sX{{\mathcal X}}
\newcommand\sY{{\mathcal Y}}
\newcommand\sZ{{\mathcal Z}}

\newcommand\vx{{\vec x}}

\newcommand{\fa}{{\mathfrak a}}
\newcommand{\fb}{{\mathfrak{b}}}
\newcommand{\fc}{{\mathfrak{c}}}

\newcommand{\fr}{{\mathfrak r}}

\newcommand{\bid}{\mathbf{1}}
\newcommand{\pt}{\partial}

\usepackage{graphicx}
\usepackage{subcaption}
\usepackage{float}
\usepackage{ragged2e}

\begin{document}

\title{Subregion-subalgebra duality: emergence of space and time in holography}

\preprint{MIT-CTP/5510}

\author{Sam Leutheusser}
\affiliation{Princeton Gravity Initiative, 
Princeton University, 
Princeton, NJ 08544, USA }

\author{Hong Liu}
\affiliation{Center for Theoretical Physics, 
Massachusetts
Institute of Technology, \\
77 Massachusetts Ave.,  Cambridge, MA 02139 }

\begin{abstract}

 \noindent 
 
In holographic duality, a higher dimensional quantum gravity system emerges from a lower dimensional conformal field theory (CFT) with a large number of degrees of freedom.  We propose a formulation of duality for a general causally complete bulk spacetime region, called subregion-subalgebra duality, which provides  a framework to describe how geometric notions in the gravity system, such as spacetime subregions, different notions of times, and causal structure, emerge from the dual CFT. 
Subregion-subalgebra duality generalizes and brings new insights into subregion-subregion duality (or equivalently entanglement wedge reconstruction). It provides a mathematically precise definition of subregion-subregion duality and gives an independent definition of entanglement wedges without using entropy. Geometric properties of entanglement wedges, including those that play a crucial role in interpreting the bulk as a quantum error correcting code, can be understood from the duality as the geometrization of the superadditivity of certain algebras. 
Using general boundary subalgebras rather than those associated with geometric subregions makes it possible to find duals for general bulk spacetime regions, including those not touching the boundary. Applying subregion-subalgebra duality to a boundary state describing a single-sided black hole also provides a precise way to define 
mirror operators.

\end{abstract}

\today

\maketitle

\tableofcontents

\section{Introduction}

Understanding the precise manner in which space, time, and the associated causal structure of a bulk gravity system arise in its boundary description has been an outstanding question in holography. 
Geometric notions of a bulk gravity system such as local spacetime regions, event horizons, and causal structure can  be sharply defined only in the $G_N \to 0$ limit,\footnote{$G_N$ is Newton's constant. In this paper we take $\hbar =1$, and by $G_N \to 0$ we mean the perturbative $G_N$ expansion to any finite order. At finite $G_N$, quantum spacetime fluctuations will make geometric concepts fuzzy.} which means that their emergence can be described in a rigorous manner only in the $N \to \infty$ limit of the boundary theory.\footnote{We use $N$ to characterize the number of degrees of freedom of the boundary theory,  which is related to the bulk Newton constant $G_N$ as  $G_N \sim {1 \ov N^2}$.  The $N \to \infty$ limit refers to perturbative expansions in $1/N^2$ to any finite order.}
An important step to understand the emergence of the bulk geometry is to pinpoint the underlying mathematical structure in the boundary theory that is responsible.  As $N \to \infty$, many states and operators of a finite $N$ theory do not have a sensible limit, and drop out of the $N=\infty$ theory. As a result, the structures of the Hilbert space and operator algebras 
undergo dramatic changes~\cite{shortPaper, longPaper,Witten:2021jzq,Witten:2021unn,Schlenker:2022dyo,Chandrasekaran:2022cip,Chandrasekaran:2022eqq,Faulkner:2022ada} ({see also~\cite{Gao:2021tzr,Chandrasekaran:2022qmq,Dabholkar:2022mxo,Sugishita:2022ldv,Gomez:2022eui,Bahiru:2022mwh,Verlinde:2022xkw,deBoer:2022zps,Seo:2022pqj,Donnelly:2022kfs,Bzowski:2022kgf}}).  
In particular, it was argued in~\cite{shortPaper, longPaper} that there is ubiquitous emergence of type III$_1$ von Neumann~(vN) algebras, which in the example of the thermal field double state was used to explain the emergence of event horizons, Kruskal-like times, and the associated causal structure in an eternal black hole geometry.

In this paper we further elaborate on the  mathematical structure of the $N \to \infty$ limit, and provide a general picture of how a local bulk spacetime region emerges from the boundary theory. We will argue that there is the following correspondence 
\medskip
\be \label{he2} 
\text{\parbox{5cm}{bulk spacetime open region}} \quad
 \Rightarrow \quad \quad \text{\parbox{5cm}{emergent type III$_1$ \\ boundary subalgebra}}   \  . 
\ee
Namely, for any open bulk spacetime region $\fb$, there exists an emergent boundary type III$_1$ vN algebra that is equivalent to the bulk operator algebra in $\fb$, i.e., 
\bega \label{he1}
\text{bulk operator algebra in $\fb$} = \text{an emergent type III$_1$ vN {(boundary)} subalgebra}    \ .
\end{gather} 
Furthermore, bulk geometric notions, such as interior times, lightcones, and horizons, can be viewed as geometrizations of algebraic properties of emergent type III$_1$ vN subalgebras. Our main message can be summarized with a slogan: 

\centerline{\it Bulk locality is a geometrization 
of emergent boundary type III$_1$ subalgebras.}

\begin{figure}[H]
        \centering
        \begin{subfigure}[b]{0.3\textwidth}
            \centering
			\includegraphics[width=\textwidth]{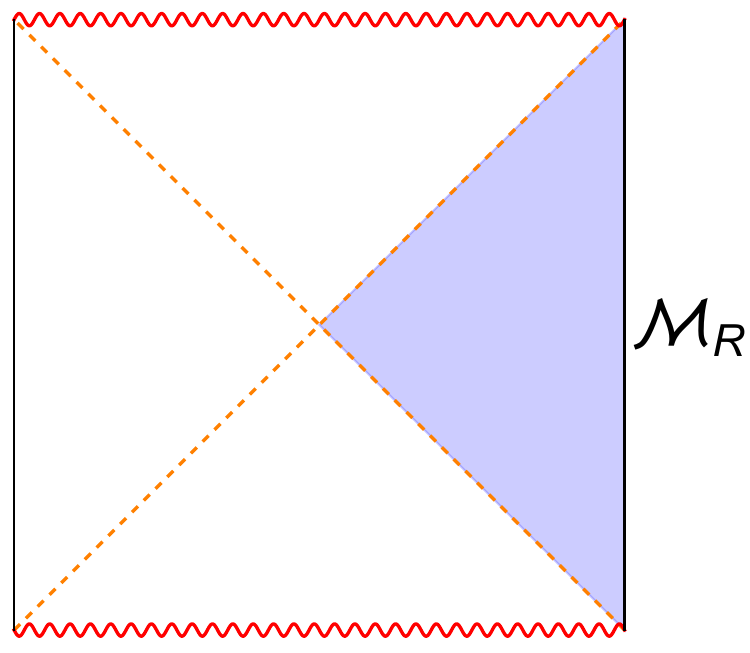} 
            \caption[]%
            {{\small }}    
        \end{subfigure}
        \hfill
        \begin{subfigure}[b]{0.3\textwidth}   
            \centering 
			\includegraphics[width=\textwidth]{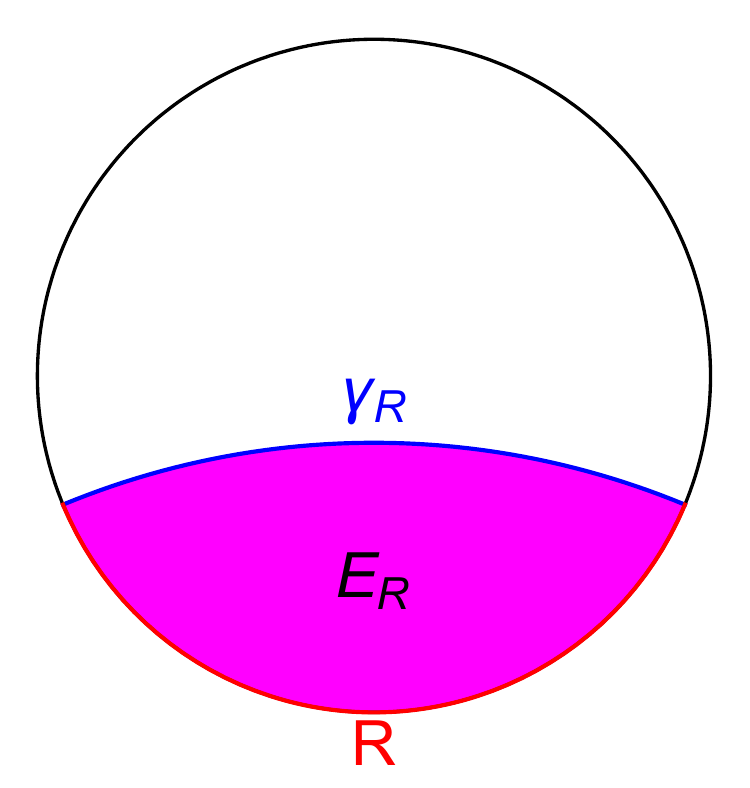}
            \caption[]%
            {{\small }}    
        \end{subfigure}
        \hfill
        \begin{subfigure}[b]{0.3\textwidth}   
            \centering 
			\includegraphics[width=\textwidth]{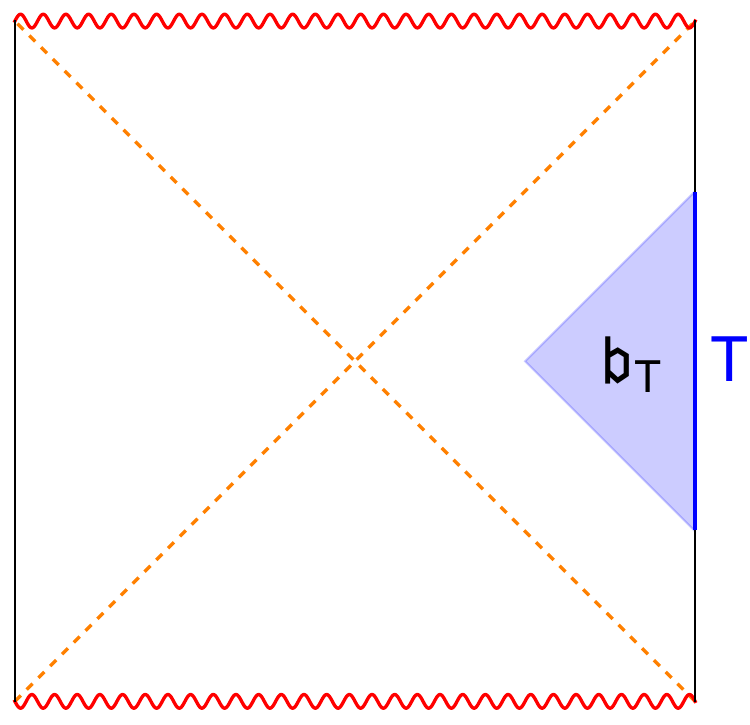}
            \caption[]%
            {{\small }}    
        \end{subfigure}
        \vskip\baselineskip
        \begin{subfigure}[b]{0.3\textwidth}
            \centering
			\includegraphics[width=\textwidth]{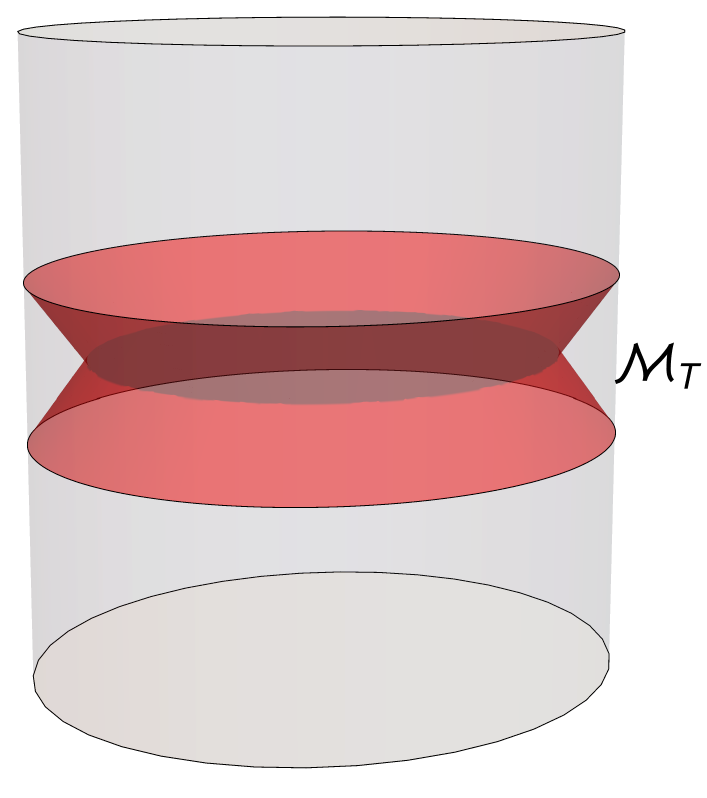} 
            \caption[]%
            {{\small }}    
        \end{subfigure}
        \hfill
        \begin{subfigure}[b]{0.3\textwidth}   
            \centering 
			\includegraphics[width=\textwidth]{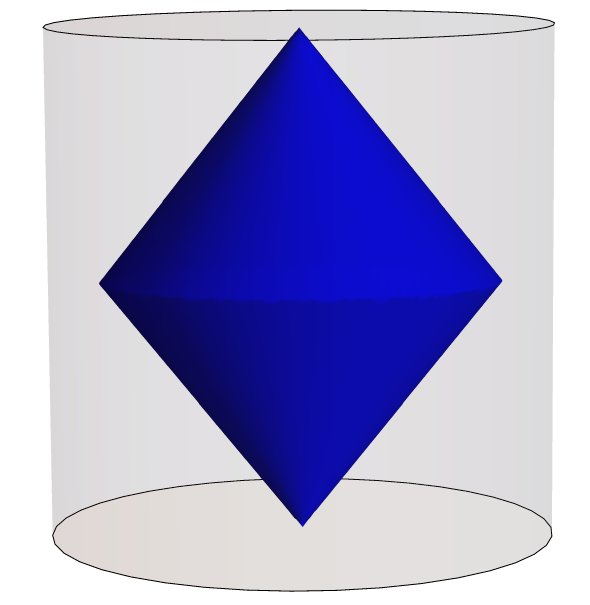}
            \caption[]%
            {{\small }}    
        \end{subfigure}
        \caption[  ]
        {\justifying\small Some examples of subregion-subalgebra duality. (a) The right region of an eternal black hole (shaded) is dual to the single-trace operator algebra $\sM_R$ of CFT$_R$ (i.e. the one defined on the right boundary) in a thermal field double state.
 (b) Duality between the entanglement wedge of a boundary spatial region $R$ and an algebra defined in $\fb_R$ (only a time slice of the bulk is shown).
 (c) A wedge region (shaded) in an eternal black hole geometry dual to the boundary subalgebra  associated with a time band indicated by $T$ in the CFT$_R$ in a thermal field double state.  (d) A radial wedge region (between the red surfaces and the conformal boundary) in global AdS dual to the boundary subalgebra $\sM_T$ associated with a time band $T$ for the CFT in the vacuum state. (e) A causal diamond in global AdS that does not touch the boundary is dual to the commutant of $\sM_T$ where $T$ is a time band. } 
\label{fig:exam}
\end{figure}

The correspondence~\eqref{he2}--\eqref{he1} gives a formulation of the duality for a general local bulk spacetime subregion, to which we will refer as subregion-subalgebra duality. See Fig.~\ref{fig:exam} for some examples. We stress that this duality for a bulk subregion can be precisely formulated {\it only} in the $1/N$ perturbation theory.
Nevertheless, it should have important implications for understanding the bulk theory at a small but finite $G_N$ (dual to a boundary theory at a large finite $N$). An analogy is provided by a phase transition in statistical physics, which can be sharply defined only in the thermodynamic limit, but does control essential physics at a large but finite volume.

Subregion-subalgebra duality generalizes the previously formulated subregion-subregion duality~\cite{VanRaamsdonk:2009ar,Czech:2012bh,Czech:2012be,Wall:2012uf,Lewkowycz:2013nqa,Faulkner:2013ana,Headrick:2014cta,Almheiri:2014lwa,Jafferis:2014lza,Pastawski:2015qua,Jafferis:2015del,Hayden:2016cfa,Dong:2016eik,Harlow:2016vwg,Faulkner:2017vdd,Cotler:2017erl}. 
The Ryu-Takayanagi (RT) formula and its covariant generalization~\cite{Ryu:2006bv,Hubeny:2007xt}  say that
\be \label{hrn}
S (\rho_R)= \frac{{\rm Area}(\ga_R)}{4\hbar G_N} \ ,
\ee
where $\rho_R$ is the reduced density matrix associated with a {spatial} boundary region $R$ with $S (\rho_R)$ the corresponding von Neumann entropy. 
$\ga_R$ is the minimal surface which ends on $\p R$ on the boundary,  see Fig.~\ref{fig:exam}(b).
Various arguments have suggested that bulk physics in the corresponding entanglement wedge $\hat E_R$, i.e. the causal completion of the region $E_R$ between $\ga_R$ and $R$, is equivalent to {boundary physics} in $R$.  

\begin{figure}[!h]
        \centering
        \begin{subfigure}[b]{0.45\textwidth}
            \centering
			\includegraphics[width=6.5cm]{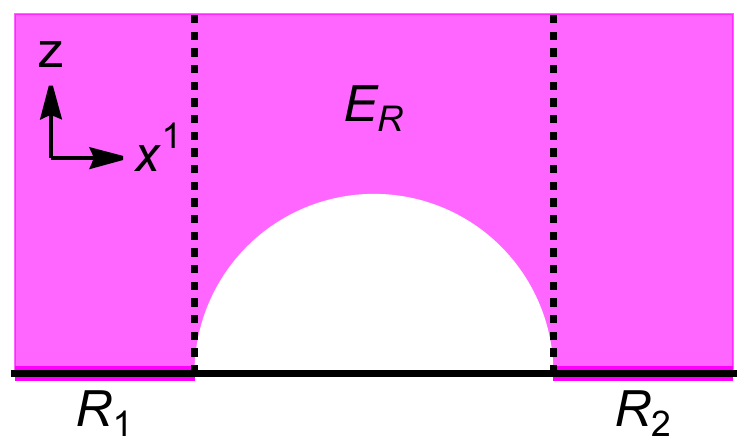} 
            \caption[]%
            {{\small }}    
        \end{subfigure}
        \hfill
        \begin{subfigure}[b]{0.45\textwidth}   
            \centering 
			\includegraphics[width=4.5cm]{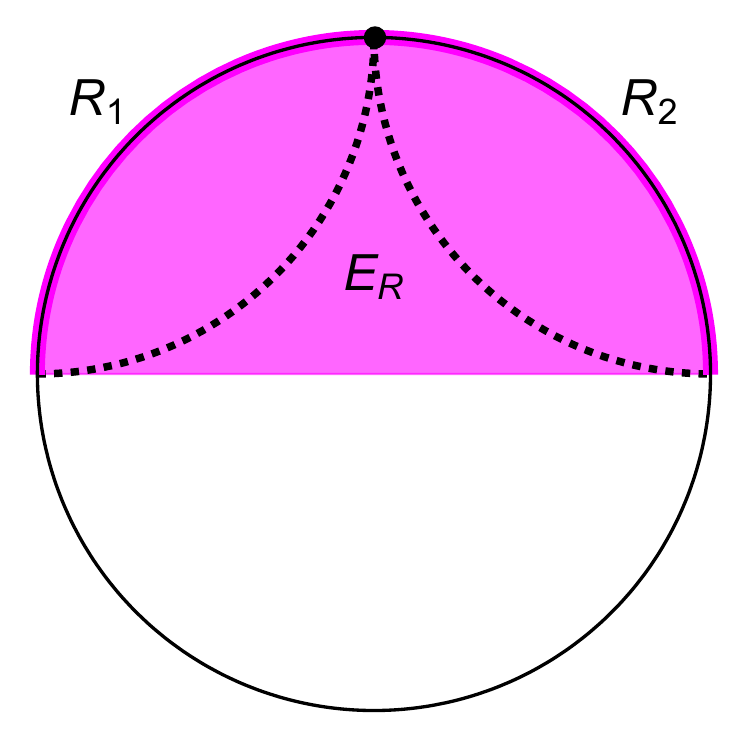}
            \caption[]%
            {{\small }}    
        \end{subfigure}
        \caption[  ]
        {\justifying\small Starting with the algebra $\sX_R$ for the boundary region $R= R_1 \cup R_2$ (where $R_{1},~R_2$ are half-lines separated from each other by an interval) in the vacuum state, it is possible (see Sec.~\ref{sec:expD}) to reconstruct the bulk spacetime region dual to $\sX_R$, i.e. the entanglement wedge of $R,$ using equivalence of algebras alone. The shaded regions are time-slices of the reconstructed entanglement wedges, and the union of the regions bounded by dashed lines gives the causal domain.  (a): a single time slice in the Poincare patch.
The region outside the dashed lines is the causal wedge. (b) a single time slice in global AdS. The union of regions between the dashed lines and the boundary is the causal domain. } 
\label{fig:etec}
\end{figure}

Even for cases like Fig.~\ref{fig:exam} (b), the prototypical situation of subregion-subregion duality, the subregion-subalgebra formulation~\eqref{he2}--\eqref{he1} brings  new insights. It provides a mathematically precise definition of subregion-subregion duality by identifying the bulk operator algebra in $\hat E_R$ with a boundary type III$_1$ subalgebra, which we will denote as $\sX_R$.  We propose that $\sX_R$ can be defined as the large $N$ limit of 
$\sB_R^{(N)}$, the operator algebra of the region $R$ in the boundary CFT at a finite $N$, i.e. 
\be\label{egq}
\sX_R = \pi_\Psi( \lim_{N \to \infty, \Psi} \sB_R^{(N)})  
\ee
where here $\Psi$ is the boundary state dual to the bulk geometry. The precise definition of the limit as well as the notation $\pi_\Psi$  will be explained in detail later in Sec.~\ref{sec:subre}. Here it is enough to note that the large $N$ limit depends on the state $\Psi$ under consideration. 
{Conversely, one can use $\sX_R$ of~\eqref{egq} to provide an alternative definition of the entanglement wedge and the associated 
RT surface without using entropy, by instead directly identifying the bulk region whose operator algebra is equivalent to $\sX_R$.} As a nontrivial explicit example, we are able to reconstruct the bulk entanglement wedge in an example  where the entanglement wedge exceeds the causal wedge. See Fig.~\ref{fig:etec}. 

\begin{figure}[!h]
\begin{center}
\includegraphics[width=4.5cm]{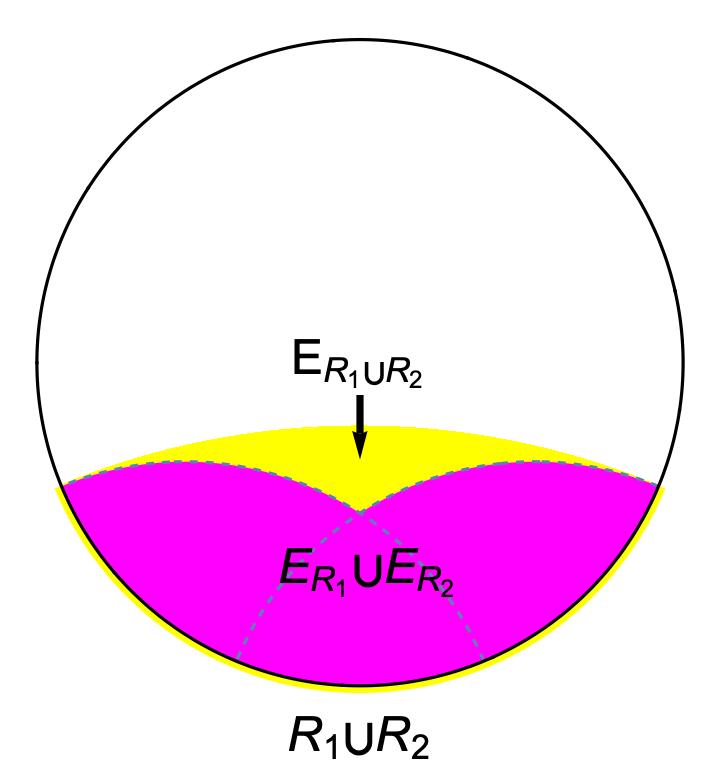}\qquad \quad\quad
\includegraphics[width=4.5cm]{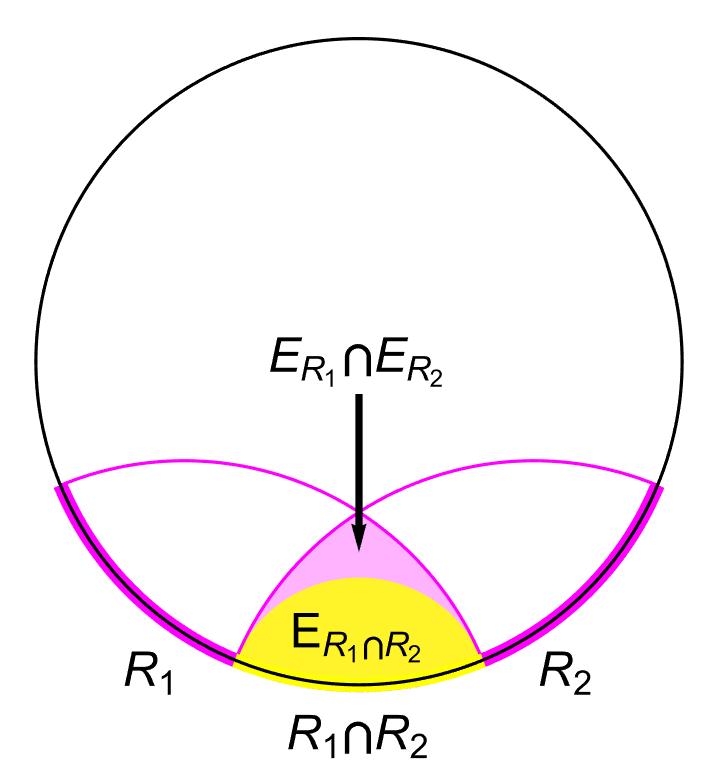}
\caption[]{\justifying\small The superadditivity~\eqref{hen03} is realized on the gravity side through geometric properties of entanglement wedges,  i.e. $E_{R_1} \cup E_{R_2} \subseteq E_{R_1 \cup R_2}$ and $E_{R_1 \cap R_2} \subseteq E_{R_1} \cap E_{R_2}$. 
These properties  played important roles in interpreting the bulk gravity system as a quantum error correcting code. }
\label{fig:error}
\end{center}
\end{figure}

A very interesting consequence of the limit~\eqref{egq} is an emergent superadditivity: while we expect $\sB_R^{(N)}$ to obey additivity, i.e. (with $R_{1,2}$ subregions of the same Cauchy slice)
\be 
\sB_{R_1}^{(N)} \lor \sB_{R_2}^{(N)} =  \sB_{R_1 \cup R_2}^{(N)}, \qquad \sB_{R_1}^{(N)} \land \sB_{R_2}^{(N)} =  \sB_{R_1 \cap R_2}^{(N)} ,
\ee
$\sX_R$ in general does not, i.e.\footnote{The fact that the algebras dual to entanglement wedges should not be additive was pointed out earlier in~\cite{Casini:2019kex}, though the interpretation there in terms of superselection sectors differs from our interpretation in terms of the large $N$ limit. (See also~\cite{Benedetti:2022aiw} for a recent discussion) } 
\be \label{hen03}
\sX_{R_1} \lor \sX_{R_2} \subseteq \sX_{R_1 \cup R_2} , \qquad \sX_{R_1 \cap R_2}   \subseteq  \sX_{R_1} \land \sX_{R_2} \ .
\ee
The superadditivity property~\eqref{hen03} underlies geometric properties of the corresponding entanglement wedges. 
See Fig.~\ref{fig:error} for an illustration. Fig.~\ref{fig:etec} is another example where $R_1 \cap R_2 = \emptyset$. 

Subregion-subregion duality has been interpreted in terms of quantum error correction since information about certain bulk operators can be recovered with only partial information in the boundary theory. Furthermore, the geometric properties of entanglement wedges in Fig.~\ref{fig:error} have been interpreted in terms of error correcting properties~\cite{Almheiri:2014lwa,Harlow:2016vwg}. Subregion-subalgebra duality and superadditivity give an explanation
of their origin (see also~\cite{Mintun:2015qda,Freivogel:2016zsb} for previous discussions).\footnote{For example, the theorems discussed in~\cite{Almheiri:2014lwa,Harlow:2016vwg} have the form of an equivalence of various statements, but it is not clear from those discussions that any of the statements actually applies to holographic systems.} 
Superadditivity implies that, in the large $N$ limit, $\sX_R$ is not locally generated on the boundary.\footnote{{For an interesting discussion of additivity violation in the context of generalized symmetries in QFT see~\cite{Casini:2020rgj}.} } This delocalization of quantum information is a crucial ingredient in the quantum error correction interpretation, and underlies 
various other quantum informational aspects of the holographic duality, which will be discussed elsewhere.

\begin{figure}[!h]
\begin{center}
\includegraphics[width=4.5cm]{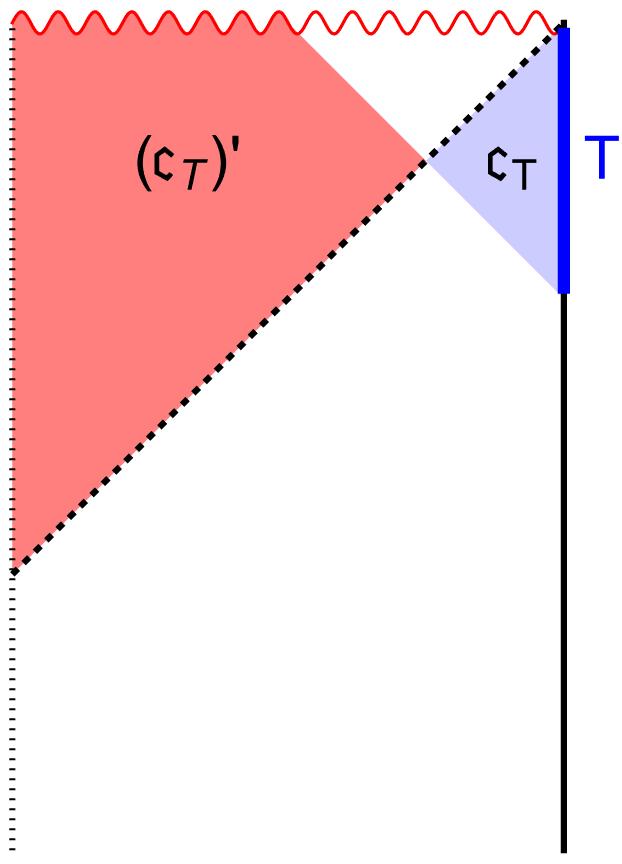}
\caption[]{\justifying\small For the boundary CFT in a state dual to a bulk single-sided black hole. The operator algebra $\sM_T$ associated with a half-infinite time band $T$ {should be} dual to the shaded bulk wedge region $\fc_T$. The commutant of $\sM_T$ gives a precise definition of mirror operators advocated for in~\cite{Papadodimas:2012aq, Papadodimas:2013jku}, and is dual to the shaded bulk region labeled by $(\fc_T)'$. }
\label{fig:sBH}
\end{center}
\end{figure}

Subregion-subregion duality covers the situations in Fig.~\ref{fig:exam} (a) and (b), but not those in (c)--(e). 
Using type III$_1$ subalgebras on the boundary rather than causally complete geometric subregions makes it possible to find duals for general bulk spacetime regions. A particularly interesting example is the causal diamond region of Fig.~\ref{fig:exam}(e), which does not touch the boundary. The corresponding dual subalgebra has no direct geometric description in the boundary theory. 
Applying subregion-subalgebra duality to a boundary state describing a single-sided black hole also provides a precise way to define the
mirror operators postulated in~\cite{Papadodimas:2012aq,Papadodimas:2013jku}. See Fig.~\ref{fig:sBH}.

In this paper we work to the leading order in the $1/N$ expansion, when the boundary theory is described by a generalized free field theory. 
The qualitative picture should survive to any finite order in the $1/N$ perturbative expansion.\footnote{When all $1/N$ corrections are included, it has been argued in~\cite{Witten:2021unn} that the type III$_1$ algebras become type II$_\infty$. The type II description leads to new ways of understanding black hole and de Sitter entropies~\cite{Witten:2021unn, Chandrasekaran:2022cip,Chandrasekaran:2022eqq}.} 
There is also a very interesting recent paper~\cite{Faulkner:2022ada} which discusses the embedding of bulk subalgebras in the $G_N \to 0$ limit to the boundary theory at a finite $N$, which in our language corresponds to the embedding of $\sX_R$ defined in~\eqref{egq} into $\sB_R^{(N)}$  at a finite $N$. Since the bulk and boundary theories are not in the same parameter regime, the embedding is not part of the duality, but provides an important alternative probe of the limit in~\eqref{egq}. 

The plan of the paper is as follows. In Sec.~\ref{sec:largN} we highlight various features of the AdS/CFT duality in the large $N$ limit that are important for subsequent discussions.  In Sec.~\ref{sec:sub} we describe our general setup and give an explicit formulation of~\eqref{he2}--\eqref{he1}. In Sec.~\ref{sec:subre} subregion-subregion duality is discussed in detail from the new algebraic perspective. In particular, we provide a boundary {\it derivation} of the superadditivity~\eqref{hen03}, and derive 
entanglement wedges for some choices of boundary regions---including that of Fig.~\ref{fig:etec}---from equivalence of algebras.
In Sec.~\ref{sec:time} we consider some more general examples of subregion-subalgebra duality for local bulk spacetime regions. We conclude with a discussion of some future directions in Sec.~\ref{sec:conc}.

\noindent{\bf Notational Conventions}
\newline We will denote boundary spatial subregions by captial Latin letters, i.e. $I, O, R, S,...$ We only discuss subregions of bulk Cauchy slices in connection to the entanglement wedge or causal domain of a boundary subregion. Such bulk spatial subregions are denoted by capital Latin letters with subscripts identifying the corresponding boundary subregion. We denote a homology hypersurface associated to the entanglement wedge of a boundary spatial subregion $R$ by $E_R.$ $E_R$ is a piece of a bulk Cauchy slice. We denote general bulk spacetime subregions by gothic letters $\fa, \fb, ...$ The entanglement wedge of a boundary spatial subregion $R$ is denoted by $\fb_R$ and the causal domain by $\fc_R.$

We will use apostrophes to denote causal complements, taken in the same spacetime in which the original region is defined. For example, $R'$ is the boundary causal complement of the boundary subregion $R,$ while $\fa'$ is the bulk causal complement of the bulk subregion $\fa.$ We use a bar to denote complements on a spatial slice. For example, $\bar R$ is the {\it spatial} subregion that is the complment to $R$ on a boundary Cauchy slice, while $\overline{E_R}$ is bulk spatial subregion that is the complement to $E_R$ on a bulk Cauchy slice.

We use hats to denote causal completions. For example, $\hat R = R''$ denotes the boundary causal completion of $R,$ while $\hat \fa = \fa ''$ denotes the bulk causal completion of $\fa.$

\section{AdS/CFT duality in the large $N$ limit} \label{sec:largN}

To set the stage for formulating the duality~\eqref{he2}--\eqref{he1}, we first
elaborate on various aspects of the AdS/CFT duality in the large $N$ limit that will be important later. 

In the AdS/CFT duality we have a bulk quantum gravity system in AdS$_{d+1}$ dual to a CFT$_d$ on $\RR\times S_{d-1}$.  
The number of degrees of freedom of the CFT$_d$ is characterized by a parameter $N$, related to the bulk Newton constant $G_N$ as  $G_N \sim {1 \ov N^2}$. While our discussion is general, it is useful to keep in mind some explicit examples, such as the duality between 
the $\sN =4$ Super-Yang-Mills~(SYM) theory with gauge group $SU(N)$ on $\RR \times S_3$ and 
the IIB superstring in AdS$_5 \times S_5$. In our explicit calculations, for technical simplicity, we will mostly work with $d=2$, with the boundary CFT$_2$ of central charge $c \propto N^2$ defined on a cylinder $\RR \times S_1$.

At a finite (but large) $N$, the boundary theory has a Hilbert space $\sH_{\rm CFT}$, which is identified with 
that of the bulk quantum gravity theory 
\be \label{hidd}
\sH_{\rm bulk} = \sH_{\rm CFT} ,
\ee
in the sense that there is a one-to-one mapping between quantum states of the bulk and boundary theories. 
Accordingly, there is also a one-to-one correspondence between boundary and bulk operator subalgebras acting on the above Hilbert space, 
and the correspondence is {\it state-independent}. For example, consider a local boundary subregion $R$, with an associated type III$_1$  local operator subalgebra  $\sB_R^{(N)}$. There must be a corresponding type III$_1$ algebra $\tilde \sB_R^{(N)}$ on the gravity side. 
Being state-independent, $\tilde  \sB_R^{(N)}$ cannot be associated with any bulk geometry. Furthermore,  $\sB_R^{(N)}$  contains very heavy operators with dimension $\De \sim N^2$ that  can create ``black holes.''\footnote{We put black holes in quotes as at a finite $G_N$, we do not really know how to define a black hole in precise terms.}  
Without a full quantum theory of gravity, there is little we could say in a precise manner about the mapping of states and subalgebras at a finite $N$.  

In the large $N$ limit, only some of the states and operators of the finite $N$ theory survive, which makes the structures of the Hilbert space and operator algebras very different~\cite{shortPaper, longPaper,Witten:2021jzq,Witten:2021unn,Schlenker:2022dyo,Chandrasekaran:2022cip,Chandrasekaran:2022eqq,Faulkner:2022ada}. In particular, the ubiquitous emergence of type III$_1$ subalgebras in the large $N$ limit will be a main theme in this paper. These emergent type III$_1$ structures are not related to the type III$_1$ nature of $\sB_R^{(N)}$ at a finite $N$. {In fact, to make the story sharper we could put the boundary theory on a spatial lattice, so that $\sB_R^{(N)}$ is type I, and the type III$_1$ structure would still emerge.}

For the convenience of later discussions, below we elaborate in detail the structures of the boundary and bulk theories in this limit, and the duality between them.

\subsection{Boundary theory} \label{sec:bdryThy}

We denote the full operator algebra of the CFT at a finite $N$ as $\sB^{(N)}$.
We say an operator has a sensible large $N$ limit if: (i) it can be defined for sufficiently large $N$; (ii)
its vacuum correlation functions have a well-defined $N \to \infty$ limit.  For the $\sN =4$ SYM theory with 
the Hamiltonian normalized as $H = N \Tr (\cdots)$, from the standard large $N$ scaling, these are operators generated by finite products of single-trace operators of the form $\sO = \Tr (\cdots)$. 
We denote the algebra generated from products of single-trace operators by $\sA_\Om$.
It can be viewed as an abstract $C^*$-algebra, {with norm inherited from a finite $N$ theory}.
We note that $\sA_\Om$ is not a subalgebra of $\sB^{(N)}$ as $\sA_\Om$ is only strictly defined in the large $N$ limit.\footnote{Previous attempts to understand low energy excitations about a fixed bulk background have defined a {\bf set} of low-energy observables that can be defined at finite $N$~\cite{Papadodimas:2012aq, Papadodimas:2013jku}. In that case, the low-energy observables do not form a closed algebra. }

An elementary point is that single-trace operators 
at different times are independent, in the sense that they cannot be expressed in terms of one another. 
 Consider, for example, a single-trace operator $V = \Tr F_{\mu \nu} F^{\mu \nu}$ in the $\sN =4$ SYM theory. 
At finite $N$, we can express it in terms of operators at $t=0$,  
\bea \label{ehn}
V (t) &=& e^{i H t} V (0) e^{- i H t} \\
&=& \sum_i c_i (t) O_i (0) 
\label{ehn1}
\eea
where the second line can (in principle) be obtained by solving the Heisenberg equation, with $\{O_i\}$ denoting the full set of operators and $\{c_i (t)\}$ being some numbers independent of $N$.   As $N \to \infty$, while $V(t)$ is well-defined through the limit of its correlation functions, neither equation~\eqref{ehn} nor~\eqref{ehn1} survives the limit.\footnote{If we work about a time-translation invariant state such as the vacuum, then a version of~\eqref{ehn} survives the large $N$ limit with $H$ being replaced by the ``charge'' $h$ generating time translations in the GFFs. While $H$ in~\eqref{ehn} is defined on a single boundary time slice, $h$ is an operator defined on the full boundary spacetime and thus cannot be used to generate equations of motion.
}  
This is because the Hamiltonian $H$ has an explicit $N$ dependence in its definition, and  the set of $O_i$'s appearing in~\eqref{ehn1} contains not just single-trace operators, but also more complicated operators that do not survive the large $N$ limit.

The loss of the relations \eqref{ehn}--\eqref{ehn1} in the large $N$ limit has some immediate implications, which have played crucial roles in the discussion of~\cite{shortPaper, longPaper} and will continue to do so in this paper: 

\ben 

\begin{figure}[!h]
        \centering
        \begin{subfigure}[b]{0.45\textwidth}
            \centering
			\includegraphics[width=4.5cm]{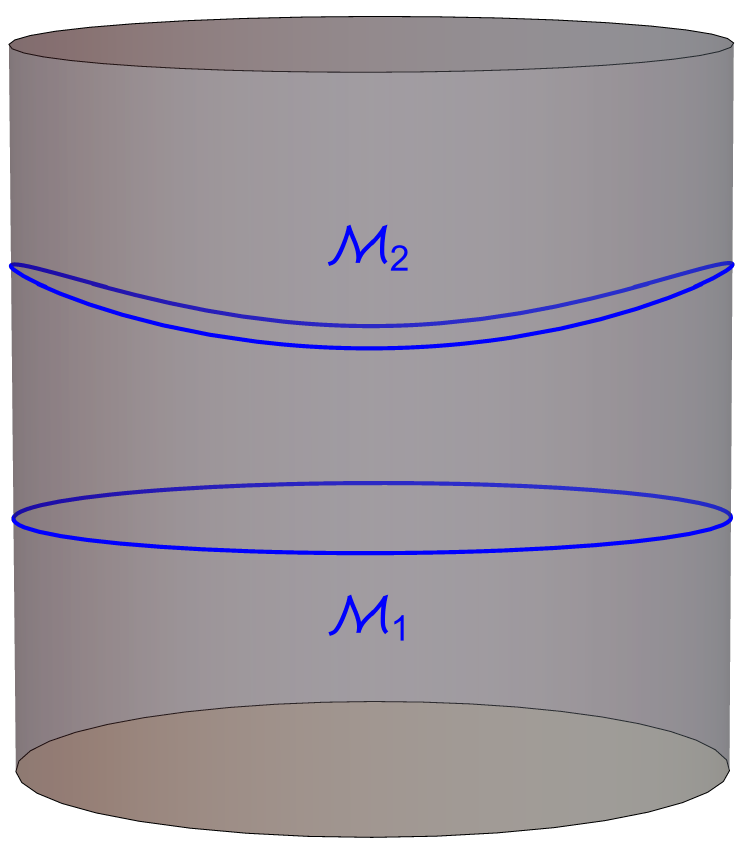} 
            \caption[]%
            {{\small }}    
        \end{subfigure}
        \hfill
        \begin{subfigure}[b]{0.45\textwidth}   
            \centering 
			\includegraphics[width=4.5cm]{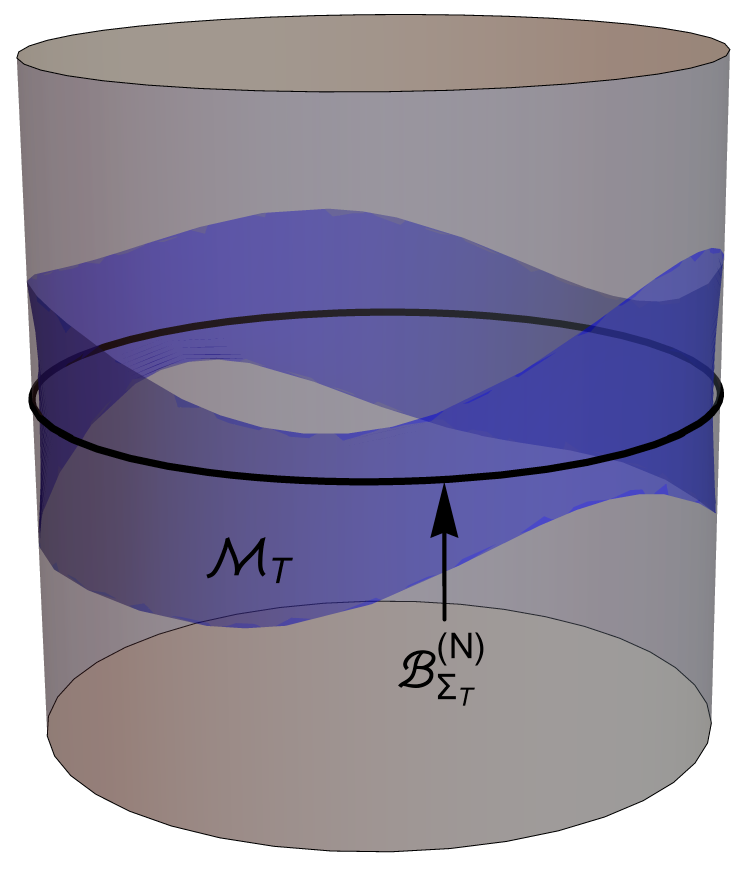}
            \caption[]%
            {{\small }}    
        \end{subfigure}
        \caption[  ]
        {\justifying\small (a) Algebras $\sM_1$ and $\sM_2$ of single-trace operators on two different Cauchy slices are inequivalent. (b) 
At finite $N$, the algebra associated with the time band indicated in the plot is equivalent to that on a single Cauchy slice. But in the large $N$ limit, it is inequivalent and is a nontrivial subalgebra of the full algebra.  } 
\label{fig:ineqCauchyAlgs}
\end{figure}

\item Single-trace operator algebras associated with different Cauchy slices cannot be expressed in terms of one another, and thus are inequivalent. See Fig.~\ref{fig:ineqCauchyAlgs} (a).

\item The algebra of single-trace operators includes operators at all boundary times, i.e. they already contain some information about the boundary time evolution for all times.

\item There are (infinitely) many more subalgebras at large $N$ that are not present at a finite $N$. For example, there are distinct {vN} subalgebras associated to {almost all} distinct {\it spacetime} subregions, even though the algebras of many of these regions would be equivalent in the finite $N$ theory. See Fig.~\ref{fig:ineqCauchyAlgs} (b). There are also subalgebras which do not have 
boundary geometric definitions. For example, the commutant of the subalgebra associated with the time band in Fig.~\ref{fig:ineqCauchyAlgs} (b) does not have any direct geometric definition.

\een

Now we turn to the large $N$ limit of $\sH_{\rm CFT}$. We will first consider pure states.  We say a state $\ket{\Psi}$ has a sensible large $N$ limit, if there exists a sequence of states, $\{\ket{\Psi^{(N)}}\},$ such that
correlation functions of single-trace operators (with their expectation values subtracted) in the states
have a well-defined $N \to \infty$ limit, e.g.
\be \label{twio}
\lim_{N \to \infty} \vev{\Psi^{(N)} |\hat \sO_1 \hat \sO_2 \cdots |\Psi^{(N)}} = {\rm finite}, \quad  \hat \sO_i \equiv \sO_i - \vev{\Psi^{(N)}|\sO_i|\Psi^{(N)}} .
\ee
Above each $\hat \sO_i$ should be understood to depend on $N$, which we have suppressed. 
$\ket{\Psi}$ is defined as the large $N$ limit of $\{\ket{\Psi^{(N)}}\}$ through the limits of correlation functions~\eqref{twio}. 
By definition the vacuum $\ket{\Om}$ has a sensible limit, but a typical state with energy $E - E_\Om \sim O(N^2) \to \infty$ may not. 

We further define a {\it semi-classical} state as one in which 
correlation functions~\eqref{twio} factorize into products of two-point functions of single-trace operators at the leading order in the large $N$ expansion.  
For a semi-classical state $\ket{\Psi}$, we use $\sO_\Psi$ to denote the $N \to \infty$ limit of $\hat \sO$,\footnote{Note that from the standard large $N$ scaling we expect $\vev{\Psi^{(N)}|\sO|\Psi^{(N)}} \sim O(N)$ as $N \to \infty$.} 
which by definition satisfies $\vev{\Psi|\sO_\Psi|\Psi} =0$. For notational simplicity, we will often suppress the subscript $\Psi$ in $\sO_\Psi$, which should be understood from the context.
We denote the algebra generated by the $\sO_\Psi$'s
as $\sA_\Psi$. Clearly for $\ket{\Psi} = \ket{\Om}$, $\sA_\Psi$ coincides with $\sA_\Om$ defined earlier.
While we have been using the notation for a pure state, the discussion can also be applied to a mixed state. The vacuum and the thermal state are known to be semi-classical. More generally, any state dual to a bulk semi-classical geometry should be a semi-classical state in the sense of factorization of correlation functions.

We
can build a Hilbert space $\sH_{\Psi}^{\rm GNS}$ ``around'' a semi-classical state  $\ket{\Psi}$ by acting products of single-trace operators on $\ket{\Psi}$,  via a procedure called the Gelfand-Naimark-Segal (GNS) construction~\cite{gelfandNaimark, segal}. More explicitly, we introduce a vector $\ket{A}$ for each operator $A$ in the algebra $\sA_\Psi$ of single-trace operators, and define an inner product between these vectors as
\be \label{hen}
\vev{A|B} = \vev{\Psi| A^\da B |\Psi} , \qquad A, B \in \sA_\Psi \ .
\ee 
Equation~\eqref{hen} does not yet define a Hilbert space as there can be  zero-norm vectors from operators $X \in \sA_\Psi$ that satisfy $\vev{\Psi|X^\da X|\Psi} =0$. The set of such operators will be denoted as $\sJ$.
To eliminate the zero-norm vectors, we introduce equivalence classes $[A]$ by the equivalence relation 
\be 
A \sim A + X , \quad A \in \sA_\Psi, \quad X \in \sJ  
\ee
and associate $\ket{A}$ instead with each $[A]$. The state $\ket{\bid} \equiv \ket{0}_{\Psi, {\rm GNS}}$ associated with the equivalence class $[\bid]$ of the identity operator is also referred to as the GNS vacuum. The GNS Hilbert space $\sH_\Psi^{\rm GNS}$ is then the completion of 
the set $\{\ket{A}\}$. The representation $\pi_{\Psi} (A)$ of an operator $A \in \sA_\Psi$ in $\sH^{\rm GNS}_\Psi$ is defined as 
\be
\pi_{\Psi} (A) [B] = [AB], \qquad A, B \in \sA_\Psi \ .
\ee
We will denote the representation of $\sA_\Psi$ on $\sH_\Psi^{\rm GNS}$ by $\sM_{\Psi} = \pi_{\Psi}(\sA_{\Psi}).$

Given the factorization property of~\eqref{twio} in a semi-classical state, 
any inner product~\eqref{hen} reduces to sums of products of two-point functions of single-trace operators 
at the leading order in $1/N$ expansion, which implies that $\sH^{\rm GNS}_\Psi$ can be associated with a Gaussian theory 
of single-trace operators. Since single-trace operators at different times are independent, it is a generalized free field theory with each single-trace operator being a generalized free field.\footnote{For earlier discussion of generalized free fields in the large $N$ limit, see e.g.~\cite{Duetsch:2002hc, El-Showk:2011yvt}.} More explicitly, we can expand the representation of a single-trace operator $\sO_\Psi$ as 
 \bega \label{hnbl}
 \pi_\Psi (\sO_\Psi (x))  
  = 
  \sum_n \le(v_n (x) a_n +  v_n^* (x) a_n^\da \ri)  \\
  \label{hnbl1} [a_m, a_n^\da] = \de_{mn}, \quad [a_m, a_n] = [a_m^\da, a_n^\da] = 0, \quad  a_{n} \ket{0}_{\Psi, {\rm GNS}} = 0  \ , 
 \end{gather} 
where $\{v_n (x)\}$ is a complete set of mode functions in the boundary {\it spacetime} 
chosen so that
the Wightman function of $\sO_\Psi$ is recovered 
\be 
\vev{\Psi|\sO_\Psi  (x) \sO_\Psi  (0)|\Psi} = {_{\Psi, {\rm GNS}}}\vev{0|\pi_{\Psi} (\sO_\Psi (x)) \pi_{\Psi} (\sO_\Psi (0))|0}_{\Psi, {\rm GNS}} 
\ .
\ee
We stress that $\pi_{\Psi} (\sO_\Psi )$ is defined only  in $\sH^{\rm GNS}_\Psi$ and 
is state-dependent, while $\sO$ is state-independent. $\sH^{\rm GNS}_\Psi$ is generated by acting $\{a_n^\da\}$ (and creation operators from other single-trace operators) on $ \ket{0}_{\Psi, {\rm GNS}}$.

\begin{figure}[!h]
\begin{center}
\includegraphics[scale=0.45]{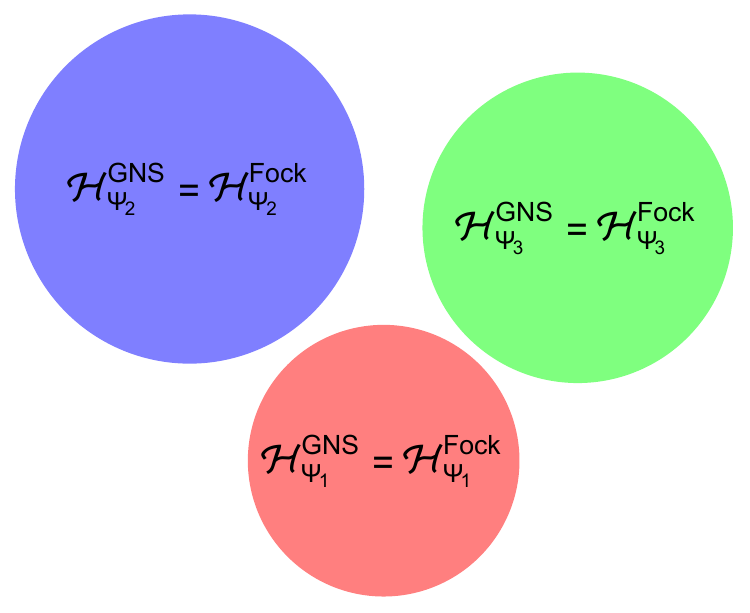} 
\end{center}
\caption[]{\justifying\small In the large $N$ limit, the full Hilbert space of the boundary CFT separates into disjoint ``continents'' of GNS Hilbert spaces around semi-classical states. As we will describe below the same structure exists on the gravity side in the $G_N \to 0$ limit, 
and there is a one-to-one correspondence between the continents.
}
 \label{fig:hilg}
\end{figure}

By definition, any product of single-trace operators has dimension $O(N^0)$ and as a result a state $\ket{\Psi'} \in \sH_{\Psi}^{\rm GNS}$ has energy $E_{\Psi'} - E_{\Psi} \sim O(N^0)$.  $\sH_{\Psi}^{\rm GNS}$ may be interpreted as the space of  ``small'' excitations around $\ket{\Psi}$. Furthermore, given the factorization property of $\ket{\Psi}$, $\ket{\Psi'}$, which is obtained by acting $a^\da_n$'s on $\ket{\Psi}$, does not satisfy the factorization property. Thus 
two semi-classical states $\ket{\Psi_1}$, $\ket{\Psi_2}$ cannot be in each other's GNS Hilbert space, i.e. their respective GNS Hilbert spaces $\sH_{\Psi_1}^{\rm GNS}$, $\sH_{\Psi_2}^{\rm GNS}$ do not overlap. In the large $N$ limit, the full space of states then has the structure of ``disjoint continents'' around semi-classical states.
See Fig.~\ref{fig:hilg}. There is no operator that can take states from one GNS Hilbert space to another.\footnote{
While in a finite $N$ theory, there exists an operator that takes $\ket{\Psi_1^{(N)}}$ to $\ket{\Psi_2^{(N)}}$, this operator does not survive the  $N \to \infty$ limit, as the operators that survive (finite products of single-trace operators) act within each GNS Hilbert space.}
We also note that $\sH^{(\rm GNS)}_\Psi$ cannot be viewed as a subspace of the $\sH_{\rm CFT}$ as
$\sH^{(\rm GNS)}_\Psi$ is precisely defined only in the large $N$ limit.

In the discussion above we considered pure states. For a mixed state, 
we can purify it first and consider the resulting pure state in the doubled theory, as in the case of thermal ensemble (which becomes the thermal field double state). In fact, the GNS construction on a mixed state $\rho$ automatically provides such a purification.

While {the set of} single-trace operators can be defined abstractly 
without referring to any state~(except the vacuum), their algebra is state-dependent.
  Since the full state space splits into disjoint GNS Hilbert spaces in the large $N$ limit, the structure of {a large $N$} operator algebra depends on the specific GNS Hilbert space it acts on. 
As an example, consider two copies of a boundary CFT in the thermal field double state discussed in detail in~\cite{longPaper}. 
The structure of the subalgebra $\sM_R$ ($\sM_L$) associated with the set of single-trace operators $\sA_R$ ($\sA_L$) for the CFT$_R$ (CFT$_L$) depends on the temperature $T$. For $T < T_{HP}$ (with $T_{HP}$ the Hawking-Page temperature), $\sM_R$ is type I, but it becomes type III$_1$ for $T > T_{HP}$.

\subsection{Bulk theory} 

In the $G_N \to 0$ limit, quantum states for a gravity system 
 can be obtained by quantizing excitations around classical solutions of bulk equations of motion. 
More explicitly, consider a bulk classical solution $\chi_\Psi$ corresponding to some state $\ket{\Psi}$ of the quantum gravity theory, where $\chi$ collectively denotes all bulk fields including the metric.
 Quantizing small metric and matter perturbations around $\chi_\Psi$, with an 
 appropriately chosen ``vacuum'' state $\ket{0}_{\chi_\Psi}$,
 results in a Fock space $\sH_{\Psi}^{\rm Fock}$. At the leading order in the $G_N$ expansion we have a free quantum field theory
 in the curved spacetime specified by $\chi_\Psi$.  
 For example, a bulk field $\Phi (X)$ (with $X$ denoting a bulk point) can be expanded in terms of 
 a complete set of mode functions $\{u_n(X)\} $  as 
\bega \label{moEX}
\Phi (X) = \Phi_0 (X) + \sum_n \le(u_n (X) a_n + u_n^* (X) a_n^\da \ri), \\
 \label{adsV}
[a_m, a_n^\da] = \de_{mn}, \qquad [a_m, a_n] = [a_m^\da, a_n^\da] = 0 , \quad a_n \ket{0}_{\chi_\Psi} = 0, 
\end{gather} 
where $\Phi_0$ denotes the background value of $\Phi$ we are expanding around, and $n$ collectively denotes all quantum numbers including continuous ones. 
$\sH_{\Psi}^{\rm Fock}$ is generated by acting $a_n^\da$'s  on the vacuum state $\ket{0}_{\chi_\Psi}$.

The classical solution $\chi_\Psi$ together with the associated vacuum $\ket{0}_{\chi_\Psi}$ can be interpreted as giving the low energy description of the quantum state $\ket{\Psi}$ in the full theory. Similarly, an excited state $\ket{\psi'} \in \sH_{\Psi}^{\rm Fock}$ along with $\chi_\Psi$ corresponds to some {quantum} state $\ket{\Psi'}$ that differs from $\ket{\Psi}$ by some low energy excitations. 
Denoting the energy expectation value of $\ket{\Psi}$ as $E_\Psi$, we have 
\be \label{uhn1}
E_{\Psi'} = E_\Psi + \de E, \quad \de E \sim O(G_N^0) \sim O(N^0) , \quad G_N \to 0  \ ,
\ee 
since the canonically normalized quadratic action of small perturbations used to construct $\sH_{\Psi}^{\rm Fock}$ is independent of $G_N$. For two distinct classical solutions $\chi_{\Psi_1}$ and $\chi_{\Psi_2}$, their Fock spaces $\sH_{\Psi_1}^{\rm Fock}$ and $\sH_{\Psi_2}^{\rm Fock}$ do not overlap, as $\chi_{\Psi_1}$ cannot be obtained from $\chi_{\Psi_2}$ by small quantum fluctuations. The space of states thus separates into disjoint ``continents'' around different classical solutions, see Fig.~\ref{fig:hilg}.

\subsection{AdS/CFT duality at large $N$} \label{sec:duality}

Under the AdS/CFT duality, a semi-classical state $\ket{\Psi}$ {in the CFT} can be mapped to a bulk quantum state which has a low energy description in terms of a classical geometry $\chi_\Psi$ together with a vacuum state $\ket{0}_{\chi_\Psi}$ for matter and perturbative metric excitations. 
Single-trace operators are in one-to-one correspondence with elementary fields on the gravity side. 
We should then identify (see Fig.~\ref{fig:hilg})
\be
\sH_{\Psi}^{\rm Fock} = \sH_{\Psi}^{\rm GNS} , \quad \ket{0}_{\chi_\Psi} = \ket{0}_{\Psi, {\rm GNS}}  \ ,
\ee
which also imply that the set of creation/annihilation operators in~\eqref{hnbl} and~\eqref{moEX} must be equivalent and is the reason that we have used the same notation for them. 
With $\sM_\Psi$ the algebra of the single-trace operators on $\sH_{\Psi}^{\rm GNS}$,
and $ \widetilde{\sM}_{\chi_\Psi} $ the operator algebra of bulk fields on $\sH_{\Psi}^{\rm Fock}$, we should then have 
\be \label{globRecon}
\sM_\Psi = \widetilde{\sM}_{\chi_\Psi}  \ .
\ee
In particular, there should a one-to-one correspondence between subalgebras of $\sM_\Psi $ and $\widetilde{\sM}_{\chi_\Psi} $.

The extrapolate dictionary between bulk and boundary operators can be viewed as a manifestation of~\eqref{globRecon}, which also underlies global bulk reconstruction~\cite{Banks:1998dd,Bena:1999jv,Hamilton:2006az}.  We stress that the boundary operators in global reconstruction should be understood as those acting on $\sH^{\rm GNS}_\Psi$, i.e. in $\sM_\Psi$, not in $\sB$ or $\sA$.

An example of $\chi_\Psi$ is the empty AdS, with $\ket{\Psi} = \ket{\Om}$ being the vacuum state of the CFT,
  and $\ket{0}_{\rm AdS} \in \sH_{\Om}^{\rm Fock} $ being the vacuum of bulk perturbations around empty AdS. 
Other examples include the thermal field double state or any state which has a classical gravity  description.\footnote{Since here we are mainly concerned with the large $N$ limit, by classical gravity description we also include those geometries which may not be describable by the Einstein gravity, but still have a classical string theory description.}

\begin{figure}[!h]
\begin{center}
\includegraphics[width=10cm]{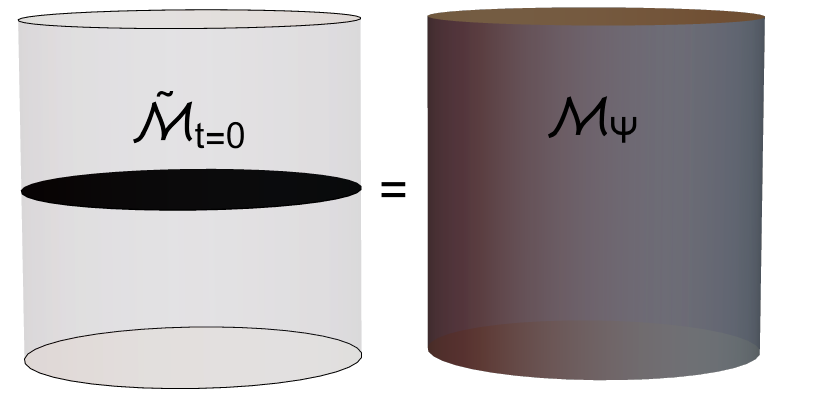}
\caption[]{\justifying\small The algebra of a bulk Cauchy slice is mapped to the algebra of single-trace operators on the entire boundary spacetime.}
\label{fig:time}
\end{center}
\end{figure}

To leading order in the $G_N \to 0$ limit, the bulk theory is that of free quantum fields on a (fixed) curved spacetime. Bulk fields obey bulk equations of motion and thus the algebra $\widetilde{\sM}_{\chi_\Psi}$ of bulk operators restricted to a single time slice is the full algebra of operators in the bulk. In contrast, $\sM_\Psi$ is defined on the full boundary spacetime.  See Fig.~\ref{fig:time}.
This implies that bulk time evolution should be viewed as an automorphism of $\sM_\Psi$, which can be used to study the boundary emergence of bulk times~\cite{shortPaper,longPaper}.

Bulk causality associates  distinct bulk subalgebras to distinct causally complete subregions of the bulk.
In particular,  we expect that {the bulk effective field theory} obeys additivity and Haag duality~\cite{Araki:1964}. Namely, for any bulk causally complete subregions $\fr_1,~\fr_2$ we should have 
\begin{align}
	\widetilde{\sM}_{(\fr_1 \cup \fr_2)''} &= \widetilde{\sM}_{\fr_1} \vee \widetilde{\sM}_{\fr_2} \\
	\widetilde{\sM}_{\fr_1'} &= \le(\widetilde{\sM}_{\fr_1}\ri)' \ ,
\end{align}
with $\fr'$ denoting the bulk causal complement of $\fr$ and $\le(\widetilde{\sM}_{\fr}\ri)'$ denoting the commutant of $\widetilde{\sM}_{\fr}.$  We will see these properties impose nontrivial constraints on the identifications of bulk subregions with boundary subalgebras.

\section{Formulation of subregion-subalgebra duality} \label{sec:sub}

In this section we introduce the formulation of subregion-subalgebra duality~\eqref{he2}--\eqref{he1}. 
The discussion of this section will be somewhat abstract; we leave more detailed discussions in specific cases to Sec.~\ref{sec:subre}--Sec.~\ref{sec:time}.

Consider a semi-classical state $\ket{\Psi}$ dual to some bulk geometry $\chi_\Psi$. 
The  single-trace operator algebra $\sM_\Psi$ in the GNS Hilbert space $\sH_{\Psi}^{\rm GNS}$  is identified with the bulk operator algebra $\widetilde \sM_{\chi_\Psi}$.
This identification implies that there is a one-to-one correspondence between subalgebras on two sides. 
Now consider a causally complete spacetime subregion $\fa$ in the bulk. The bulk operator algebra $\tilde \sY_\fa$ associated with $\fa$
is a type III$_1$ algebra in the $G_N \to 0$ limit. There must therefore be a corresponding type III$_1$ boundary subalgebra $\sY \in \sM_\Psi$ 
that can be identified with $\tilde \sY_\fa$, i.e.
\be 
\sY = \tilde \sY_\fa \ .
\ee
This implies that, in the large $N$ limit, $\sM_\Psi$ must be rich enough to contain all the subalgebras associated to local bulk subregions.

In Fig.~\ref{fig:exam} we gave some examples, which include subregion-subregion duality when the boundary type III$_1$ algebra $\sX_R$ is associated with a boundary spatial subregion $R$.
Other examples include the bulk regions that are dual to the boundary subalgebra associated with a time band, and also 
bulk regions that do not touch the boundary, which we will discuss in more detail in Sec.~\ref{sec:time}.

Here we make some general comments on how properties of a type III$_1$ boundary subalgebra translate into geometric properties of the corresponding bulk region, including local time evolutions and causal structure. For this purpose, we first briefly review the story of the example of Fig.~\ref{fig:exam}(a) where the $R$ region of an eternal black hole is dual to the  single-trace operator  subalgebra $\sM_R$  of CFT$_R$ in the thermal field double state: 

\ben 

\item Modular flow of $\sM_R$ can be identified as the Schwarzschild time evolution of  the $R$ region, which is an internal time flow, i.e. a time evolution which takes $\tilde \sM_R = \sM_R$ to itself. 

\item Kruskal-like times which can take an element of $\sM_R$ into the $F$ or $P$ regions of the black hole spacetime are generated by half-sided modular translations. They are obtained by identifying a subalgebra of $\sM_R$ that satisfies the half-sided modular inclusion properties.

\item The event horizon can be ``seen'' in the boundary theory from non-analytic behavior of an operator in $\sM_R$ under half-sided modular translations. 

\een

\begin{figure}[!h]
\begin{center}
\includegraphics[width=6cm]{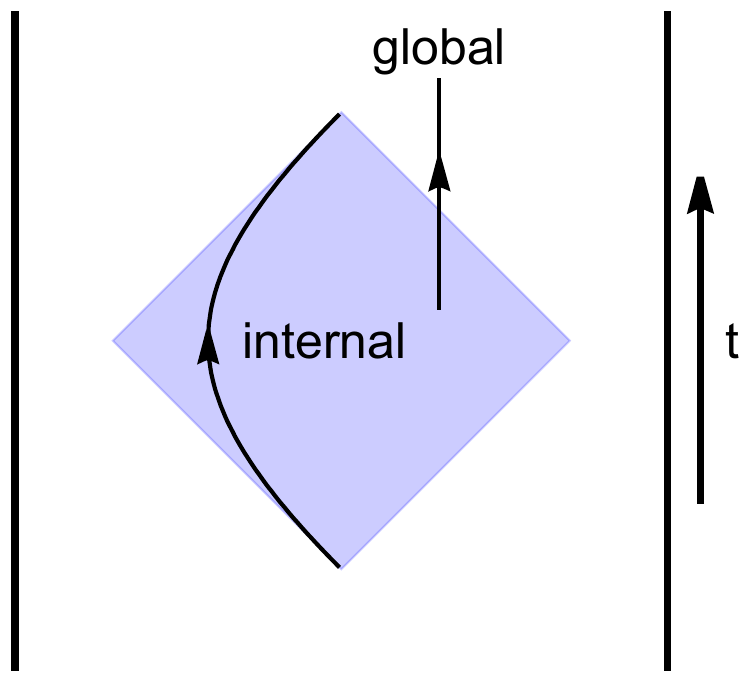}
\end{center}
\caption[]{\justifying\small Internal and global flows of a causally complete bulk spacetime region $\fb$ in the global AdS.}
\label{fig:bulkR}
\end{figure}

We believe that this example provides a general paradigm for understanding emergent local properties in the bulk. 
 More explicitly, 
consider a causally complete bulk subregion $\fb$, whose algebra $\tilde \sY_\fb$ is equivalent to some boundary subalgebra $\sY$. 
Then we should have (see Fig.~\ref{fig:bulkR}):

\ben 

\item The identification of the bulk operator algebra $\tilde \sY_{\fb}$ that is equivalent to $\sY$ {completely} determines the region $\fb$.

\item Modular flows of $\sY$ can be used to describe internal time flows in $\fb$ (though they may not be local).  

\item While it is not clear whether the half-sided modular inclusion structure exists for a general bulk region $\fb$, 
an analogous emergent ``global'' time  that can take one outside of $\fb$ may be generated by considering some appropriate subalgebra.

\item The light-cone boundary of $\fb$ and the corresponding causal structure can be ``seen'' in the boundary theory from 
non-analytic properties of the flow of an operator under 
the ``global time'' translations.

\een 
We mention that, except in some simple cases, it may be technically challenging to construct the dual algebra $\sY$ for a given $\fb$, as is finding the corresponding modular flows or half-sided modular translations if they exist. The above discussion provides a conceptual framework where such questions can be precisely formulated.

\section{Reformulation of subregion-Subregion duality} \label{sec:subre}

 In this section we give a reformulation of subregion-subregion duality~\cite{VanRaamsdonk:2009ar,Czech:2012bh,Czech:2012be,Wall:2012uf,Headrick:2014cta,Almheiri:2014lwa,Jafferis:2014lza,Pastawski:2015qua,Jafferis:2015del,Hayden:2016cfa,Dong:2016eik,Harlow:2016vwg,Faulkner:2017vdd,Cotler:2017erl}
based on the more general subregion-subalgebra duality outlined in Sec.~\ref{sec:sub}. 
 
 \subsection{Formulation}

Consider a local boundary spatial subregion $R$, with $\hat R$ being its {(boundary)} causal completion. 
At finite $N$, the boundary operator algebra $\sB_R^{(N)} = \sB_{\hat R}^{(N)}$ associated with $R$ is {state-independent} and type III$_1$, but not all operators in $\sB_R^{(N)}$  have a sensible large $N$ limit.

Now consider the system in a semi-classical state $\ket{\Psi}$ dual to some bulk geometry $\chi_\Psi$. 
Subregion-subregion duality says that bulk physics in the entanglement wedge $\fb_R$ 
can be obtained from boundary physics in $\hat R$. The entanglement wedge $\fb_R$ is defined as the domain of dependence of the 
region $E_R$ that satisfies $\p E_R = \ga_R \cup R$ where $\ga_R$ is the RT surface for $R$. 
In particular, it has been proposed that any bulk operator $\phi$
in $\fb_R$ acting on $\ket{\Psi}$ can be represented by some operator $\sO \in \sB_R^{(N)}$, which is called entanglement wedge reconstruction~\cite{Almheiri:2014lwa,Pastawski:2015qua,Hayden:2016cfa,Dong:2016eik,Harlow:2016vwg,Faulkner:2017vdd,Cotler:2017erl}.

From the perspective of subregion-subalgebra duality of Sec.~\ref{sec:sub}, in the large $N$ limit there should be a boundary subalgebra 
$\sX_R \subset \sM_\Psi$ that is {\it identified} with the bulk operator algebra $\widetilde \sM_{\fb_R} = \widetilde \sM_{E_R} \subset \widetilde \sM_{\chi_\Psi}$ in the entanglement wedge $\fb_R$. Below we will give a proposal for $\sX_R$, discuss its properties, and comment on the connection of our formulation with entanglement wedge reconstruction formulated in~\cite{Almheiri:2014lwa,Dong:2016eik,Harlow:2016vwg,Faulkner:2017vdd} toward the end in Sec.~\ref{sec:summary}.

Another bulk region often discussed in connection with the entanglement wedge is the
causal wedge~\cite{Hubeny:2012wa}, which is defined as the intersection of bulk causal future and causal past of $\hat R$. The causal wedge as defined, however, may not be causally complete.\footnote{We thank Netta Engelhardt for discussions on this.} We define the causal domain $\fc_{\hat R}$ of $\hat R$, as the smallest bulk domain of dependence enclosing the causal wedge. 
There should also be a boundary subalgebra $\sY_{\hat R}$ that is identified with the bulk operator subalgebra $\widetilde \sM_{\fc_{\hat R}} \subset \widetilde \sM_{\chi_\Psi}$ for the causal domain $\fc_{\hat R}$. 
Since the entanglement wedge always encloses the causal wedge~\cite{Hubeny:2012wa,Wall:2012uf,Hubeny:2013gba}, we have $\fc_{\hat R} \subseteq \fb_R$, and thus $\widetilde \sM_{\fc_{\hat R}} \subseteq \widetilde \sM_{\fb_R}$. 

We propose that, in the boundary theory,  $\sX_R$ and  $\sY_{\hat R}$ can be defined as follows: 
\bega \label{xr}
\sX_R = \pi_\Psi( \lim_{N \to \infty, \Psi} \sB_R^{(N)}) , 
\\
\sY_{\hat R} =  \le(\sM_{\hat R}\ri)'' \ .  
\label{xr1}
\end{gather} 
Here $\lim_{N \to \infty, \Psi} \sB_R^{(N)}$ is the subalgebra of $\sA_\Psi$ obtained from $ \sB_R^{(N)}$ 
in the large $N$ limit. It is defined in the same manner as $\sA_\Psi$. 
It depends on $\ket{\Psi}$.
$\sM_{\hat R}$ denotes the restriction of $\sM_{\Psi}$ to $\hat R$ and $\sY_{\hat R}$ is its completion under the weak operator topology.
By definition, both $\sX_R$ and $\sY_{\hat R}$ are subalgebras of $\sM_\Psi$. Introducing a restricting operation $\sP_R$ to region $R$, we can also write~\eqref{xr}--\eqref{xr1} as 
\be 
\sX_R =  \pi_\Psi (\lim_{N \to \infty, \Psi} \sP_R \sB^{(N)}) , \quad \sY_{\hat R} =  \sP_{\hat R} \pi_\Psi (\lim_{N \to \infty, \Psi} \sB^{(N)})  \ .
\ee
Since $\sP_R \sB^{(N)} = \sP_{\hat R} \sB^{(N)}$, the difference in the definitions of $\sX_R$ and $\sY_{\hat R}$ 
lies only in the order of taking the restriction to $R$ and the large $N$ limit. The two procedures in general do not commute, as we will see in an explicit example in Sec.~\ref{sec:expD}.

An underlying assumption of the proposal is that both $\sX_R$ and $\sY_{\hat R}$ should be type III$_1$. We also expect $\ket{0}_{\Psi, {\rm GNS}}$ to be cyclic and separating with respect to both of them. 
Both algebras depend on $\ket{\Psi}$, and can in principle have very different properties for different 
$\ket{\Psi}$.

\subsection{Entanglement wedge algebra}

We now list properties of the entanglement wedge algebra $\sX_R$, most of which are immediate:

\ben

\item Since $\lim_{N \to \infty, \Psi} \sB_R^{(N)}$ includes all the single-trace operators in $\hat R$, we have 
$\sY_{\hat R} \subseteq \sX_R$.

\item $\sX_R$ depends only on the {\it spacetime} subregion $\hat R$; however, it can be understood as being associated with $R$, which is a subregion of a Cauchy slice. In particular, if $\tilde R$ is a subregion of another Cauchy slice such that $\hat{\tilde R} = \hat R,$ then we can equally well think of $\sX_R$ as being associated to $\tilde R$. At finite $N$, we have $\sB_R^{(N)} = \sB_{\tilde R}^{(N)}$, which implies $\sX_R = \sX_{\tilde R}$. In contrast, $\sY_{\hat R}$ is only defined for $\hat R$ and cannot be associated to any spatial subregion.

\item Nesting:  for $R_1 \subset R_2$,  we have $\sX_{R_1} \subset \sX_{R_2}$, as $\sB_{R_1}^{(N)} \subset \sB_{R_2}^{(N)}$.

\begin{figure}[!h]
\begin{center}
\includegraphics[width=3.5cm]{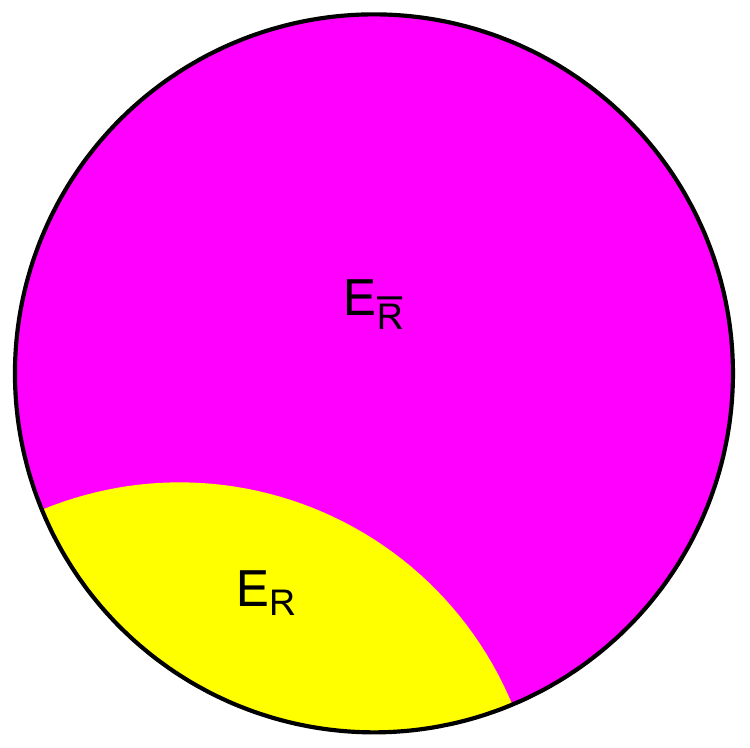}
\caption[]{\justifying\small Since bulk effective field theory obeys Haag duality, we should have $\overline{E_R} = E_{\bar R}$.}
\label{fig:compEW}
\end{center}
\end{figure}

\item Suppose $\sB_R^{(N)}$ satisfies Haag duality, i.e. $\sB_R^{(N)} = (\sB_{\bar R}^{(N)})'$, where $\bar R$ denotes the complement of $R$ on a Cauchy slice. {\it Assuming} that the procedure of taking $N \to \infty$ and the commutant can be exchanged,  i.e. 
\be\label{exli}
\lim_{N \to \infty, \Psi}  ( \sB_{\bar R}^{(N)})'  = ( \lim_{N \to \infty, \Psi} \sB_{\bar R}^{(N)})' , 
\ee
where the commutant on the right hand side should be understood as being taken within $\sA_\Psi$, 
we then have 
\be\label{hdt}
\sX_R = \sX_{\bar R}' \ .
\ee
i.e. $\sX_R$ satisfies Haag duality. 
Whether~\eqref{exli} is indeed correct should clearly be carefully understood.
Since in the supergravity limit, the bulk effective field theory is expected to satisfy Haag duality (see Fig.~\ref{fig:compEW}), for our proposal to work,~\eqref{exli} should at least hold  in the strong coupling limit.
 In~\cite{longPaper} it was found~\eqref{hdt} holds for a two-dimensional CFT when $R$ is given by a half space,\footnote{That discussion is independent of couplings and should work even in the free theory.} which we will discuss further in Sec.~\ref{sec:expD}.

Since $\sY_{\hat R}$ is defined in a generalized free field theory, in general it does not satisfy Haag duality, i.e. in general $\sY_{\hat R} \neq \sY_{\hat R'}'$.

 \item The definition~\eqref{xr} is not explicit, as we currently do not know how to precisely describe $\lim_{N\to \infty, \Psi} \sB_R^{(N)}$. 
$\sX_R$ can be given an alternative definition in terms of modular flows using the results of~\cite{Jafferis:2015del,Faulkner:2017vdd}.  
Denote the modular operator for $\sX_R$ with respect to $\ket{0}_{\Psi, {\rm GNS}}$ by $\De_0$. 
Now consider
\be\label{modT0}
\sO (s; \vx) \equiv \De^{is}_0 \pi_{\Psi} (\sO (\vx) ) \De^{-is}_0  \in \sX_R , \quad  s \in (-\infty, \infty) , \quad \vx \in R  ,
\ee
where $\sO (\vx)$ is a single-trace operator. We should emphasize that 
$\sO (s;\vx)$, is, by definition, part of the single-trace operator algebra; however, in general it may not be expressible in terms of local single-trace operators in $\hat R$. We will discuss an explicit example in Sec.~\ref{sec:expD}.
Denote the algebra generated by $\sO (s; \vx)$ as $\sZ_R$. 
The discussion of~\cite{Jafferis:2015del,Faulkner:2017vdd} can be translated to our language as 
\be \label{im2}
\sX_R = \sZ_R  \ .   
\ee

Denoting the modular operator of $ \sB^{(N)}_R$ with respect to $\ket{\Psi}$ at finite $N$ by $\De_\Psi$,
the statement of~\cite{Jafferis:2015del} of the identification of the boundary and bulk modular operators can be written 
in our language as 
\be \label{yehk}
\pi_\Psi \le(\lim_{N \to \infty, \Psi} \De^{-i s}_\Psi \sO (\vx) \De^{is}_\Psi \ri) =  \De_0^{-is} \pi_\Psi (\sO (\vx))  \De_0^{is} \ .
\ee
Equation~\eqref{yehk} is a striking statement, as it implies that the modular flow $\De^{-i s}_\Psi \sO (\vx) \De^{is}_\Psi $ 
of a single-trace operator has a well-defined large $N$ limit, i.e. remains part of the single-trace operator algebra. 
This is the property that connects the two definitions.

The identification~\eqref{im2} implies that for a CFT in the vacuum state, and for $R$ given by a half-space or a sphere 
 \be \label{ygd1}
 \sX_R = \sY_{\hat R}
 \ee
 as in these cases, modular flows are geometric, and thus $\sO(s; \vx)$ can be expressed in terms of single-trace operators in $\hat R$.

 \item Let $R_1$ and $R_2$ be subregions on the same boundary Cauchy slice. While $\sB_R^{(N)}$ obeys additivity, i.e. 
\be 
\sB_{R_1}^{(N)} \lor \sB_{R_2}^{(N)} =  \sB_{R_1 \cup R_2}^{(N)}, 
\ee
 it can be checked that $\sX_R$ in general does not, i.e. 
\be \label{hen3}
\sX_{R_1} \lor \sX_{R_2} \subseteq \sX_{R_1 \cup R_2} \ .
\ee
The inequality in~\eqref{hen3} should then be a consequence of the large $N$ limit, and will be referred as superadditivity. 
Superadditivity has its origin again in Fig.~\ref{fig:ineqCauchyAlgs}, i.e. single-trace operators at different times are independent. 
We will give some explicit examples in Sec.~\ref{sec:expD} and~\ref{sec:union}. 

Superadditivity also leads to an anomaly in the intersection of $\sX_{R_1}$ and $\sX_{R_2}$
\be \label{intViol}
	\sX_{R_1 \cap R_2}   \subseteq  \sX_{R_1} \land \sX_{R_2} \ .
\ee
while finite-$N$ algebras $\sB_R^{(N)}$ should satisfy 
\be 
\sB_{R_1}^{(N)} \land \sB_{R_2}^{(N)} =  \sB_{R_1 \cap R_2}^{(N)} \ .
\ee
To see~\eqref{intViol}, we start with~\eqref{hen3} for $\bar R_1$ and $\bar R_2$
\be \label{intl1}
	\sX_{\bar{R_1}} \lor \sX_{\bar{R_2}}    \subseteq \sX_{\bar{R_1} \cup \bar{R_2}}  
\ee
and take the commutant on both sides of~\eqref{intl1}. Equation~\eqref{intViol} follows by using~\eqref{hdt}. Note 
$\overline{\bar{R_1} \cup \bar{R_2}} = R_1 \cap R_2$, and $(\sX_{\bar{R_1}} \lor \sX_{\bar{R_2}})' = \sX_{\bar{R_1}}' \land \sX_{\bar{R_2}}' =  \sX_{R_1} \land \sX_{R_2}$.

\een
The above properties are consistent with the bulk dual of $\sX_R$ being the entanglement wedge. 
For example, the superadditivity~\eqref{hen3}--\eqref{intViol} describes precisely what was observed geometrically in the bulk for entanglement wedges~\cite{Casini:2019kex}, see Fig.~\ref{fig:error}.
Haag duality,~\eqref{hdt}, also implies that there is no superadditivity when $R_1 = \overline{R_2}$, i.e. 
\be 
\sX_R \lor \sX_{\bar R} = \sM_\Psi \ .
\ee

 \subsection{RT surfaces without entropy} \label{sec:expD}

Given the identifications~\eqref{xr}--\eqref{xr1}, we can in principle derive the entanglement wedge $\fb_R$
and the causal domain $\fc_{\hat R}$ from the boundary theory. In particular, in the case of~\eqref{xr}, subregion-subalgebra duality 
can be used to  provide 
an independent definition of the RT~(or HRT) surface without using entropy. 
It should be possible to derive the minimal surface prescription of the RT~(or HRT) surface, and the quantum extremal surface prescription~(including islands), from equivalence of the bulk and boundary subalgebras. These are ambitious goals and they will not be 
attempted here. 
 
 Here we discuss three explicit examples in which the entanglement wedges (and the associated RT surface) and causal wedges can be found by brute force from the identification of algebras. 
 
 We will consider a boundary CFT$_2$ on $\RR^{1,1}$ in the vacuum state $\ket{\Om}$, with boundary coordinates $ x^\mu = (x^0, x^1)$.  The bulk theory is defined on a Poincare patch of AdS$_3$, with coordinates $X^M = (z, x^\mu)$ and boundary at $z =0$. 
 At leading order in the large $N$ limit,  $\sM_\Om$ factorizes into a product of subalgebras generated by different single-trace operators. 
  An important and highly nontrivial self-consistency condition is that the entanglement and causal wedges obtained from the subalgebra generated by any single-trace operator should be the same. 
We will consider the subalgebra $\sM_\sO \subset \sM_\Om$
generated by a scalar operator $\sO$ of dimension $\De$, dual to a bulk field $\phi$. For notational simplicity, we will simply refer to it as $\sM_\Om$, and similarly for $\sX_R$ and $\sY_{\hat R}$.  We will use the definition of $\sX_R$ based on~\eqref{im2} and~\eqref{modT0}. 
We show that the resulting entanglement and causal wedges from $\sX_R$ and $\sY_{\hat R}$ for various examples indeed agree with their standard bulk definitions. 
 By a conformal transformation~(or isometric transformation in the bulk), the examples considered can be converted to 
 those in the global AdS. 
Below when talking about a boundary spatial region $R$, we always refer to the time slice $x^0=0$.  

The examples we will consider are: 
 
\ben 

\begin{figure}[h]
        \centering
        \begin{subfigure}[b]{0.3\textwidth}
            \centering
			\includegraphics[width=\textwidth]{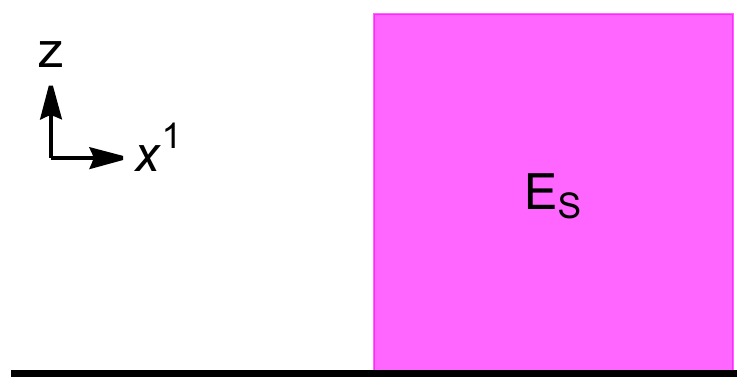} 
            \caption[]%
            {\centering{\small }}    
        \end{subfigure}
        \hfill
        \begin{subfigure}[b]{0.3\textwidth}   
            \centering 
			\includegraphics[width=\textwidth]{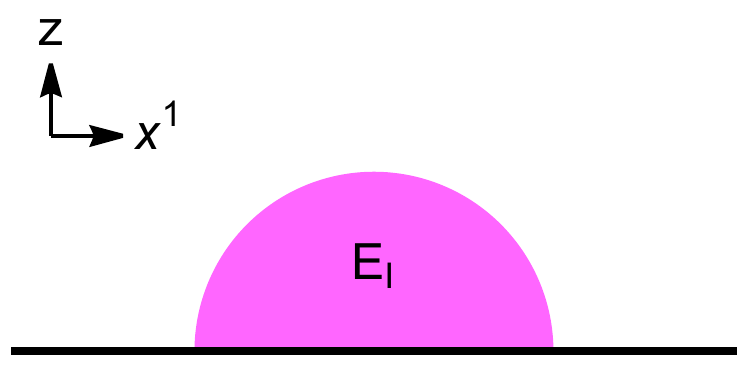}
            \caption[]%
            {\centering{\small }}    
        \end{subfigure}
        \hfill
        \begin{subfigure}[b]{0.3\textwidth}   
            \centering 
			\includegraphics[width=\textwidth]{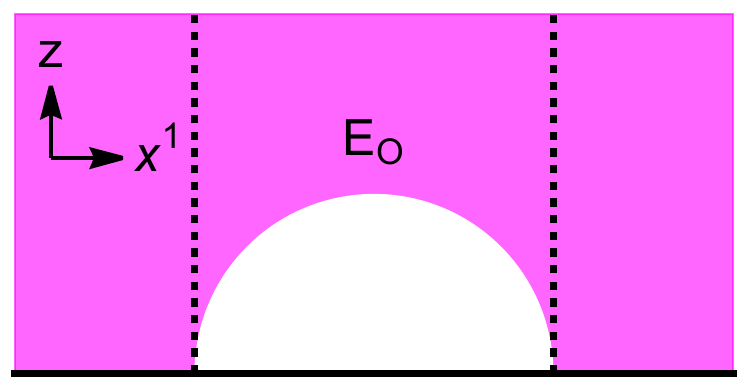}
            \caption[]%
            {\centering{\small }}    
        \end{subfigure}
        \caption[  ]
        {\justifying\small Algebraic entanglement wedges in three examples. (a) $x^0=0$ slice of the entanglement wedge of a half-space. (b) $x^0=0$ slice of the entanglement wedge of an interval. (c) $x^0=0$ slice of the entanglement wedge of two-finitely separated half-spaces. Note that the causal domain (the region outside of the dashed lines) is properly contained in the entanglement wedge. } 
\label{fig:ewExamples}
\end{figure}

\item[E1:]  $R$ is the half space $S: x^1 \in (0, \infty)$ with $\hat S$ the right Rindler wedge of $\RR^{1,1}$.  From discussion around~\eqref{ygd1}, $\sX_S$ and $\sY_{\hat S}$ coincide.  We show that $\sX_S$ is equivalent to the bulk subalgebra in the domain of dependence of the bulk region $x^1 > 0$ on the $x^0 =0$ slice.  See Fig.~\ref{fig:ewExamples} (a). 

\item[E2:] $R$ is an interval $I: x^1 \in (-a, a)$. In this $\sX_I$ and $\sY_{\hat I}$ again coincide, and can be shown to be equivalent to the bulk subalgebra in the domain of dependence of the bulk region inside the half circle $z^2 + (x^1)^2 = a^2$ 
on the $x^0 =0$ slice.  See Fig.~\ref{fig:ewExamples} (b).

\item[E3:]  Take $R = O = O_L \cup O_R  = \bar I$,  where  $O_L$ and $O_R$ are two half spaces separated by the finite interval $I: x \in (-a, a)$. In this case, $\sX_O$ and $\sY_{\hat O}$ are different, with 
\be 
\sY_{\hat O} =  \sX_{O_L} \lor \sX_{O_R} \subset \sX_{O}  \ , 
\ee
 and the corresponding bulk regions whose algebras are equivalent to $\sX_O$ and $\sY_{\hat O}$ can be shown to be those in Fig.~\ref{fig:ewExamples} (c). This is also an example of the superadditivity~\eqref{hen3}.

\een

We now outline the calculations behind the above examples, leaving details to various appendices.
We first describe the boundary side of the story, i.e. how to define $\sX_R$ and $\sY_{\hat R}$ for each example, and 
their relations. We then discuss how to obtain the bulk regions whose algebras are identified with them. 

\subsubsection{$\sX_R$ and $\sY_{\hat R}$ in various examples}

In the large $N$ limit, a scalar single-trace operator $\sO$ can be described by 
a generalized free field, with a mode expansion\footnote{Below $\sO (x)$ should be more precisely understood as $\pi_\Om (\sO (x))$. We suppress $\pi_\Om$ for notational simplicity.}  
\bega \label{hje}
\sO (x) = \int {d^2 p \ov (2 \pi)^2} \, f_p e^{i p \cdot x}  b_p , \quad 
f_p  = C  \theta (-p^2)    \ka^{\De -1},  \quad p \cdot x = - p^0 x^0 + p^1 x^1,
 \\
 \ka = \sqrt{- p^2}, 
\quad
 b_p^\da = b_{-p} , 
 \quad  [b_p, b_{p'}] = (2 \pi)^2 \ep (p^0) \de^{(2)} (p+p') ,
\end{gather}
where $C$ is a constant depending on the normalization of $\sO$.  It is convenient to choose $C = 2^{1-\De}\sqrt{\pi} \Ga(\De)^{-1}$ as this puts the extrapolate dictionary for a canonically normalized bulk field in the simplest form. We make this choice of normalization from here on.
The above equations follow from two-point correlation functions of $\sO(x)$, which are fully determined by conformal symmetry. 
$\sM_\Om$ can also be regarded as being generated by $\{b_p\}$. 
$\sY_{\hat R}$ for a spacetime region $\hat R$ is generated from $\sO (x)$ in~\eqref{hje} with the restriction $x \in \hat R$.

Now take $R$ to be the half space $S: x^1 > 0$, and $\hat S$ is given by the right Rindler region with metric 
\be 
ds^2 = e^{2 \chi} (-d \eta^2 + d \chi^2) = - e^{2 \chi} d\xi^+ d \xi^-, \quad \xi^\pm = \eta \pm \chi  \ .
\ee
We will use $\xi = (\eta, \chi) = (\xi^+, \xi^-)$ to denote the Rindler coordinates. From two-point functions of $\sO$ 
in the Rindler region we can obtain a mode expansion for the restriction of $\sO$ to $\hat S$ as 
\bega \label{ccc1}
 \sO_{S}  (\xi)  = \int \frac{d^2 k}{(2\pi)^2} \, 
 u_k^{(S)} a_{k}^{(S)} , \quad  u^{(S)}_k  = N_k e^{\bar q_+ \xi^- - q_- \xi^+}   \\
q_+ = {\De \ov 2} + i k^+, \quad q_- = {\De \ov 2} + i k^- , \quad \bar q_+ = {\De \ov 2} - i k^+, \quad \bar q_- = {\De \ov 2} - i k^- \\
   k^+ = {\om + q \ov 2}, \quad
k^- = {\om - q \ov 2} , \quad 
 N_k =  {\sqrt{\sinh \pi |\om| }\ov \sqrt{2 \pi} \Ga (\De)} \le|\Gamma\left(q_+ \right) \Gamma\left(q_- \right) \ri|\\
(a_k^{(S)})^\da = a_{-k}^{(S)} , \qquad    [a_k^{(S)} , a_{k'}^{(S)}] = \ep (\om)  (2\pi)^2 \de^{(2)} (k+k')  
   \ .
   \label{cccn}
\end{gather} 
Here $\om$ is the frequency for $\eta$, and $q$ is the momentum for $\chi$. 
The subalgebra $\sX_{S} = \sY_{\hat S}$ can be regarded as being generated by $\{a_k^{(S)}\}$. 

$\bar S$ is the half space $x^1 < 0$ at $x^0=0$, and we can similarly write down the mode expansion for $\sO_{{\bar S}} (\xi)$, where now $\xi$ refers to the Rindler coordinates in the left Rindler region, 
\be \label{ccc3}
 \sO_{ {\bar S}} (\xi)  = \int \frac{d^2 k}{(2\pi)^2} \, 
 u_k^{(\bar S)} a_{k}^{(\bar S)}  , \quad 
   u^{(\bar S)}_k  = N_k   e^{ q_+ \xi^- - \bar q_-  \xi^+} 
   \ .
 \ee 
$\{a_k^{(\bar S)}\}$ can be used to generate $\sX_{\bar S} = \sY_{\hat {\bar S}}$. 

Since a modular flow for operators in $S$ and $\bar S$ corresponds to a translation in $\eta$, $\om$ can be interpreted as the modular frequency, i.e. conjugate to the modular time.

It can be shown that $\sO(x)$ for any $x \in \RR^{1,1}$ can be expressed in terms of $\{a_k^{(S)}\}$ and 
$\{a_k^{(\bar S)}\}$~\cite{longPaper}, 
\be 
\sO(x) = \int  \frac{d^2 k}{(2\pi)^2} \, \sum_{\al = S, \bar S} f^{(\al)} (x; k) a^{(\al)}_k ,  
\ee
which is the statement 
\be 
\sX_S \lor \sX_{\bar S} = \sM_\Om \ .
\ee
More explicitly, there exists an (invertible) transformation matrix $A_{kp}^{(\al)}$ between two sets of oscillators $\{b_p\}$ and $\{a^{(S)}_k, a^{(\bar S)}_k\}$, 
\be\label{bogo00}
 b_p =\sum_{k, \al} A_{kp}^{(\al)} a_k^{(\al)},  \quad \al = S, \bar S \ .
\ee
The explicit form of $A_{kp}^{(\al)}$ as well as its inverse are given in Appendix~\ref{app:rindBaseChange}.

\begin{figure}[h]
        \centering
        \begin{subfigure}[b]{0.45\textwidth}
            \centering
			\includegraphics[width=\textwidth]{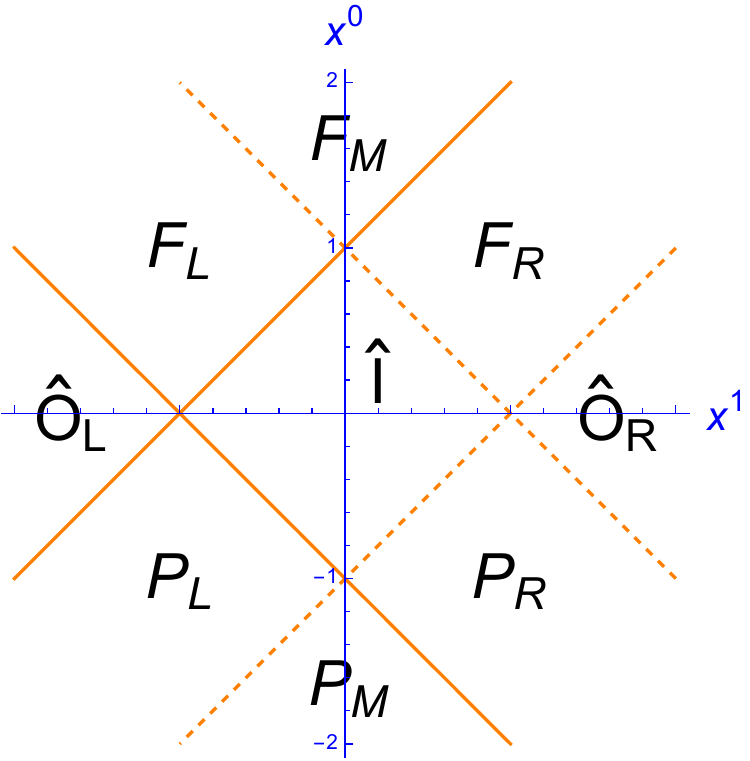} 
            \caption[]%
            {\justifying{\small The decomposition of the Minkowski $x-$plane with respect to the causal structure of the interval and its complement. The solid orange lines are the image of the $y-$plane Rindler horizon, while the dashed lines are the singularities of the inverse of the conformal transformation.}}    
        \end{subfigure}
        \hfill
        \begin{subfigure}[b]{0.45\textwidth}   
            \centering 
			\includegraphics[width=\textwidth]{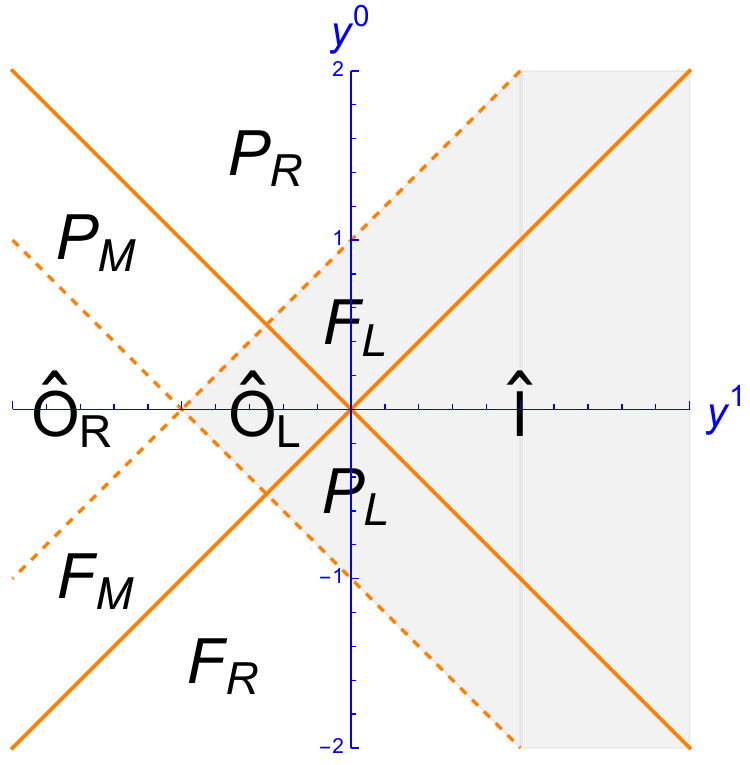}
            \caption[]%
            {\justifying{\small The decomposition of the Minkowski $y$-plane into the nine regions that map to the nine regions of the $x$-plane shown in the left plot, with $b = c = 1$. The solid orange lines are the Rindler horizons in $y,$ while the dashed lines are $y^- = b$ and $y^+ = -c$, where the transformation is singular.}}    
        \end{subfigure}
        \caption[  ]
        {\justifying\small The decomposition of the Minkowski $x-$ and $y-$planes with respect to the causal structure of the $x-$plane decomposition. Note that while in the right plot, the region $\hat I$ is spacelike separated from $P_M, F_M$, they are timelike separated in the left plot. This is due to the singularity in the transformation.  } 
\label{fig:nine0}
\end{figure}

Now consider examples E2 and E3. Since the regions in E2 and E3 are complements on a single time slice,
we will treat them together. $\sY_{\hat I}$ and $\sY_{\hat O}$ are defined  by restricting~\eqref{hje} to 
$\hat I$, which is the causal diamond for $I$, and $\hat O = \hat O_R \cup \hat O_L$, respectively. In particular, since $O_R$ and $O_L$ are disconnected, we have 
\be
\sY_{\hat O} = \sY_{\hat O_R} \lor \sY_{\hat O_L} = \sX_{O_R} \lor \sX_{O_L} \ .
\ee

We can obtain a mode expansion for an operator in $\hat I$ or $\hat O$ by conformally mapping the mode expansions discussed above for the right and left Rindler regions to $\hat I$ and $\hat O$. 
 Consider another spacetime spanned by $y^\mu = (y^0, y^1)$ with $y^\pm = y^0 \pm y^1$, and  the following conformal transformation that maps smoothly from the right Rindler wedge in the $y$-plane to $\hat I$ in $x$-spacetime
\be \label{ToDi}
	x^- = a  {y^- + b \ov b - y^-}, \qquad x^+ = a  {y^+ - c \ov y^+ + c} \ ,
\ee 
where $b, c$ are arbitrary real {\it positive} constants. The above transformation is singular at null lines $y^-= b$ and $y^+ = -c$, which lie outside the right Rindler quadrant in the $y$-plane. Below we will take $b=c=1$. 
Depending on their causal relations with $\hat I$ and $\hat O_R, \hat O_L$, the $x$-plane can be separated into nine different regions. 
How various regions in the $y$-plane are mapped to these nine regions is shown in Fig.~\ref{fig:nine0}. 
 Due to the singular behavior at $y^- =b$ and $y^+ = -c$, the mapping preserves casual structure only for the regions lying within the quadrant to the right of the dashed lines in Fig.~\ref{fig:nine0}(b)~(the shaded region). For example, while the region $\hat I$ is causally disconnected from regions $F_M$ and $P_M$ in plot (b), it is causally connected with them in plot (a). We 
 can then reliably conformally transform the mode expansions in the $y$-plane to the $x$-plane only for the shaded region in the plot (b).

Using the mode expansions~\eqref{ccc1} and~\eqref{ccc3} for $\sO (y)$ in the right and left Rindler regions of the $y$-plane, we can use the conformal transformation to obtain 
the mode expansion for $\sO(x)$ in the $\hat I$ and $\hat O_L$ regions (after expressing $\xi^\pm$ in terms of $y^\pm$ and setting $b=c=1$) 
\bega\label{oi1}
\sO (x \in \hat I)  = \int {d^2k \ov (2\pi)^2} u^{(I)}_k(x)~ a^{(I)}_k \ ,  \\
\sO(x \in \hat O_L) 
	= \int {d^2k \ov (2\pi)^2} u^{(O)}_k(x)~ a^{(O)}_k \ ,
	\label{ol1}\\
u^{(I)}_k   (x) = N_k (2a)^{\De} 
(a - x^-)^{-\bar q_+} (x^- + a)^{- q_+} (a - x^+)^{-\bar q_-} (x^+ + a)^{- q_-} ,
\label{iMT} \\	
\label{eMT}
	u^{(O)}_k (x) = N_k (2a)^{\De} 
	(x^- - a)^{-q_+} (x^- + a)^{-\bar q_+} (a - x^+)^{- q_-} (-x^+ - a)^{-\bar q_-} \ .
\end{gather} 
We can obtain the expressions for $\sO$ in other regions of the $x$-plane by analytic continuations of~\eqref{oi1}--\eqref{eMT}. The analytic continuation procedure is a generalization of that for a generalized field in Rindler introduced in~\cite{longPaper} (which is in turn a generalization of the standard Unruh procedure). The details are given in Appendix~\ref{app:intAnalCont}. 
We find that equations~\eqref{ol1} and~\eqref{eMT} (negating the arguments of all complex powers) in fact also apply to $\hat O_R$ region and thus the full $\hat O$ region, i.e. 
\be 
\sO(x) \label{uhn}
	= \int {d^2k \ov (2\pi)^2} u^{(O)}_k(x)~ a^{(O)}_k ,  \quad x \in \hat O = \hat O_L \cup \hat O_R \ .
\ee
In other regions of the $x$-plane, $\sO(x)$ can be expressed in terms of both $a_k^{(I)}$ and $a_k^{(O)}$. Again there is an invertible 
basis transformation between $\{b_p\}$ and $\{a_k^{(I)}, a_k^{(O)}\}$, 
\be  \label{saix}
b_p =  \int {d^2k \ov (2 \pi)^2}\, \le(P^{(I)}_{kp}  a^{(I)}_k + P^{(O)}_{kp} a^{(O)}_k \ri) \ .
\ee
The explicit expressions for $P^{(I)}_{kp}$ and $P^{(I)}_{kp}$ are given in appendix~\ref{app:intBaseChange}.

Since~\eqref{oi1} is obtained from conformal transformation of the right Rindler region $\hat S$, whose associated algebra $\sX_S = \sY_{\hat S}$, we should have $\sX_I = \sY_{\hat I}$, which can also be considered as being generated by $\{a^{(I)}_k\}$.

We will now interpret $\sX_O$ as being generated by $\{a^{(O)}_k\}$. We note that the quantum numbers $k = (\om, q)$ have no obvious geometric meaning in the $O$ region; they went along for the ride with the conformal transformation. Since before the conformal transformation, $\om$ corresponds to the modular frequency of Rindler regions, $\om$ in~\eqref{oi1} and~\eqref{uhn} can be interpreted as the modular frequency for the $I$ and $O$ regions. 
From~\eqref{uhn}, we have $\sX_{O_L} \lor \sX_{O_R} \subseteq \sX_{O}$, but now 
an important difference from the $I$ region is that equation~\eqref{uhn} is not invertible, i.e. it cannot be inverted to solve for
$a_k^{(O)}$ in terms of $\sO (x)$ in $\hat O$.
More explicitly, it is not possible to find a set of functions $\{f_k(x), k \in \mathbb{R}^2\}$ supported for $x \in \hat O$ such that
\be \label{cantWriteAOk}
	a^{(O)}_k = \int_{x \in \hat O} d^2x~ f_k(x) \sO(x) , \quad \forall k\  .
\ee
 See Appendix~\ref{app:nonInvO} for a proof. This non-invertibility implies that 
\be \label{hda}
\sX_{O_R} \lor \sX_{O_L} \subset \sX_{O}  \quad \Lra \quad \sY_{\hat O} \subset \sX_O \ .
\ee
We can also describe $\sX_O$ in terms of a ``coordinate'' basis by introducing operators defined in terms of the modular time $s$ conjugate to $\om$
\be 
\sO (s; x^1) =  \int {d^2k \ov (2 \pi)^2} \,  u^{(O)}_k(x^1, x^0=0) e^{- i \om s} a^{(O)}_k , \qquad x^1 \in O, 
\ee
which is invertible 
\bega \label{i1u}
	a^{(O)}_k = \int_{x^1 \in O} dx^1 ds ~ f_k(x^1, s) \sO(s; x^1) , \\
	f_k(x^1, s) = {1 \ov N_k} \le({(x^1)^2 - a^2 \ov 2a}\ri)^{\De -1} \le({x^1-a \ov x^1 + a}\ri)^{-iq} e^{i\om s}  \ .
\end{gather} 
Thus $\sX_O$ can also be viewed as being generated by the set $\{\sO (s; x^1)\}$. We also note that 
 there does  not exist a conformal transformation that can smoothly map a Rindler region in the $y$-plane to $\hat O$. Otherwise, we would have $\sX_O$ equivalent to $\sY_{\hat O}$.

\subsubsection{Entanglement and causal wedges}

We now turn to finding the bulk regions corresponding to $\sX_R$ and $\sY_{\hat R}$ for the examples discussed above. 

A bulk scalar field $ \phi (X)$ in AdS$_3$  
can be expanded in modes as 
\bega \label{phex}
\phi (X) = \int {d^2 p \ov (2 \pi)^2} v_p (X) b_p  , \quad v_p (X) = \theta (-p^2)  \sqrt{\pi} z^{d \ov 2} J_{\De -1} (\ka z)  e^{i p \cdot x } , 
\end{gather}
where $X = (x^\mu, z)$ are the Poincare coordinates of a bulk point, and the rest of the notation is the same as in~\eqref{hje}. 
Note that $\phi (X)$ is expanded in terms of the same set of $\{b_p\}$ as $\pi_{\Om}(\sO(x))$. This follows from the extrapolate dictionary and gives the identification~\eqref{globRecon}. With the identification, $\phi (X)$ as written in~\eqref{phex}  can be regarded as a boundary operator.

Consider first the case the boundary region  is given by half space $R = S: x^1 > 0$. Using the basis change~\eqref{bogo00}
we can write~\eqref{phex} as 
\be 
\phi (X) = \sum_{k} \le(\tilde v^{(S)}_k(X) a^{(S)}_k + \tilde v^{(\bar S)}_k(X) a^{(\bar S)}_k \ri)  , \quad   \tilde v^{(\al)}_k(X) =  \int {d^2 p \ov (2 \pi)^2} A_{kp}^{(\al)}  v_p (X) \ . 
\ee 
The entanglement wedge $\fb_R$ for $R$ is then obtained by finding the collection of $X$ such that $\tilde v^{(\bar S)}_k(X) =0$ for all $k$.  We then find the entanglement wedge (which coincides with the causal domain as $\sX_S = \sY_{\hat S}$) is given by 
that in Fig.~\ref{fig:ewExamples}(a). For calculation details see Appendix~\ref{app:rindRT}. 

The same discussion applies to the $I$ and $O$ regions. Using~\eqref{saix} we can write~\eqref{phex} as
\be 
\phi (X) = \sum_{k} \le(\tilde v^{(I)}_k(X) a^{(I)}_k + \tilde v^{(O)}_k(X) a^{( O)}_k \ri)  , \quad   \tilde v^{(\al)}_k(X) =  \int {d^2 p \ov (2 \pi)^2} P_{kp}^{(\al)}  v_p (X) \ . 
\ee 
The entanglement wedge of $I$ is  obtained by identifying the collection of $X$ such that $\tilde v^{(O)}_k(X) =0$ for all $k$, which 
gives the region  in Fig.~\ref{fig:ewExamples}(b) (the causal domain is the same as $\sX_I = \sY_{\hat I}$). Similarly, 
the entanglement wedge of $O$ is  obtained by identifying the collection of $X$ such that $\tilde v^{(I)}_k(X) =0$ for all $k$, which 
gives the region  in Fig.~\ref{fig:ewExamples}(c). In this case the entanglement wedge is larger than the causal domain.  
 For calculation details see Appendix~\ref{app:int}.

\subsection{Superadditivity: union and intersection of wedge algebras} \label{sec:union}

We have already seen an example of superadditivity in~\eqref{hda}.
We will now explore it further by examining unions and intersections of 
$\sX_R$'s in a few situations. We start by considering a simple example purely from the boundary perspective using the generalized free field theory description of a scalar operator, which already illustrates well various surprising and unintuitive features associated with such unions and intersections.  We then discuss these features from the bulk perspective using the duality with entanglement wedges. We see that the bulk description can be interpreted as providing  a geometrization of these unintuitive properties.

\begin{figure}[h]
        \centering
        \begin{subfigure}[b]{0.4\textwidth}
            \centering
			\includegraphics[width=\textwidth]{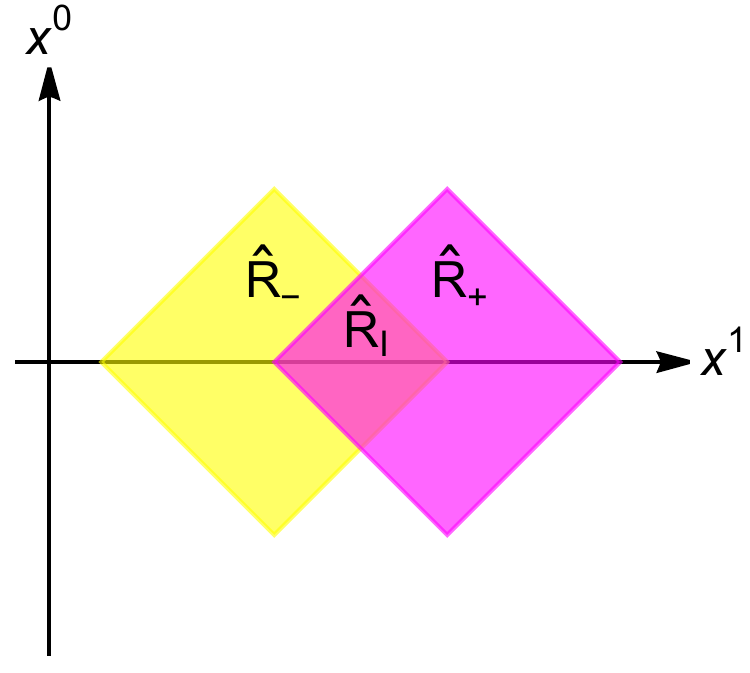} 
            \caption[]%
            {{\small }}    
        \end{subfigure}
        \hfill
        \begin{subfigure}[b]{0.4\textwidth}   
            \centering 
			\includegraphics[width=\textwidth]{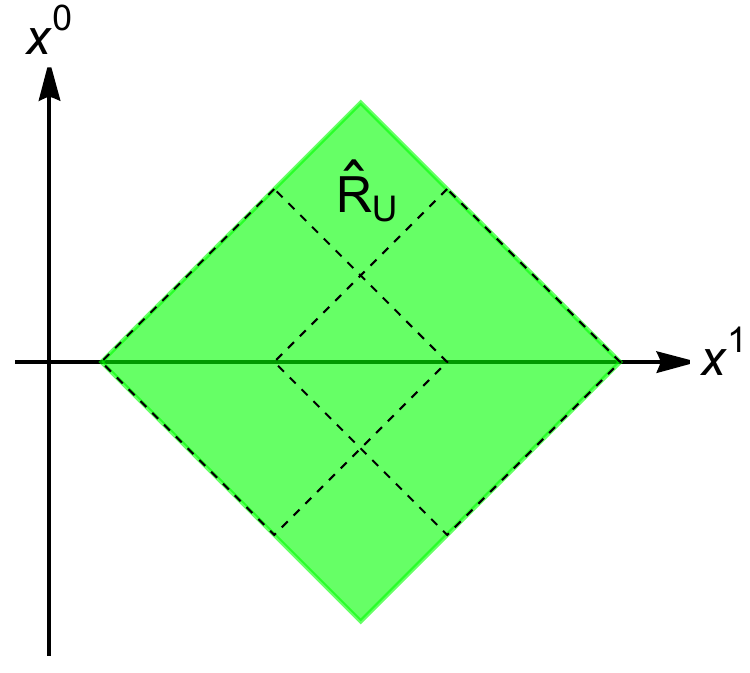}
            \caption[]%
            {{\small }}    
        \end{subfigure}
        \caption[  ]
        {\justifying\small Causal completions of various intervals on the boundary. (a) The regions $\hat{R}_-,~\hat{R}_+$ and $\hat{R}_I.$ (b) The region $\hat{R}_U.$ } 
\label{fig:addIntDiamonds}
\end{figure}

Consider a boundary CFT$_2$ in the vacuum state, and two intersecting intervals $R_- : x^1 \in (-b-a, -b+a)$ and $R_+ : x^1 \in (b-a, b+a)$ (with $a > b > 0$) on the $x^0 = 0$ slice.
$R_U = R_- \cup R_+$ is an interval of width $2(a+b)$ while $R_I = R_- \cap R_+$ is an interval of width $2(a-b)$. 
These regions and their causal completions are shown in Fig.~\ref{fig:addIntDiamonds}. Note that $\hat R_- \cup \hat R_+$ is not causally complete. 
We would like to understand the relations among
 $\sX_{R_-} \lor \sX_{R_+}$, $\sX_{R_-} \land \sX_{R_+}$, $\sX_{R_U}$, and $\sX_{R_I}$. 
 If we were at finite $N$, i.e. dealing with $\sB_{R}$ (suppressing $N$ below) we would have 
 \be 
 \sB_{R_-} \lor \sB_{R_+} = \sB_{R_U}, \quad \sB_{R_-} \land \sB_{R_+} = \sB_{R_I} \ .
 \ee

Recall that in the large $N$ limit and the vacuum state $\ket{\Om}$, $\sX_R$ for $R$ given by an interval is 
generated by single-trace operators in $\hat R$, i.e. $\sX_R = \sM_{\hat R}$. Since $R_-, R_+, R_U, R_I$ are all intervals 
we have $\sX_{R_i} = \sM_{\hat R_i},$ for $i=+,-,U, I$.  Recall that $\sM_A$ denotes the restriction of single-trace algebra $\sM_\Om$ to a spacetime subregion $A$.

Since single-trace operators at different times are independent, {\bf naively} we may expect that 
\be 
\sX_{R_-} \lor \sX_{R_+} = \sM_{\hat R_-} \lor \sM_{\hat R_+} = \sM_{\hat R_- \cup \hat R_+} \subset \sM_{\hat R_U} = \sX_{R_U}
 , \quad \sX_{R_-} \land \sX_{R_+} = \sM_{\hat R_I}   = \sX_{R_I} \ .
\ee
This expectation turns out to be incorrect. Instead, we find 
\bega \label{uha1}
\sM_{\hat R_- \cup \hat R_+} \subset \sX_{R_-} \lor \sX_{R_+} \subset \sM_{\hat R_U} = \sX_{R_U}, 
\\
\sX_{R_I} =   \sM_{\hat R_I}  \subset \sX_{R_-} \land \sX_{R_+} \ .
\label{uha2}
\end{gather} 

\begin{figure}[h]
        \centering
        \begin{subfigure}[b]{0.45\textwidth}
            \centering
			\includegraphics[width=\textwidth]{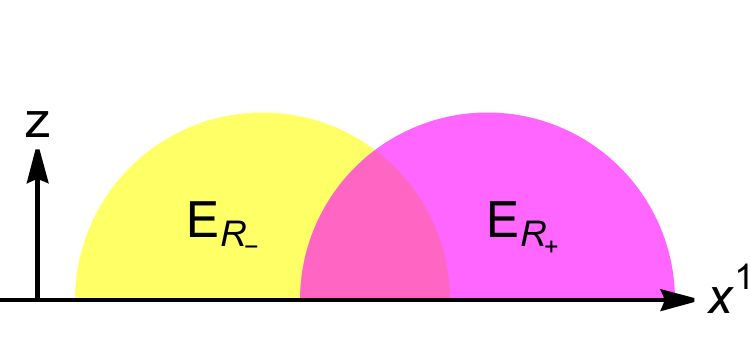} 
            \caption[]%
            {\justifying{\small The $x^0=0$ slices of $\fb_{R_-}$ and $\fb_{R_+}.$}}    
        \end{subfigure}
        \hfill
        \begin{subfigure}[b]{0.45\textwidth}   
            \centering 
			\includegraphics[width=\textwidth]{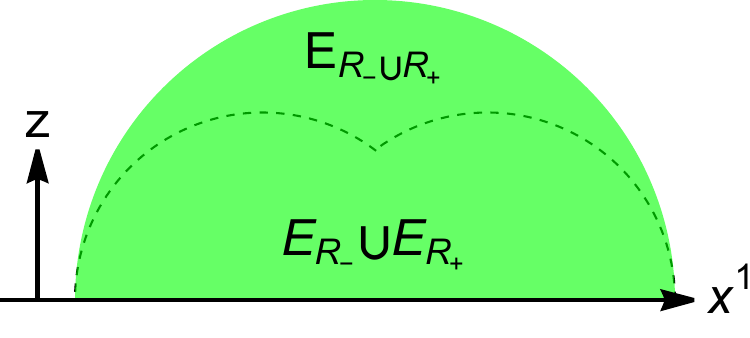}
            \caption[]%
            {\justifying{\small The $x^0=0$ slice of $\fb_{R_- \cup R_+}$.}}    
        \end{subfigure}
        \caption[  ]
        {\justifying\small Bulk time slices of the entanglement wedges of intervals.  } 
\label{fig:addIntBulk}
\end{figure}

To see~\eqref{uha1}, recall that $ \sX_{R_-} \lor \sX_{R_+} $ is defined as 
\be 
 \sX_{R_-} \lor \sX_{R_+} \equiv ( \sX_{R_-} \cup \sX_{R_+} )''  \ .
 \ee
The first relation in~\eqref{uha1} implies that there are operators outside $\sM_{\hat R_- \cup \hat R_+}$ that commute 
with $( \sX_{R_-} \cup \sX_{R_+} )'$, while the second relation says that there are operators in $\sM_{\hat R_U}$ that do not commute with 
$( \sX_{R_-} \cup \sX_{R_+} )'$. Equation~\eqref{uha2} says that there exist operators which can be expressed separately in terms of single-trace operators in $\hat R_-$ and $\hat R_+$, but not those in $\hat R_I$. See Fig~\ref{fig:addIntDiamonds} (a).
The second and third properties can be demonstrated explicitly by using the mode expansion of $\sO$ in terms of an interval and its complement. For details, see Appendix~\ref{app:addAnom}. Their bulk geometric descriptions are given in Fig.~\ref{fig:error} and Fig.~\ref{fig:addIntBulk} (b).

The first relation of~\eqref{uha1} is more subtle, and so far we can only show it explicitly with the help of the bulk description (see Appendix~\ref{app:hbs} for a pure boundary discussion).\footnote{After publication of this paper we found that an argument due to Araki~\cite{ArakiTT} can be applied to this case to obtain this relation using only properties of the boundary theory. The argument 
is reminiscent of AdS/CFT despite being written down in the 1960's. See~\cite{addPaper} for a simplified discussion.}
In the bulk, $\sX_{R_-} \lor \sX_{R_+}$ is given by operators lying inside the domain of dependence of $E_{R_-} \cup E_{R_+}$ 
in Fig.~\ref{fig:addIntBulk} (a). From the duality,  $\sX_{R_-} \lor \sX_{R_+}$ is given by $\sM_{R_e}$ where $R_e$ is the largest boundary spacetime subregion that has $\widehat{E_{R_-} \cup E_{R_+}}$ as its causal domain. $R_e$ is found explicitly in Appendix~\ref{app:bdrySuperadd}, and is shown in Fig.~\ref{fig:twoDiamSuperAdd1}. It encloses 
$\hat R_- \cup \hat R_+$.

\begin{figure}[h]
\begin{centering}
\includegraphics[width=4.5cm]{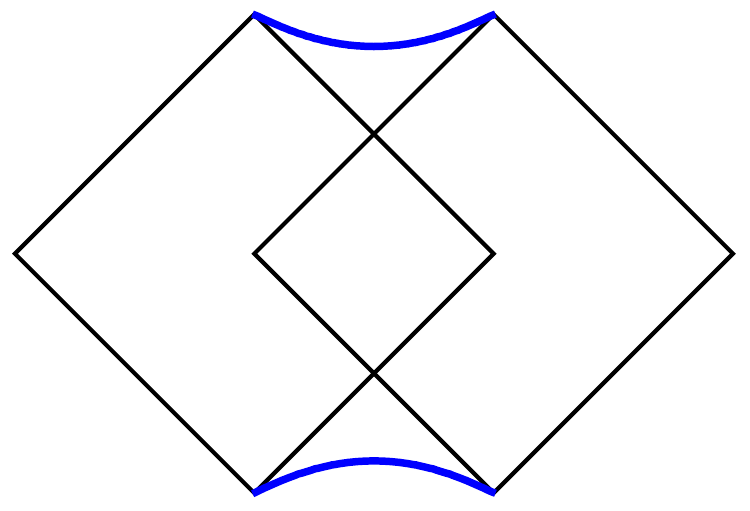}
\end{centering}
\caption[]{\justifying\small 
The region $R_e$ defined by $\sX_{R_+} \lor \sX_{R_-} = \sM_{R_e}$. It is given by the largest boundary subregion whose causal domain is $\widehat{E_{R_-} \cup E_{R_+}}$. The points between the black and blue curves lie in $R_e$ but not in $\hat{R}_- \cup \hat{R}_+.$ The blue curves are given by $x^+x^- = a^2-b^2$ for $x^1 \in (-b,b)$.
}
\label{fig:twoDiamSuperAdd1}
\end{figure}

We note that when $R_-$ and $R_+$ are non-intersecting (i.e. if $b>a>0$), we have 
\be 
 \sX_{R_-} \lor \sX_{R_+} = \sM_{\hat R_- \cup \hat R_+}  \ .
 \ee
Depending on the sizes of $R_-, R_+$ and the distance between them, we can have superadditivity 
\be 
\sX_{R_-} \lor \sX_{R_+}  \subset \sX_{R_- \cup R_+} 
\ee
when the entanglement wedge is connected, of which~\eqref{hda} is a special example.  
Thus the ``phase transition'' of the entanglement wedge for two intervals can be considered as the onset of superadditivity. 

In contrast to the intersecting interval case, superadditivity does not have a geometric description in the boundary theory, as 
$\widehat{R_- \cup R_+} = \hat R_- \cup \hat R_+$. We also note that the violation of the intersection property is {\it never geometrically realized} on the boundary. This is because intersections of causally complete regions are also causally complete themselves.

\begin{figure}[h]
        \centering
        \begin{subfigure}[b]{0.22\textwidth}
            \centering
			\includegraphics[width=\textwidth]{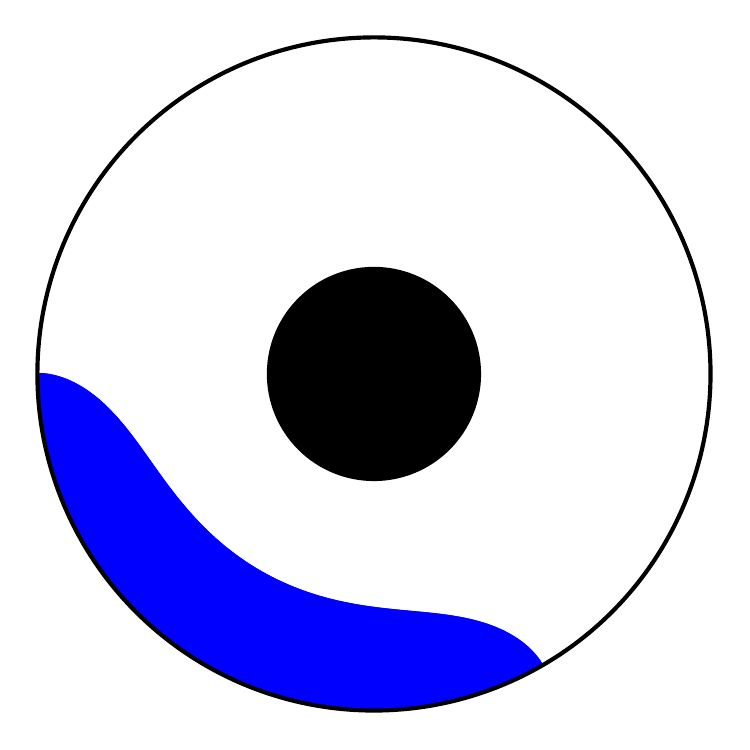} 
            \caption[]%
            {{\small }}    
        \end{subfigure}
        \hfill
        \begin{subfigure}[b]{0.22\textwidth}   
            \centering 
			\includegraphics[width=\textwidth]{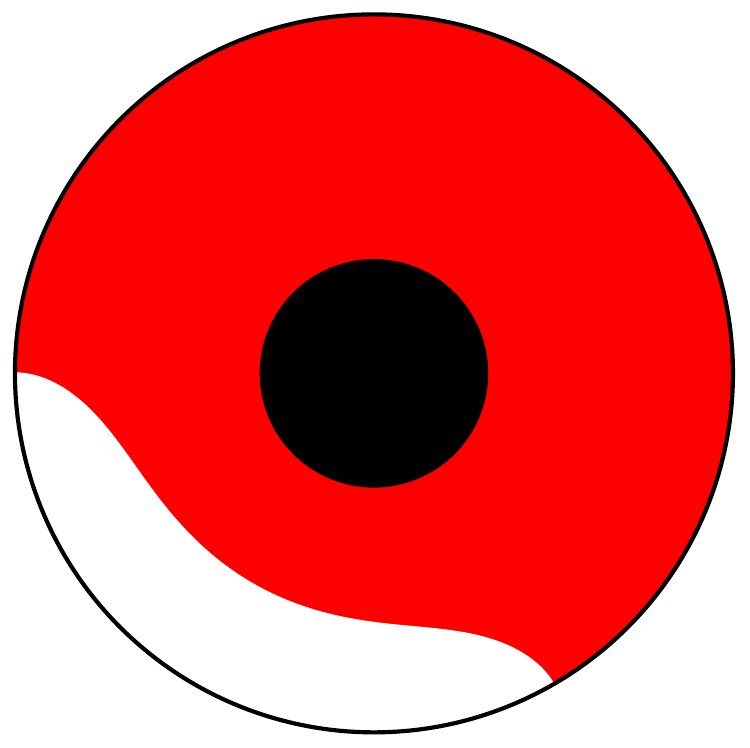}
            \caption[]%
            {{\small }}    
        \end{subfigure}
        \hfill
          \begin{subfigure}[b]{0.22\textwidth}  
            \centering 
            \includegraphics[width=\textwidth]{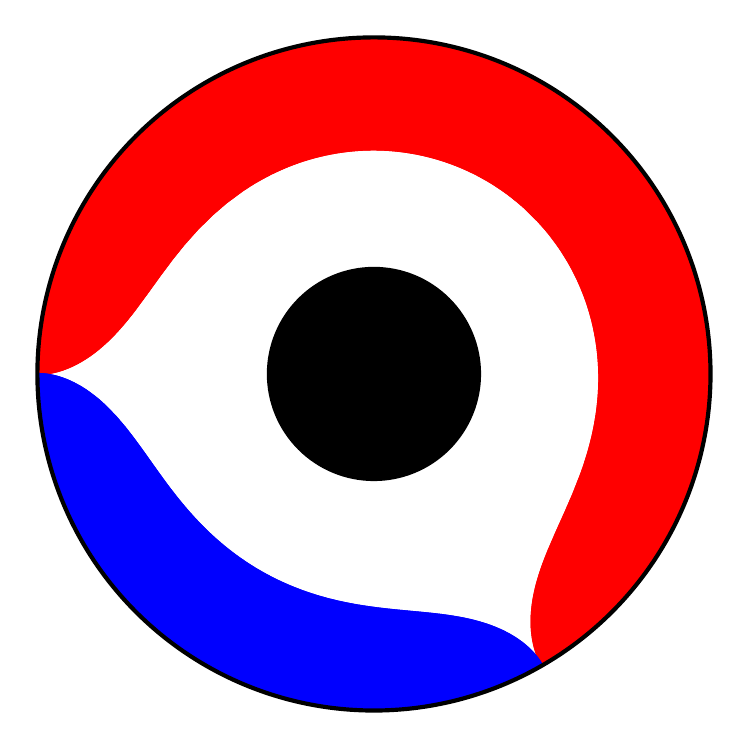}
            \caption[]%
            {\centering {\small }}    
        \end{subfigure}
        \caption[  ]
        {\justifying\small The ``phase transition'' in the entanglement wedge for an interval of angular half-width $w$ at finite temperature. (a) the $t=0$ slice of the entanglement wedge for $w<w_c$. (b) the $t=0$ slice of the entanglement wedge for $w>w_c$. 
        (c) The red shaded region gives the $t=0$ slice of the causal domain for $w>w_c$. The blue shaded region is the causal domain for its complement (which coincides with its entanglement wedge).
        } 
\label{fig:repeatDiamonds}
\end{figure}

\begin{figure}[h]

        \begin{subfigure}[b]{0.3\textwidth}   
            \centering 
			\includegraphics[width=\textwidth]{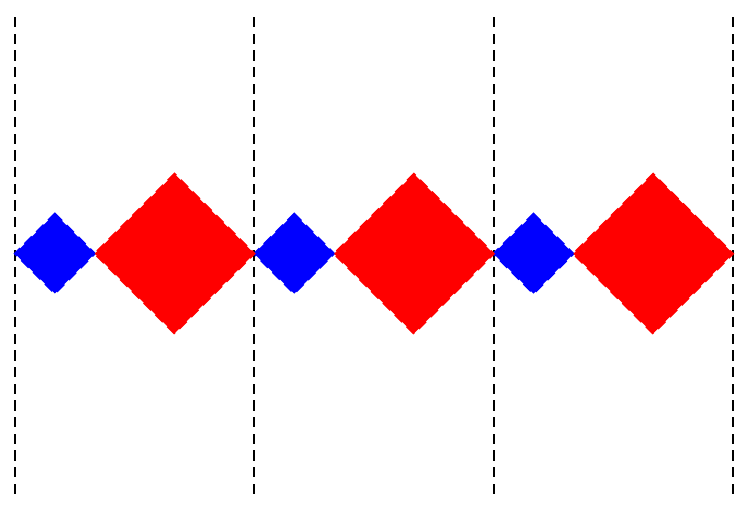}
            \caption[]%
            {{\small }}    
        \end{subfigure}
\hfill
        \centering
        \begin{subfigure}[b]{0.32\textwidth}
            \centering
            \includegraphics[width=\textwidth]{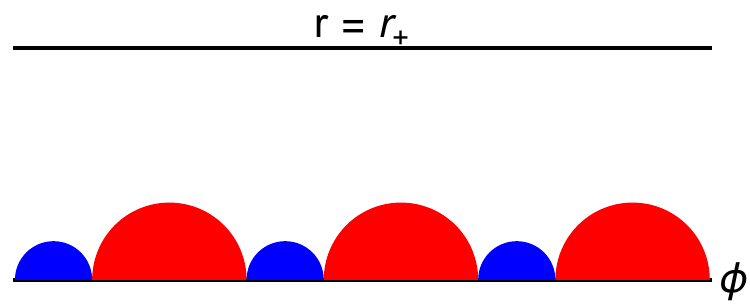}
            \caption[]%
            {\centering {\small }}    
        \end{subfigure}
        \hfill
        \begin{subfigure}[b]{0.32\textwidth}   
            \centering 
            \includegraphics[width=\textwidth]{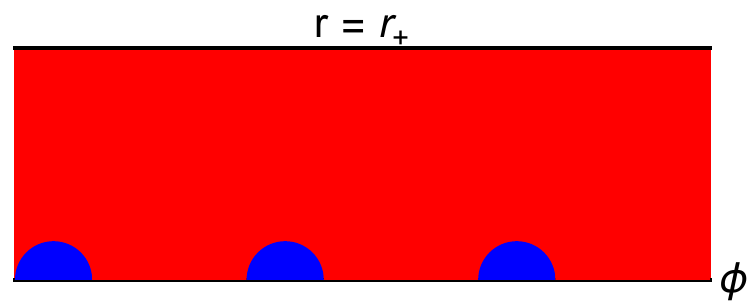}
            \caption[]%
            {\centering {\small }}    
        \end{subfigure}
	        \caption[  ]
        {\justifying\small Causal domains and entanglement wedges of diamonds on the covering space (the Minkowski plane) of a Lorentzian cylinder. (a)  Image of the causal completion of an interval (red) and its causal complement (blue) on the cylinder in the covering space. 
(b) Causal domain of the unions of red or blue diamonds. (c) Entanglement wedge of the union of red or blue diamonds. } 
        \label{fig:CWEWfromBlackBrane}
\end{figure}

 A ``phase transition'' in the entanglement wedge is also encountered for a single interval at finite temperature, with the bulk geometry described by a BTZ black hole. See Fig.~\ref{fig:repeatDiamonds}. This transition can again be understood as the onset of violation of the additivity property, as indicated in figures~\ref{fig:repeatDiamonds} and~\ref{fig:CWEWfromBlackBrane}. For boundary diamonds of half-widths $w < w_c < \pi,$ the entanglement wedge and causal domain coincide and are given in Fig.~\ref{fig:repeatDiamonds} (a). When $w > w_c$ (but still $w < \pi$) the entanglement wedge undergoes a ``phase transition'' to suddenly include points all the way up to the black hole horizon, see Fig.~\ref{fig:repeatDiamonds} (b). The entanglement wedge now exceeds the causal domain which is given by the red region in Fig.~\ref{fig:repeatDiamonds} (c) and does not include points near the horizon.

We can understand this ``phase transition'' in the entanglement wedge for an interval at finite temperature from superadditivity. Going to the covering space of the boundary cylinder, the domains of dependence of the interval and its complement are mapped to the repeated diamond regions, $\cup_i D_i,~ \cup_i \tilde{D}_i,$ shown in red and blue respectively in Fig.~\ref{fig:CWEWfromBlackBrane} (a). We can consider two possible algebras for the union of red diamonds in Fig.~\ref{fig:CWEWfromBlackBrane} (a): $\lor_i \sX_{D_i}$ or $\sX_{\cup_i D_i}$ (with respect to the thermal state). The phase transition of the entanglement wedge at $w = w_c$ can then be interpreted as the onset of superadditivity at $w=w_c$: these two algebras become different for $w > w_c$. 
 The algebra $\lor_i \sX_{D_i}$ has bulk dual shown in red in Fig.~\ref{fig:CWEWfromBlackBrane} (b) while $\sX_{\cup_i D_i}$ has bulk dual shown in red in Fig.~\ref{fig:CWEWfromBlackBrane} (c). With $\lor_i \sX_{D_i}$ in the covering space, upon mapping back to the cylinder we obtain the bulk dual shown in red in Fig.~\ref{fig:repeatDiamonds} (c), the causal domain of the boundary causal completion of the interval. With $\sX_{\cup_i D_i}$ in the covering space, upon mapping back to the cylinder we instead obtain the bulk dual shown in red in Fig.~\ref{fig:repeatDiamonds} (b), the entanglement wedge of the interval.

Superadditivity occurs when the algebra associated to a subregion, $R,$ is larger than the algebra that is `built up' from smaller subregions whose union is $R.$ It reflects some degree of non-locality in the field theory. Unlike a local field theory where we expect that the algebras of a subregion $R$ can be constructed from those associated to infinitesimal diamond-shaped subregions whose union covers $R,$ this is not possible in a theory with superadditivity. In such a theory, the algebra of a region cannot necessarily be decomposed into those of subregions covering it.
Physical implications of superadditivity will be further discussed elsewhere~\cite{addPaper}.

\subsection{Summary} \label{sec:summary}

To conclude this section, we give a summary of the main insights into subregion-subregion duality from the lens of 
equivalence of bulk and boundary operator algebras:

\ben

\item Subregion-subalgebra duality provides a mathematically precise definition of subregion-subregion duality. 
It highlights that subregion-subregion duality can be precisely formulated {\it only} in the large $N$ limit. 
Previous formulations~\cite{Almheiri:2014lwa,Pastawski:2015qua,Hayden:2016cfa,Dong:2016eik,Harlow:2016vwg,Cotler:2017erl} assumed a bulk Hilbert space embedded into the boundary Hilbert space and that both the code subspace corresponding to the bulk Hilbert space and the boundary Hilbert space could be factorized into Hilbert spaces corresponding to subregions. These assumptions are not realized in (field theoretic) examples of the duality.\footnote{Some of these assumptions have been relaxed in~\cite{Kang:2018xqy,Faulkner:2020hzi}, but the formulations are still incompatible with the Reeh-Schlieder theorem~\cite{Kelly:2016edc,Faulkner:2020hzi}.}  
 Properly generalized, such formulations may be interpreted at a finite $N$. But note that at a finite $N$, the bulk geometric picture becomes murky (for example geometric concepts such as the RT surface and the entanglement wedge cannot be precisely defined), and such formulations can at most be approximate. 
Recently,~\cite{Faulkner:2022ada} discussed embeddings of bulk subalgebras in the $G_N \to 0$ limit to boundary theory at a finite $N$, which helps to clarify how subregion-subregion duality emerges in the large $N$ limit. Note that since the boundary and bulk theories are not defined in the same regime of $N$, the embedding cannot be viewed as a duality map.

\item  The new formulation provides an alternative way to define the entanglement wedge and the associated 
RT or HRT surfaces without using entropy associated to a boundary subregion. It also provides a boundary explanation of why the entanglement wedge always encloses causal domain.

Since the boundary operator algebra in a local subregion is type III$_1$, entanglement entropy is  not well defined: a regularization is needed and the answer depends on the regularization. Furthermore, in the $N \to \infty$ limit, even the regularized entropies go to infinity, which means that they are not intrinsically defined objects in the large $N$ limit. In contrast, entanglement wedges and RT or HRT surface are well-defined geometric objects in the $G_N \to 0$ limit---there is no ambiguity in their definitions. 
The new algebraic formulation resolves this tension: the objects on both sides of the duality are well-defined in the same limit.

\item The equivalence of algebras immediately implies that, in the large $N$ limit, the relative entropies of the bulk and boundary theory in the corresponding regions must agree~\cite{Jafferis:2015del}. The conclusion not only applies to entanglement wedges, but also causal 
domains.\footnote{Note that the notions of relative entropies associated to causal domains have no analogue at finite $N.$ Such quantities can thus only be used to probe properties of states in the large $N$ limit.}

\item A surprising result following from~\cite{Jafferis:2015del} and global reconstruction~\cite{Banks:1998dd,Bena:1999jv,Hamilton:2006az} is that modular flows of single-trace operators in a spatial subregion $R$ are part of the associated single-trace algebra in the large $N$ limit, although they may not be written in terms of local single-trace operators in $\hat R$. This underlies our proposal~\eqref{im2}. We have now been able to see this in an explicit example in Sec.~\ref{sec:expD}. It may be expected that operators generated from non-geometric modular flows 
can be exponentially complex when viewed  at a large, but finite $N$. This means that certain exponentially complex computational tasks at a finite but large $N$, may be achieved ``easily'' using operators in the algebra of single-trace operators. An example is the reconstruction of bulk operators in the region lying outside the causal domain, but still in the entanglement wedge, i.e. what is called Python's lunch~\cite{Brown:2019rox,Engelhardt:2021mue,Engelhardt:2021qjs}.

\een

\section{Duality for general bulk regions} \label{sec:time} 

Entanglement wedge reconstruction and the RT or HRT prescription allow us to study bulk subregions bounded by extremal surfaces that contain near-boundary points. In order to study more general bulk subregions, subregion-subregion duality is not sufficient, and we must pass to the more general subregion-subalgebra duality. In this section we discuss some explicit examples.

\subsection{Bulk dual regions of time bands in the boundary}

\begin{figure}
        \centering
        \includegraphics[width=0.3\textwidth]{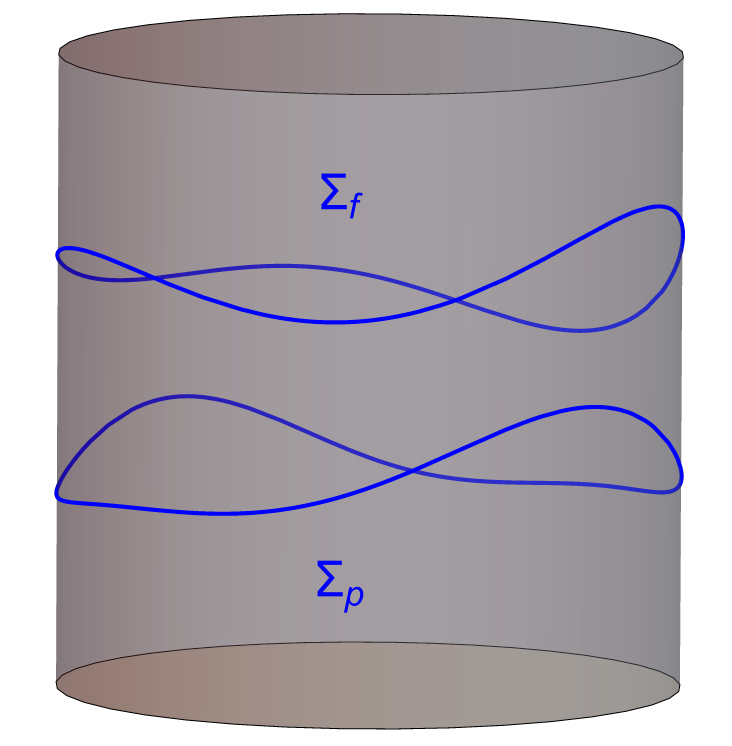}
        \caption[  ]
        {\justifying\small A general time band for $X = S^1$. $T_{p,f}$ is the {\it spacetime} region on the cylinder between the two Cauchy slices $\Sig_f$ and $\Sig_p$.} 
        \label{fig:genTimeBand}
\end{figure}

Consider the boundary with topology $\mathbb{R} \times X,$ for some spatial manifold $X.$ A time-band\footnote{{For other discussions of the algebras associated to time-bands, see~\cite{Banerjee:2016mhh,Bahiru:2022oas}}} on the boundary is a {\it spacetime} subregion that contains an entire Cauchy slice. It can be specified as the region between two non-intersecting Cauchy slices, $\Sigma_p$ and $\Sigma_f.$ See figure~\ref{fig:genTimeBand}. We will denote the time band bounded by $\Sigma_p$ to the past and $\Sigma_f$ to future as $T_{p,f}.$

\begin{figure}
        \centering
        \includegraphics[width=0.3\textwidth]{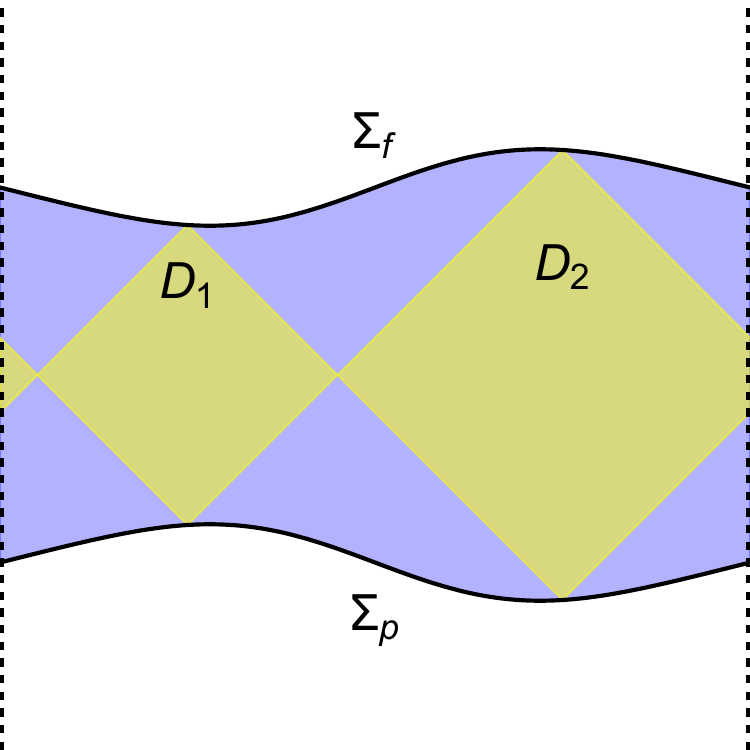}
        \caption[  ]
        {\justifying\small A boundary time-band (blue) can be constructed from the union of an infinite number of diamond-shaped subregions (yellow). The left and right dashed lines in the figure are identified. Here the time band is assumed to be not too long and we only show two diamonds.} 
        \label{fig:diad}
\end{figure}

While $T_{p,f}$ is not a causally complete boundary spacetime subregion (its causal completion is the entire boundary) we can still associate to it a subalgebra,  
$\sM_{T_{p,f}}$, by 
restriction of $\sO(x)$ to $x \in T_{p,f}$. Any $T_{p,f}$ can be constructed from a union of an infinite number of diamond shaped regions $\{D_i\}$,
with $\cup_i D_i = T_{p,f}$, see Fig.~\ref{fig:diad} for an example. 

Since $\sM_A$ is defined as the restriction of $\sO(x)$ to $x \in A$ and the union of the diamonds exactly covers the time-band, i.e. $\cup_i D_i = T_{p,f},$ we have 
\be \label{TBAddDiamonds}
	\le(\sM_{T_{p,f}}\ri)'' = \lor_{i} \sM_{D_i}  \ . 
\ee
Notice that the resulting algebra $\le(\sM_{T_{p,f}}\ri)''$ is independent of the choice of diamonds $\{D_i\}$ used to build $T_{p,f}$.

From the duality for diamond-shaped boundary subregions $\sM_{D_i} = \widetilde{\sM}_{\fc_{D_i}}$ and additivity of the bulk field theory we have 
\be \label{addBulkDiams}
	\lor_{i} \sM_{D_i} = \lor_{i} \widetilde{\sM}_{\fc_{D_i}} = \widetilde{\sM}_{\le(\cup_i \fc_{D_i}\ri)''} \ .
\ee
The causal domain of the time band $\fc_{T_{p,f}}$ is defined by
\be \label{simpWTimeBand}
	\fc_{T_{p,f}} = \le(J^-(T_{p,f}) \cap J^+(T_{p,f})\ri)'' \ ,
\ee
i.e. the bulk causal completion of the intersection of the bulk past and future of $T_{p,f}.$ Since $\cup_i D_i = T_{p,f},$ from the bulk causal structure we have 
\be \label{equalCausalDomains}
\fc_{T_{p,f}} = \le( \cup_i  \fc_{D_i} \ri)'' \Rightarrow \widetilde{\sM}_{\fc_{T_{p,f}}} = \widetilde{\sM}_{\le(\cup_i \fc_{D_i}\ri)''}\ .
\ee
Thus, from~\eqref{equalCausalDomains},~\eqref{addBulkDiams}, and~\eqref{TBAddDiamonds} we have
\be
	 \le(\sM_{T_{p,f}}\ri)'' = \widetilde{\sM}_{\fc_{T_{p,f}}} \ ,
\ee
showing that $\fc_{T_{p,f}}$ defined in~\eqref{simpWTimeBand} is indeed the bulk dual of $\le(\sM_{T_{p,f}}\ri)''.$

\begin{figure}
        \centering
        \includegraphics[width=0.3\textwidth]{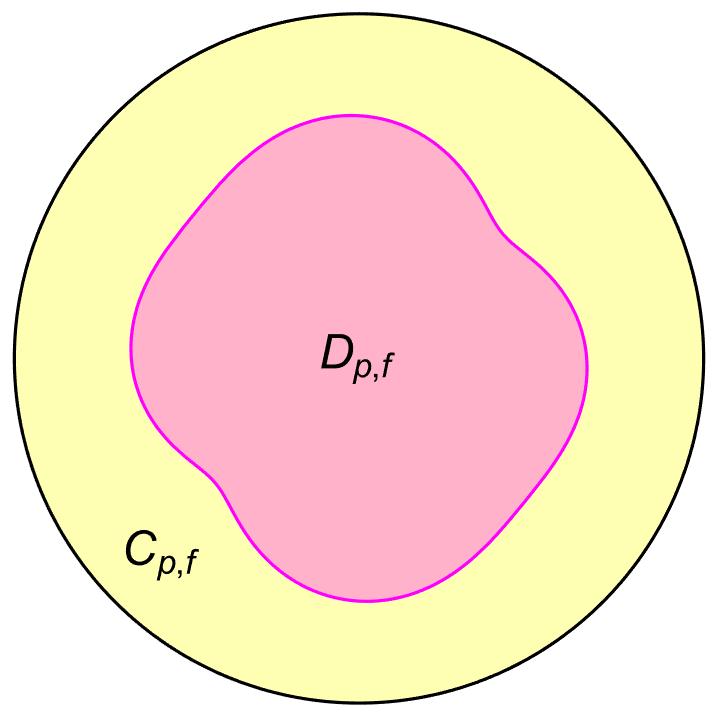}
        \caption[  ]
        {\justifying\small The bulk dual of $\le(\sM_{T_{p,f}}\ri)''$ intersects a bulk Cauchy slice in the yellow region, $C_{p,f}.$ By bulk Haag duality the complementary region $D_{p,f} = \overline{C_{p,f}}$ shown in pink must be described by the commutant algebra, $\le(\sM_{T_{p,f}}\ri)'$.} 
        \label{fig:gBand}
\end{figure}

Since the bulk theory obeys Haag duality, the commutant, $\le(\sM_{T_{p,f}}\ri)',$ of $\sM_{T_{p,f}}$ 
should then be dual to the bulk spacetime region $(\fc_{T_{p,f}})'$. The bulk dual of the time-band is a causally complete bulk subregion and thus is the domain of dependence of a {\it spatial} bulk subregion on some Cauchy slice. Consider that bulk Cauchy slice with a subregion $C_{p,f}$ and its complement $D_{p,f} = \overline{C_{p,f}}$ such that $\widehat{C_{p,f}} = \fc_{T_{p,f}}$ and $\widehat{D_{p,f}} =  (\fc_{T_{p,f}})'$, see Fig.~\ref{fig:gBand}. $D_{p,f}$ and its boundary $b_{p,f} = \p D_{p,f}$ lie completely in the bulk, with no near-boundary points, but we can still identify the boundary algebra dual to $\widehat{D_{p,f}}$ as $\le(\sM_{T_{p,f}}\ri)'.$

In~\cite{Balasubramanian:2013lsa, Headrick:2014eia}, there was a very interesting proposal to associate an entropy to bulk surfaces like $b_{p,f}$ by connecting it to a collection of boundary diamonds analogous to the set $\{D_i\}$ discussed above. The discussion there  was in the reverse direction: given a co-dimension two bulk surface, they used co-dimension two extremal surfaces tangent to the surface at each point to give a collection of diamond-shaped boundary regions, $\{\tilde{D}_i\}$.  {In our language, their procedure describes the bulk dual\footnote{{In the case that the co-dimension two bulk surface is nowhere-concave.}} of $\lor_i \sX_{\tilde{D}_i},$ which differs from $\lor_i \sM_{\tilde{D}_i}$ for general non-vacuum states.}\footnote{{The bulk dual of $\lor_i \sM_{\tilde{D}_i} = \le(\sM_{\tilde{T}_{p,f}}\ri)''$ with $\tilde{T}_{p,f} \equiv \cup_i \tilde{D}_i,$ will be a bulk subregion contained within the causal completion of the exterior of the co-dimension two surface studied in~\cite{Balasubramanian:2013lsa, Headrick:2014eia}.}}
Despite these differences, it is of great interest to see whether the entropy discussed there can be considered as being associated with the algebra $\le(\sM_{T_{p,f}}\ri)''$ or $(\sM_{T_{p,f}})'$ {when working in the vacuum}.

We will now study some examples in which an explicit description of the causal domain of the time-band can be given.

\subsection{Vacuum}
Take the boundary spatial slice to be a circle, i.e. $X = S^1.$  In the vacuum state, the bulk dual is pure AdS$_3,$ which can be described by coordinates $(t, \phi, \rho)$ with metric 
\be \label{globAdSMet}
	ds^2 = {l^2 \ov \cos^2\rho} \le(-dt^2 + d\rho^2 + \sin^2\rho ~d\phi^2 \ri) \ ,
\ee
and coordinate ranges $t \in \mathbb{R},~ \phi \in (-\pi,\pi],~ \rho \in (0, {\pi \ov 2}).$ The boundary is at $\rho = {\pi \ov 2}$.\footnote{Note that we have compactified the radial coordinate in this discussion.}

\begin{figure}
        \centering
        \begin{subfigure}[b]{0.35\textwidth}
            \centering
            \includegraphics[width=5cm]{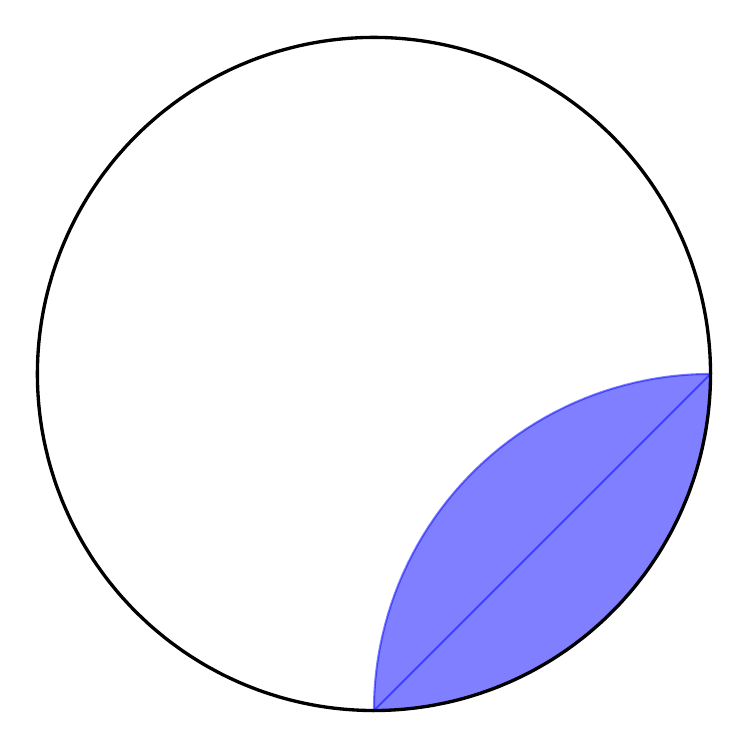}
            \caption[]%
            {\centering{\small }}    
        \end{subfigure}
        \qquad \qquad 
        \begin{subfigure}[b]{0.35\textwidth}   
            \centering 
            \includegraphics[width=5cm]{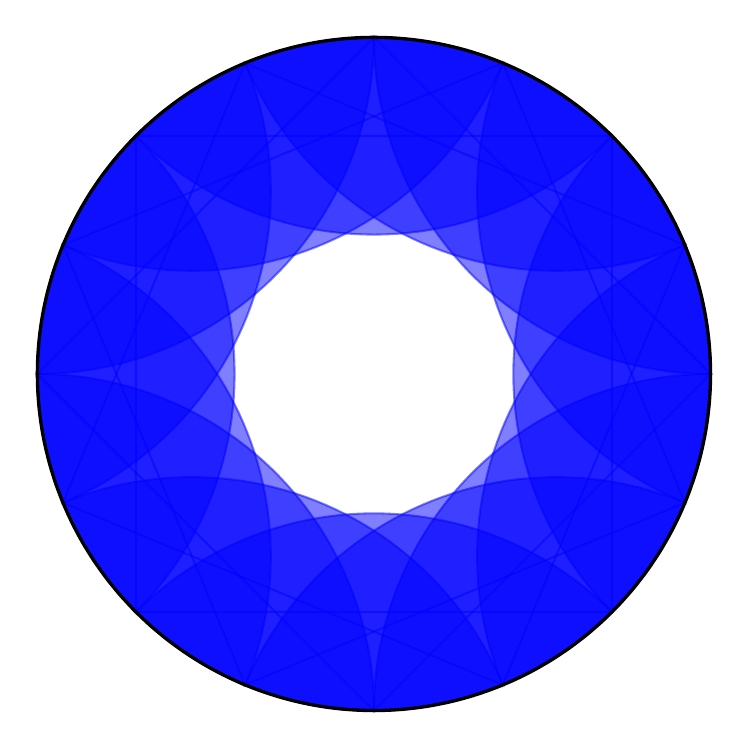}
            \caption[]%
            {\centering{\small }}    
        \end{subfigure}
        \caption[  ]
        {\justifying\small Bulk time slice of the causal domains for diamond-shaped regions and the time-band. (a) Intersection of the causal domain (and entanglement wedge) of a diamond-shaped region of half-width $w={\pi \ov 4}$ with the $t=t_0$ bulk slice. (b) Intersection of the union of causal domains of sixteen diamond-shaped regions, each of half-width $w={\pi \ov 4},$ with the $t=t_0$ bulk slice. In the limit of the continuous union of diamonds around the boundary circle, the causal wedge covers $H_{t_0,w}$ on the $t=t_0$ slice. } 
        \label{fig:vacuumTimeBand}
\end{figure}

We consider a uniform time band of total width $2w,$ i.e. we take $\Sig_p$ to be described by $t = t_0 - w$ and $\Sig_f$ by $t = t_0 + w$ for some fixed $w > 0.$ When $w<\pi,$ for each $\phi_0 \in (-\pi, \pi],$ the diamond of angular (and temporal) width $w$ centered at $(t=t_0, \phi = \phi_0)$ is contained in the time-band $T_{p,f}$ and the union of such diamonds over all $\phi_0$ is $T_{p,f}$. 
When $w < {\pi \ov 2},$ the intersection of the entanglement wedge (or, equivalently in this case, the causal domain) of such a diamond with the $t=t_0$ bulk slice is bounded by a bulk spacelike geodesic, $\rho = \rho_{\phi_0,w}(\phi),$ reaching a minimum value of the (compactified) radial coordinate $\rho_{\min}(w) = {\pi \ov 2} - w$ at $\phi = \phi_0$. See figure~\ref{fig:vacuumTimeBand} (a). The union of the entanglement wedges of all such diamonds then intersects the $t=t_0$ slice in the subregion $H_{t_0,w} = \{(\phi, \rho) ~|~ \rho > \rho_{\min}(w)\}$. See figure~\ref{fig:vacuumTimeBand} (b). The bulk dual of the uniform time-band of half-width $w < {\pi \ov 2}$ is then the bulk causal completion of $H_{t_0,w}.$

When ${\pi \ov 2} < w < \pi,$ the time-band can still be covered by a one-parameter family of diamond-shaped regions centered on the $t=t_0$ slice. However, the entanglement wedge of such a diamond now covers more than half of the bulk $t=t_0$ slice, including the `centre' of AdS, $\rho = 0.$ The union of entanglement wedges of such diamond-shaped regions thus covers the entire $t=t_0$ Cauchy slice of the bulk, and therefore the bulk dual of a time band of total width between $\pi$ and $2\pi$ is the entire bulk spacetime. When $w > \pi,$ there are no longer diamond-shaped regions of half-width $w$ on the boundary, however, since such a time-band clearly contains another time-band of half-width larger than ${\pi \ov 2},$ the bulk dual of this larger time band is also the entire bulk spacetime.

This is precisely as one would expect purely from the boundary GFF theory. Around the vacuum, the GFF $\pi_{\Om} (\sO (x^\mu))$ can be expanded as 
\bega
\label{yhn2}
\pi_{\Om} (\sO (t, \phi)) = \sum_{nl }  C_{nl}  \left[ a_{nl } e^{- i \om_{nl} t + i l \phi}  + a_{nl }^\da e^{ i \om_{nl} t - i l \phi}  \ri] ,
 \\
 \om_{nl} = \De + l + 2n, \quad n =0,1,\cdots, \quad l \in \ZZ,
 \end{gather} 
where $C_{nl}$ are some constants. Due to quantized nature of $\om_{nl}$ we can express $a_{nl}$ in terms of 
$\pi_{\Om} (\sO (t, \phi)) $ within any time band with half width $w \geq {\pi \ov 2}$. Explicitly, we have (as in~\cite{Hamilton:2006az})
\be 
	a_{nl} = {1 \ov C_{nl}} \int_{t_0 - {\pi \ov 2}}^{t_0 + {\pi \ov 2}} {dt \ov \pi} \int {d\phi \ov 2\pi} e^{i\om_{nl}t - il\phi} ~\pi_{\Om} (\sO (t, \phi))_+ \ ,
\ee
where $\pi_{\Om} (\sO (t, \phi))_+$ denotes the positive frequency part $\pi_{\Om} (\sO (t, \phi)).$

\subsubsection{Temporally separated time bands}
\begin{figure}
        \centering
        \begin{subfigure}[b]{0.2\textwidth}
            \centering
            \includegraphics[width=\textwidth]{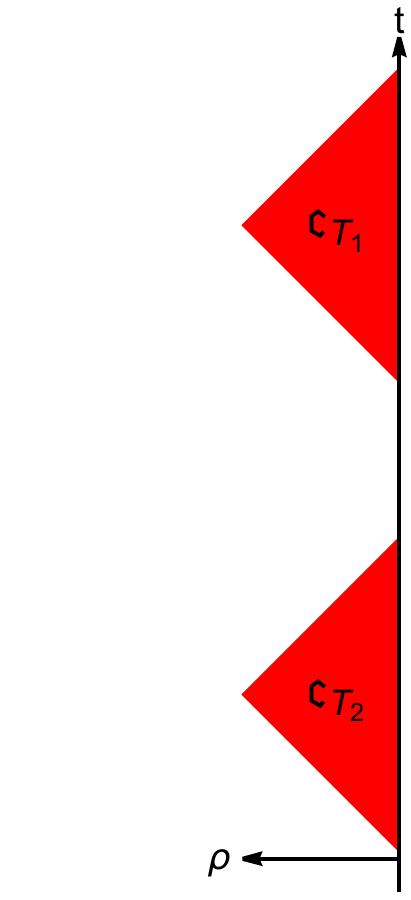}
            \caption[]%
            {{\small }}    
        \end{subfigure}
        \qquad \qquad \qquad \qquad
        \begin{subfigure}[b]{0.2\textwidth}   
            \centering 
            \includegraphics[width=\textwidth]{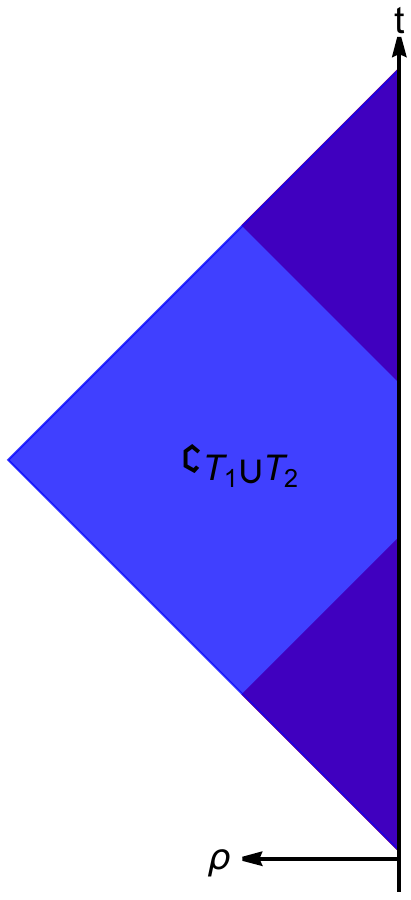}
            \caption[]%
            {{\small }}    
        \end{subfigure}
        \caption[  ]
        {\justifying\small (a) Bulk duals of $\sM_{T_1}$ and $\sM_{T_1}$. (b) Bulk dual of $\le(\sM_{T_1 \cup T_2}\ri)''$ (with $t_1 + w_1 - (t_2-w_2) < \pi$). Both figures are a constant $\phi$ slice of global AdS$_3$.} 
        \label{fig:sepTimeBands}
\end{figure}

Now consider two temporally separated time-bands, $T_1 : t \in (t_1 -w_1, t_1 + w_1)$ and $T_2 : t \in (t_2 -w_2, t_2 + w_2).$\footnote{We thank Ahmed Almheiri, Venkatesa Chandrasekaran, and Henry Lin for discussions on this.} Without loss of generality we take $t_1 > t_2.$ 
If $w_1 > {\pi \ov 2}$ or $w_2 > {\pi \ov 2}$ then the corresponding time-band algebra is equal to the algebra on the entire boundary spacetime. In this case the algebra of the union of time-bands $\le(\sM_{T_1 \cup T_2}\ri)'' \equiv \sM_{T_1} \lor \sM_{T_2}$ is equal to the full algebra, $\sB(\sH^{\rm GNS}_\Om)$ as well. We will thus restrict to $w_1, w_2 < {\pi \ov 2}$. While we can in principle 
deduce $\le(\sM_{T_1 \cup T_2}\ri)''$ by working with boundary GFF, we will do so by using the bulk duals of $T_1$ and $T_2$, which 
can be directly read from the bulk geometry. From Fig.~\ref{fig:sepTimeBands}, we find that $\le(\sM_{T_1 \cup T_2}\ri)''$ is given by the time band 
formed by joining the two time bands and the spacetime region between them.

Explicitly, we find that $\le(\sM_{T_1 \cup T_2}\ri)'' = \sM_{T_{12}},$ where
\be
	T_{12} = \{(t,\phi) ~|~ t_m - w_m < t < t_m + w_m \} \ , 
\ee
with $t_m = \ha \le(t_1 + w_1 + t_2 - w_2\ri)$ and $w_m \equiv {t_1 + w_1 -(t_2 - w_2) \ov 2}.$ i.e. The algebra of the union of two temporally separated time-bands is equal to the algebra of the smallest single (connected) time-band that contains the original time-bands.

\subsubsection{More general boundary spacetime regions}
Consider the case of the boundary spatial manifold $X = \mathbb{R},$ so that the boundary spacetime is $(1+1)-$dimensional Minkowski space. With the boundary theory in the vacuum state, the bulk dual is Poincar\'e AdS$_3$. Consider a time-reflection symmetric spacetime subregion. Such time-reflection symmetric regions are of interest since their bulk duals are domains of dependence of bulk subregions contained within a time-reflection symmetric Cauchy slice of the bulk. For simplicity, we consider a connected subregion with axis of time reflection symmetry being the $t=0$ boundary slice. Such a subregion can then be described by $T_w = \{(t,x) ~|~ -w(x) < t < w(x),~ x \in I\},$ where $I$ is an interval on the $x-$axis and $w(x) > 0$ on $I$.

In this case, the causal domain of $T_w$ is the domain of dependence of a bulk spatial subregion $C_w$ on the bulk $t=0$ slice. 
Since Poincar\'e AdS$_3$ is conformally flat, a point $(t,x,z)$ is in the past of $T_w$ only if $(t-w(x_0))^2 > (x-x_0)^2 + z^2,~ t < w(x_0)$ for some $x_0 \in I.$ Focusing on points on the bulk $t=0$ slice, one finds that the only points in the bulk past of $T_w$ are
\be 
	\{(t,x,z) ~|~ t=0,~ 0 < z < z_w(x)\} \ ,
\ee
where
\be \label{t0TBCurve}
	z_w(x) = \max_{x_0 \in I} \sqrt{w(x_0)^2 - (x-x_0)^2} \ ,
\ee
and for each $x,$ the maximum is over all $x_0$ such that the argument of the square root is positive. If no such $x_0$ exists for a given $x,$ we have $z_w(x) = 0.$ By time-reflection symmetry, the exact same set of points on the $t=0$ bulk slice are in the bulk future of $T_w,$ and thus the intersection of the causal domain of $T_w$ with the $t=0$ slice is exactly this set, 
\be 
	C_w = \{(t,x,z) ~|~ t=0,~ 0 < z < z_w(x)\} \ .
\ee

\begin{figure}
        \centering
        \begin{subfigure}[b]{0.3\textwidth}
            \centering
            \includegraphics[width=\textwidth]{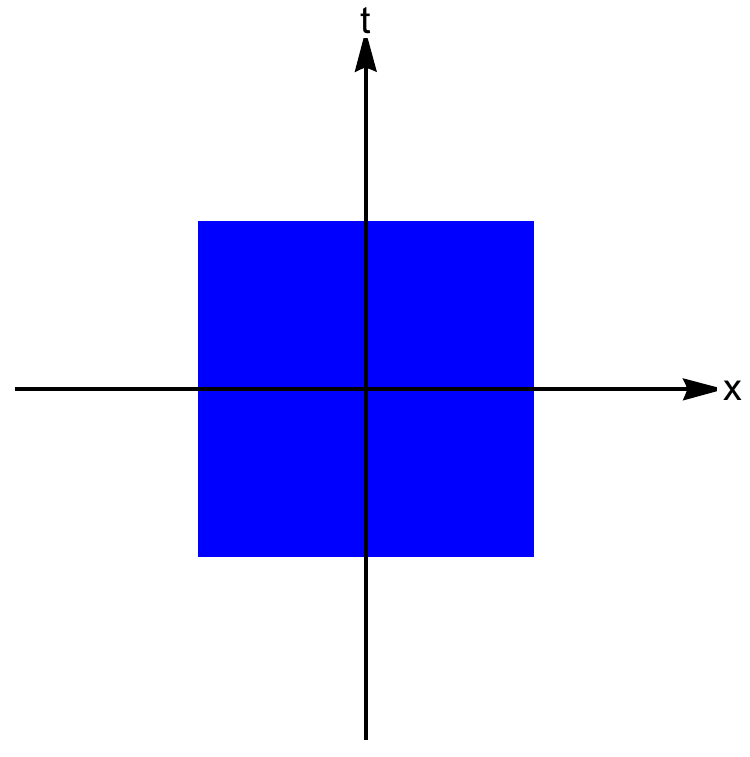}
            \caption[]%
            {{\small }}    
        \end{subfigure}
        \hfill
        \begin{subfigure}[b]{0.3\textwidth}   
            \centering 
            \includegraphics[width=\textwidth]{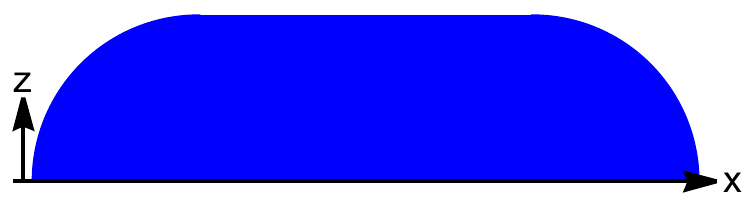}
            \caption[]%
            {{\small }}    
        \end{subfigure}
        \hfill
        \begin{subfigure}[b]{0.3\textwidth}   
            \centering 
            \includegraphics[width=\textwidth]{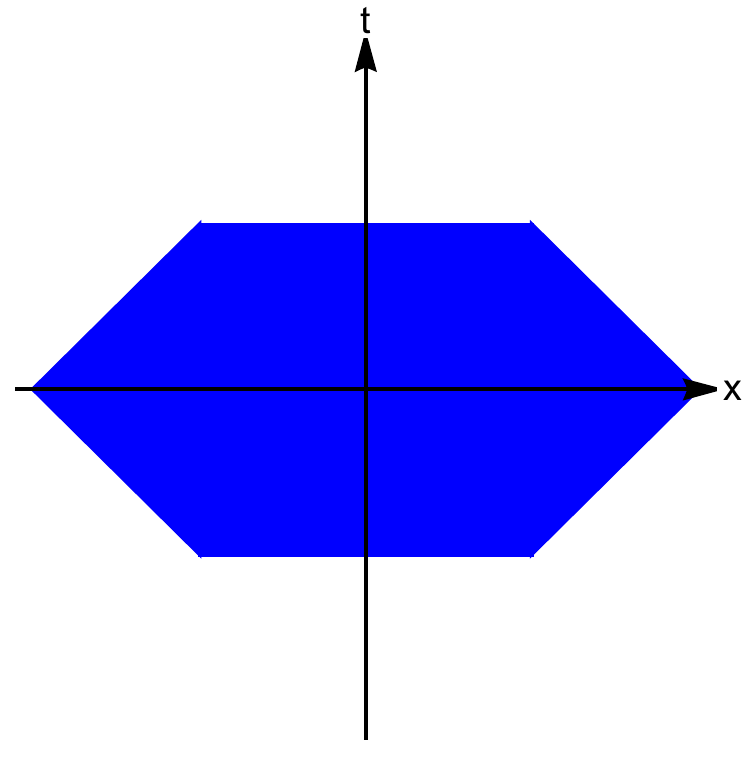}
            \caption[]%
            {{\small }}    
        \end{subfigure}
        \caption[  ]
        {\justifying\small (a) A `square' boundary {\it spacetime} subregion. (b) $t=0$ slice of causal domain of the `square' boundary subregion. (c) A boundary subregion with the same bulk causal domain. } 
        \label{fig:squareSubReg}
\end{figure}

For example, the choice of a `square' boundary subregion, see figure~\ref{fig:squareSubReg} (a), of half-width $w>0,$ i.e.
\be
	w(x) = w,~ x \in I = [-w,w]
\ee
leads to a causal domain whose intersection with the bulk $t=0$ slice, $C_w$, is shown in figure~\ref{fig:squareSubReg} (b). The causal domain of this square region turns out to be identical to the causal domain of the larger boundary subregion shown in figure~\ref{fig:squareSubReg} (c).\footnote{We thank Ahmed Almheiri, Venkatesa Chandrasekaran, and Henry Lin for discussions on this.}

An interesting feature of the bulk spatial subregions found in this procedure is that their complements on the $t=0$ bulk slice, $D_w \equiv \overline{C_w},$ are {\bf geodesically convex}, i.e. for any two points $p_1,~p_2 \in D_w,$ there is a unique minimizing geodesic, $\ga_{12},$ on the $t=0$ bulk slice that connects them and, moreover, this geodesic lies entirely in $D_w.$ For a proof of this, see appendix~\ref{app:geodConvex}.
It would be interesting to study if the geodesic convexity observed here can be understood in terms of the entanglement wedges for gravitating regions recently discussed by Bousso and Penington~\cite{Bousso:2022hlz}.

\subsection{An interior causal diamond} 
\begin{figure}
        \centering
        \includegraphics[width=8cm]{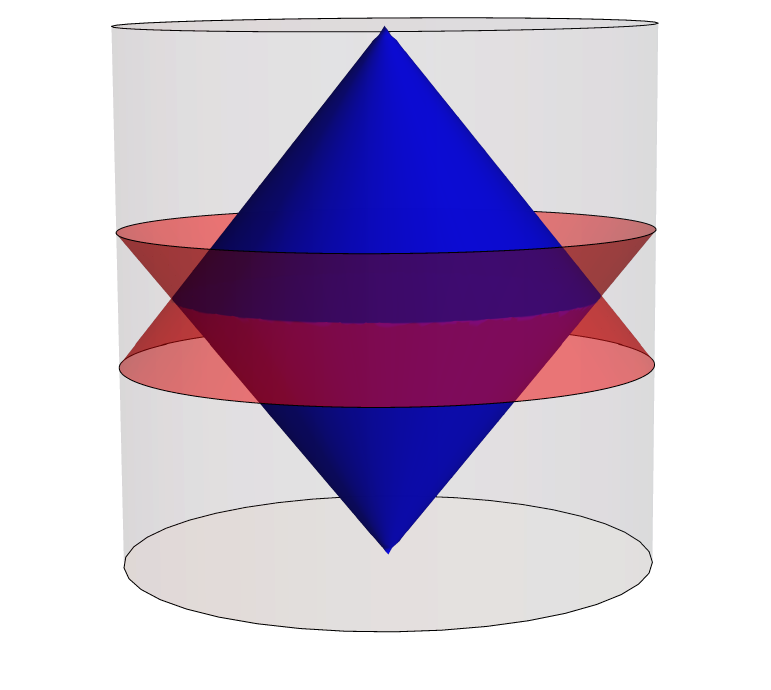}
        \caption[  ]
        {\justifying\small An interior causal diamond is shown in blue. Its causal complement is the region between the red surfaces and the conformal boundary. The algebra in the causal complement region is dual to a boundary time-band.} 
        \label{fig:wdw}
\end{figure}

Using Haag duality of the bulk field theory we may use the results of the previous section to understand which boundary algebras describe bulk operators within an interior causal diamond of global AdS. In particular, the algebra of operators in the bulk domain of dependence of a spherically symmetric bulk spatial subregion $H_{t_0, \rho_m} = \{(t,\phi,\rho) ~|~ t=t_0, \rho < \rho_m (< {\pi \ov 2})\}$ in AdS$_3$ is the commutant of the algebra of operators on the domain of dependence of $\overline{H_{t_0, \rho_m}} = \{(t,\phi,\rho) ~|~ t=t_0, \rho > \rho_m\}.$ By the discussion of the previous section, this algebra of operators is dual to the algebra of boundary operators in the uniform time-band, $T,$ of width $\pi - 2\rho_m$ centered at $t_0.$ See figure~\ref{fig:wdw}.

From bulk Haag duality, we then obtain an explicit boundary characterization of the algebra of bulk operators in the interior causal diamond that is the domain of dependence of the spatial subregion $H_{t_0, \rho_m}.$ The boundary dual of this interior diamond is $\le(\sM_{T}\ri)',$ the commutant of a time-band algebra.

Notice that the algebra of such an interior causal diamond, which contains no near-boundary points, is the commutant of the algebra of a time-band. Recalling that time-band algebras are only non-trivial subalgebras in the large $N$ limit of the boundary CFT, we see that there is no finite $N$ analogue of the algebra dual to bulk operators in an interior causal diamond. At finite $N$ the commutant of a ``time-band algebra'' is trivial. Thus, bulk subregions with no near-boundary points are {\it purely emergent in the large $N$ limit}. 
The finite $N$ theory can at most describe such bulk subregions approximately, and at the moment we do not have the proper language for such an approximate description.

For more general interior causal diamonds in the bulk the story should be the same. Bulk operators in those patches are described by the commutant of a time-band algebra, though the corresponding time-bands will no longer be spherically symmetric. One interesting subtlety is that, in the examples we have studied, time-reflection symmetric boundary time-bands always describe a subregion of a time-reflection symmetric bulk Cauchy slice that is the complement of a geodesically convex subregion on that slice. Thus, it appears that only interior causal diamonds associated to geodesically convex subregions of a bulk time-reflection symmetric slice can be described by the commutant of the algebra of some boundary time-reflection symmetric time-band. It would be very interesting to understand the boundary dual of interior causal diamonds associated to non-geodesically convex spatial subregions, though there may not be a simple geometric understanding of the boundary subalgebra in this case. 

\subsection{Finite temperature}

Consider again the case of boundary spatial manifold $X = S^1,$ but now with the boundary in a thermal state above the Hawking-Page temperature. The bulk dual is then the BTZ black hole, whose right exterior can be described
by coordinates $(t,\phi,r)$ with metric
\be \label{BTZExtMet}
	ds^2 = -l^2  f(r)  dt^2 + {dr^2 \ov f(r)} + r^2 d\phi^2, \qquad f(r) = {r^2 - r_+^2 \ov l^2} \ ,
\ee
and coordinate ranges $t \in \mathbb{R},~ \phi \in (-\pi,\pi],~ r \in (r_+, \infty).$ The boundary is at $r = \infty$ and the horizon at $r = r_+.$ The AdS length scale is denoted by $l$. The temperature with respect to $t$ (which is dimensionless) is then $T = {r_+ \ov 2\pi l}.$

 Consider a uniform time-band, $T_w : \{(t,\phi) ~|~ -w < t < w\},$ of half-width $w$ centered at $t = 0$ on the boundary.
When $w < \pi,$ we may decompose the time-band into diamond-shaped subregions of angular (and temporal) half-width $w$ centered at a point $(t=0, \phi = \phi_0),$ which we denote by $D_{\phi_0,w}.$ We then have $\cup_{\phi_0} D_{\phi_0,w} = T_{ w}.$ The causal and entanglement wedges of such diamond-shaped boundary subregions in the BTZ black hole geometry were discussed in 
Fig.~\ref{fig:repeatDiamonds}. The subalgebra of the time-band is then $\le(\sM_{T_{w}}\ri)'' = \lor_{\phi_0} \sM_{D_{\phi_0,w}}.$

\begin{figure}
        \centering
        \begin{subfigure}[b]{0.35\textwidth}
            \centering
            \includegraphics[width=\textwidth]{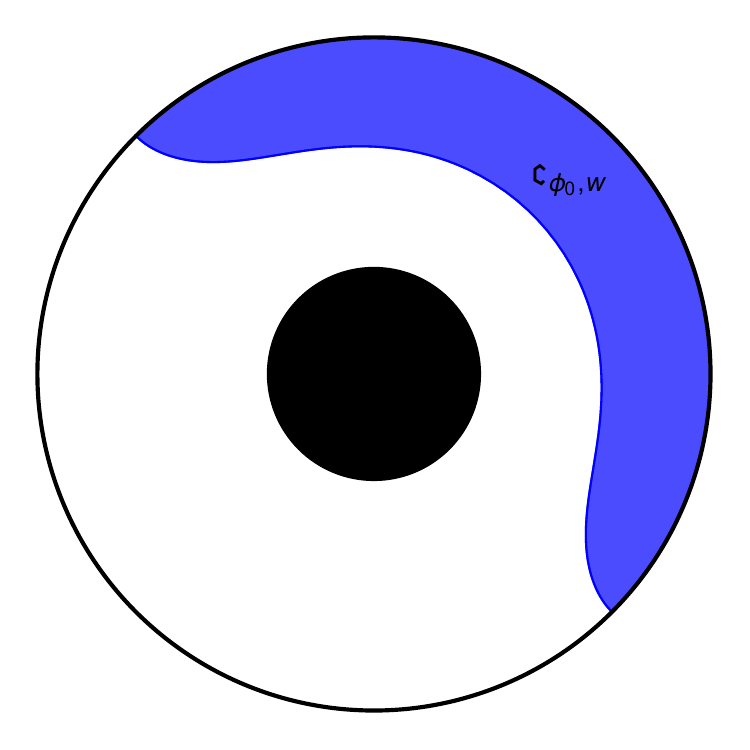}
            \caption[]%
            {{\small }}    
        \end{subfigure}
        \hfill
        \begin{subfigure}[b]{0.35\textwidth}   
            \centering 
	        \includegraphics[width=\textwidth]{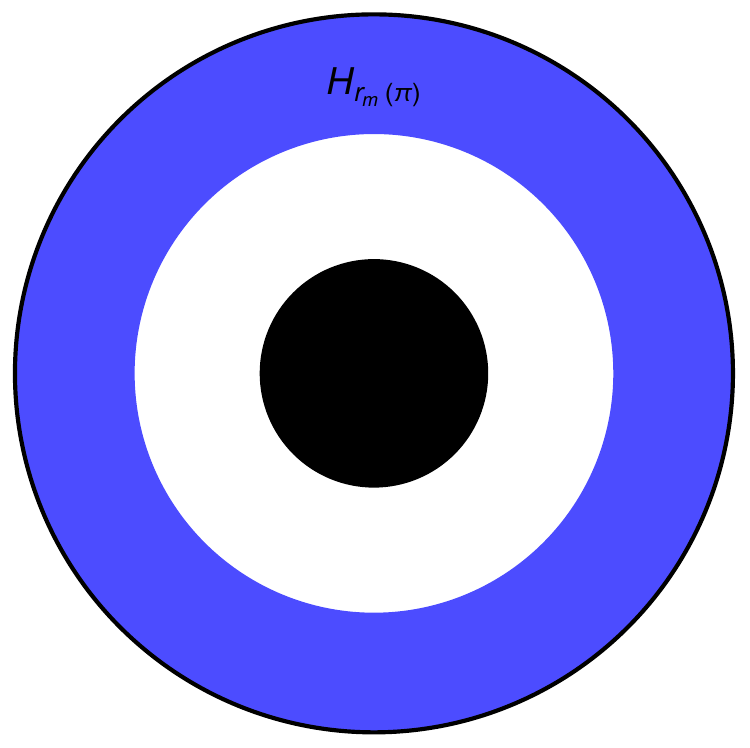}
            \caption[]%
            {{\small }}    
        \end{subfigure}
        \caption[  ]
        {\justifying\small The $t=0$ slice of the right exterior of the BTZ black hole. The black hole is the black region $r \in (0,r_+).$ (a) Bulk dual of a diamond boundary subegion ($\phi_0 = {\pi \ov 4},~ w = {\pi \ov 2}$). (b) Bulk dual of the boundary time band of half-width $w=\pi.$ } 
        \label{fig:cwTimeBandBTZ}
\end{figure}

The bulk region dual to $\le(\sM_{T_{w}}\ri)''$ can then be obtained by the union of causal domains for  $D_{\phi_0,w}$, i.e. 
\be 
\fc_{T_{w}} = \le( \cup_{\phi_0} \fc_{{\phi_0,w}} \ri)'',
\ee
where $\fc_{{\phi_0,w}}$ is the causal completion of the region shown in Fig.~\ref{fig:cwTimeBandBTZ} (a). 
The intersection of $\fc_{T_{w}}$ with the $t=0$ slice is then
\be 
	H_{r_m(w)} = \{(t,\phi,r) ~|~ t=0, r > r_m(w) \} \ ,
\ee
with
\be \label{rMinBTZTimeBand}
	r_m(w) = r_+~ \coth{w r_+ \ov l} \ .
\ee
The bulk dual of such a time-band is then the bulk domain of dependence of $H_{r_m(w)}$. See figure~\ref{fig:cwTimeBandBTZ} (b).
$r_m (w)$ can be obtained from the intersection with the $t=0$ slice of the radial null geodesics from $t=w$ given in~\cite{Cruz:1994ir}. 

\begin{figure}
        \centering
        \begin{subfigure}[b]{0.45\textwidth}
            \centering
            \includegraphics[width=\textwidth]{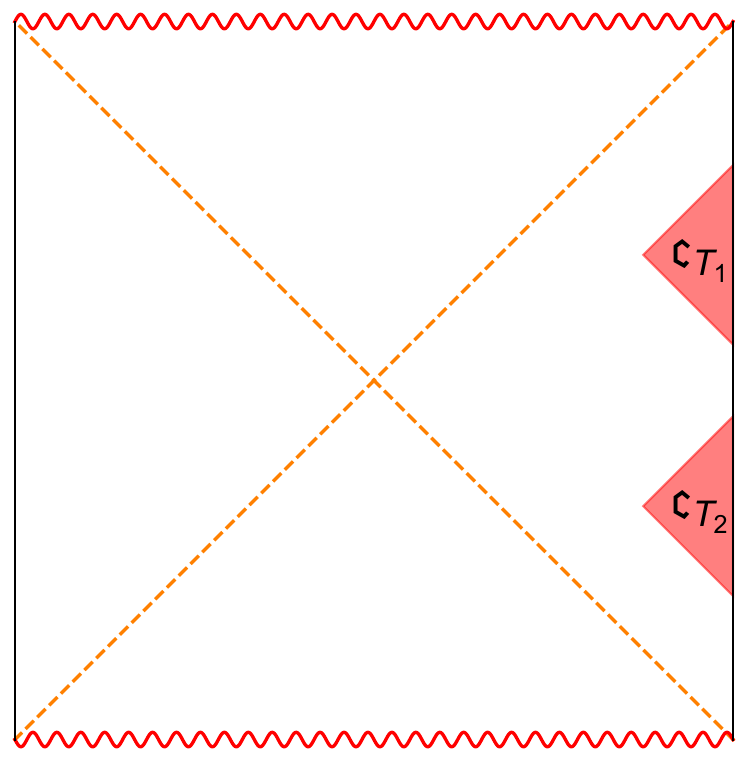}
            \caption[]%
            {\justifying{\small Bulk duals of time-bands, $T_1,~T_2$.}}    
        \end{subfigure}
        \hfill
        \begin{subfigure}[b]{0.45\textwidth}   
            \centering 
            \includegraphics[width=\textwidth]{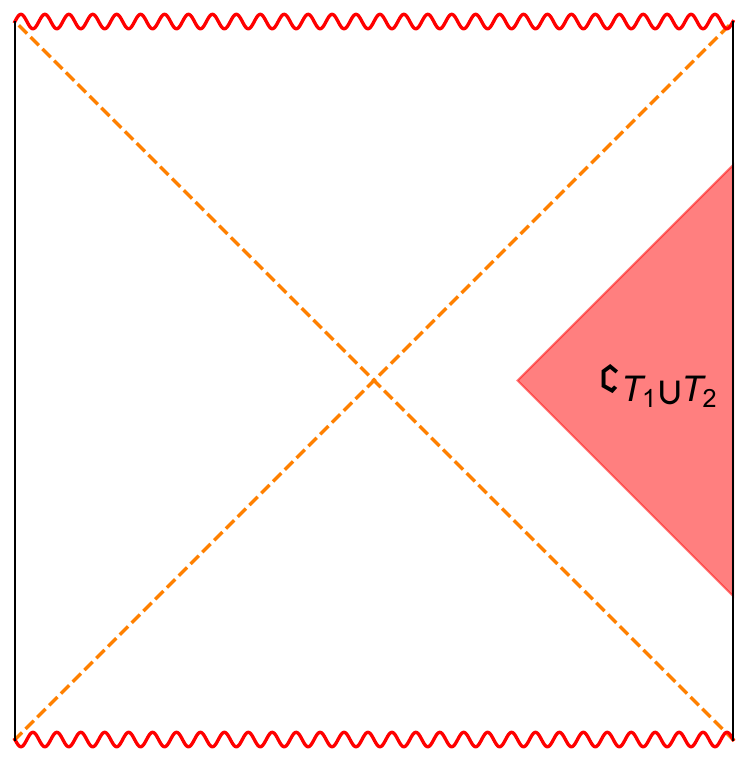}
            \caption[]%
            {\justifying{\small Bulk dual of the union $T_1 \cup T_2$.}}    
        \end{subfigure}
        \caption[  ]
        {\justifying\small Bulk duals of spherically symmetric time-bands at finite temperature.} 
        \label{fig:sepTB-BTZ}
\end{figure}

When $w > \pi,$ we can no longer build the time-band from diamond-shaped boundary subregions centered on a single boundary time-slice. We can, however, construct such a time-band from the union of time-bands each with half-width $w < \pi.$ 
Consider for example, two time bands $T_1$ (centered around $t_1$ with half width $w_1$) and $T_{2}$ (centered around $t_2$ with half width $w_2$)  as indicated in Fig.~\ref{fig:sepTB-BTZ}. From the bulk geometry we find that  
the dual bulk region for $\le(\sM_{T_1 \cup T_2}\ri)'' \equiv \sM_{T_1} \lor \sM_{T_2} $ is given by the causal completion of the region $H_{r_m (w_{12})}$ where $w_{12}$ is the half width of the time band $T_{12}$ obtained by including both $T_1$ and $T_2 $, and the spacetime region between them. Here $H_{r_m (w_{12})}$ lies on the bulk time-slice at the value of $t$ about which $T_{12}$ is time-reversal symmetric. It is thus natural to propose that $\le(\sM_{T_1 \cup T_2}\ri)'' = \le(\sM_{T_{12}}\ri)''$. 
Thus we conclude that the dual region of the boundary time-band $T_w$ is always given by~\eqref{rMinBTZTimeBand} for any $w > 0$. 

It is convenient to describe the bulk subregion dual to a time of width $w$ in Kruskal coordinates, defined in the right exterior of the BTZ black hole by
\be
	U = - \sqrt{r - r_+ \ov r + r_+} e^{- {r_+ \ov l} t},~ V = \sqrt{r - r_+ \ov r + r_+} e^{{r_+ \ov l} t} \ .
\ee
The `innermost' points of $H_{t_0, r_m(w)}$  are then described by Kruskal coordinates
\be \label{kruskalTBBounds}
	U_m = - e^{-{wr_+ \ov l}} e^{- {r_+ \ov l} t_0},~ V_m = e^{-{wr_+ \ov l}} e^{ {r_+ \ov l} t_0} \ .
\ee

\begin{figure}
        \centering
        \includegraphics[width=0.45\textwidth]{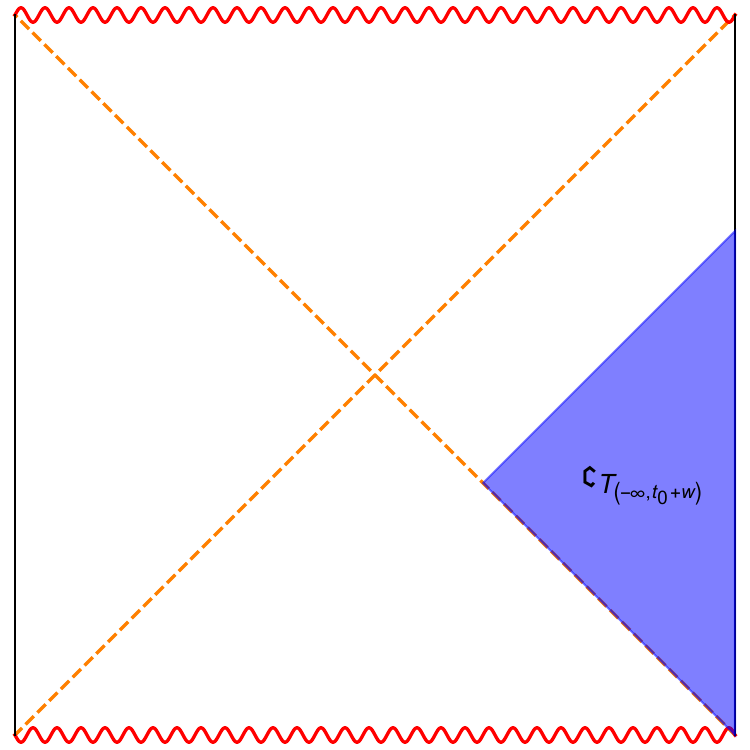}
        \caption[  ]
        {\justifying\small Bulk dual of the algebra of the semi-infinite time-band $T_{-\infty, t_0+w}$ at finite temperature.} 
        \label{fig:cwSemiInf}
\end{figure}

Considering a semi-infinite time-band obtained by the limit $w, (-t_0) \to \infty$ with $t_0 + w$ held fixed, i.e. the time-band $t \in (-\infty, t_0 + w)$. From~\eqref{kruskalTBBounds} we then have the bulk dual to be
\be 
	U < U_m = - e^{-{r_+ \ov l} (t_0 + w)},~ V > V_m = 0,~ UV < -1 \ ,
\ee
which is exactly the bulk dual argued for in~\cite{longPaper}, see Fig.~\ref{fig:cwSemiInf}. 

\begin{figure}
        \centering
        \begin{subfigure}[b]{0.45\textwidth}
            \centering
            \includegraphics[width=\textwidth]{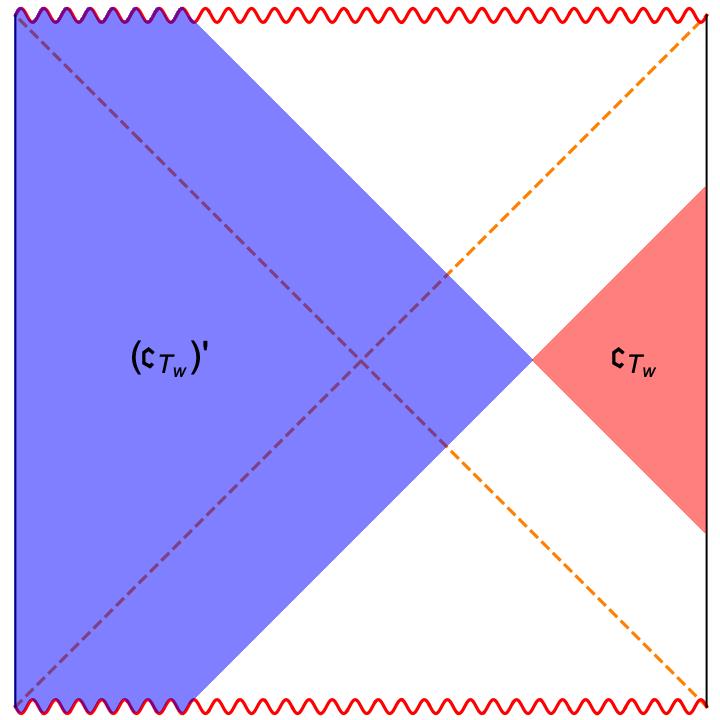}
            \caption[]%
            {{\small }}    
        \end{subfigure}
        \hfill
        \begin{subfigure}[b]{0.45\textwidth}   
            \centering 
            \includegraphics[width=\textwidth]{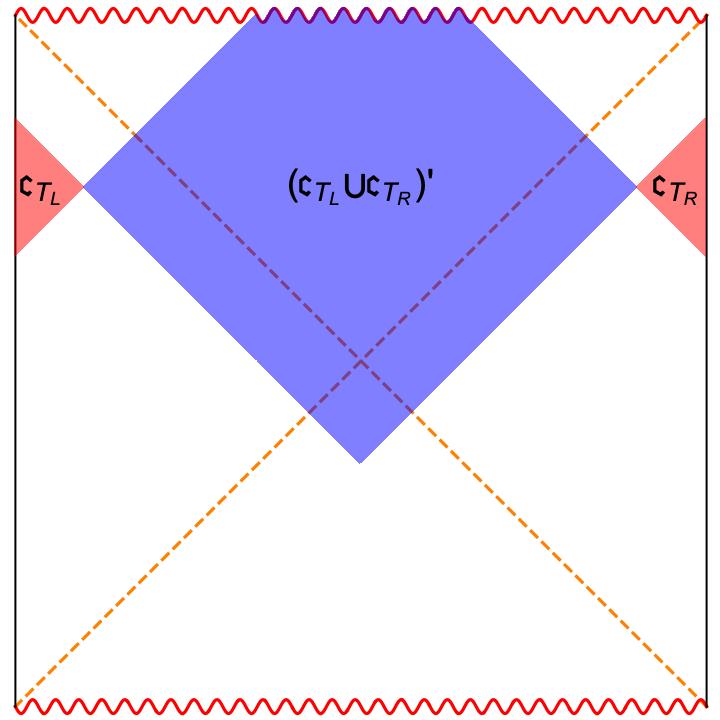}
            \caption[]%
            {{\small }}    
        \end{subfigure}
        \caption[  ]
        {\justifying\small (a) Bulk dual of the algebra of a finite time-band at finite temperature (red) and its commutant (blue). (b) Bulk dual of the addition of the algebras of finite time-bands on the left and right in the thermal field double (red) and the commutant (blue).} 
        \label{fig:compsBTZ}
\end{figure}

The commutant $(\sM_{T_w})'$ of the time-band subalgebra $\le(\sM_{T_w}\ri)''$ is dual to the blue bulk region in Fig.~\ref{fig:compsBTZ} (a).  
This is a bulk subregion that contains points in the right and left exterior regions as well as the future and past interior regions. In particular, it contains the entire left exterior region as well as near singularity points. Since the corresponding algebra is defined as the commutant of an algebra that has no non-trivial analogue at finite $N,$ this algebra is necessarily emergent in the large $N$ limit. 

Interestingly though, in the case that we work with the thermal field double on two copies of the system, this equivalence of the algebra of two disjoint time-bands with that of a larger time-band only occurs when those time-bands are on the same copy of the boundary. The addition of the algebras of two time-bands on different copies of the boundary is not associated with any larger boundary subregion, i.e. for left/right time bands $T_{L/R}$ we have $\fc_{T_L \cup T_R} = \fc_{T_L} \cup \fc_{T_R}$. The commutant of this algebra is interesting though. See the blue region of Fig.~\ref{fig:compsBTZ} (b). 
This bulk subregion includes the bifurcation surface ($U=V=0$) and may include points near the singularity. It does not contain any near boundary points, and exactly as for such subregions in the vacuum case, it therefore is emergent in the large $N$ limit as there is no finite $N$ analogue of its associated algebra.

\subsection{Single-Sided Black Holes}

In the case of a high-energy\footnote{Note that $\ket{\Psi}$ should not be an energy eigenstate but we should have $\bra{\Psi} H \ket{\Psi} = O(N^2).$} semiclassical pure state, $\ket{\Psi},$ on the boundary, the bulk dual should be a single-sided black hole. Since we are working with a pure state on the boundary, the GNS vacuum $\ket{0}_{\Psi, {\rm GNS}}$ should not be separating for $\sM_{\Psi},$ the single-trace algebra on the entire boundary, and moreover $\sM_{\Psi}$ should have a trivial commutant on $\sH^{\rm GNS}_{\Psi}.$ This implies that $\sM_{\Psi}$ should be of type $I.$ 

In the large $N$ limit, the boundary theory should still be described by a GFF theory where operators on different time-slices are independent. In particular, this implies that we can assign non-trivial subalgebras to time-bands. We expect that the algebra of a semi-infinite time-band from some late time to infinity should be dual to bulk field operators supported at late times outside the horizon of the black hole.\footnote{See~\cite{Chandrasekaran:2022eqq} for an interesting discussion of the relationship between these late time algebras and the generalised second law of black hole entropy.} See the blue region of fig.~\ref{fig:sBH}. On the field theory side, two-point functions of these operators should look approximately thermal. On the bulk side, the two-point functions look like those in the exterior of the eternal black hole. Such late time boundary operators should not be able to annihilate the GNS vacuum and thus their algebra $\sM_{\rm late},$ should have a non-trivial commutant on $\sH^{\rm GNS}_{\Psi}.$ This commutant is precisely the algebra of mirror operators whose existence was argued for in~\cite{Papadodimas:2012aq, Papadodimas:2013jku}. In particular, $\le(\sM_{\rm late}\ri)'$ should describe excitations of the bulk field in the causal complement of the dual of $\sM_{\rm late},$ i.e. $\le(\sM_{\rm late}\ri)'$ should describe bulk excitations in the red region of figure~\ref{fig:sBH}. 

Similar to what was found in the case of the interior causal diamond in global AdS, these mirror operators are emergent in the large $N$ limit. They do not have any analogue at finite $N.$ At finite $N,$ the algebra of late time operators is equal to the full boundary algebra and its commutant is trivial. {Nevertheless, we expect some approximate version of such ``mirror" operators to play important roles at a large but finite $N$}

\section{Discussion} \label{sec:conc}

Subregion-subalgebra duality opens up new conceptual perspectives and powerful technical tools for understanding the emergence of space and time in holography. There are many further questions to investigate, which include: (i) working out an explicit construction for a bulk diamond region (with no near-boundary points) in the vacuum state of bulk gravity, (ii) deriving the Wheeler-de Witt equation for such a local bulk region from the boundary theory, (iii) understanding how the construction works for a black hole formed from gravitational collapse such as the Vaidya geometry, and (iv) understanding the algebraic structures of boundary theories dual to AdS$_2$ and Jackiw-Teitelboim (JT) gravity theories~\cite{Lin:2022zxd,Lin:2022rzw,Lin:2022rbf}, where the boundary is dynamical and leads to new elements beyond those considered in~\cite{shortPaper, longPaper}. JT gravity also provides a laboratory where we may be able to probe the operator algebraic structure in the full quantum gravity regime. 

Since physics in a sufficiently small local spacetime region (which does not touch the boundary) 
of an AdS spacetime is essentially the same as that in an asymptotically flat or de Sitter spacetime, the boundary description of such a region can potentially give important hints on how to formulate holographic duality for asymptotically flat and cosmological spacetimes. 
See~\cite{Chandrasekaran:2022cip} for recent progress for understanding operator algebras in de Sitter spacetime. 

Subregion-subalgebra duality also leads to new insights into and perspectives on entanglement wedge reconstruction and its interpretation in terms of quantum error correction~\cite{Almheiri:2014lwa,Dong:2016eik}.
It can also be used to obtain entanglement wedges directly from equivalence of algebras, rather than using 
entropies as in~\cite{Ryu:2006bv,Hubeny:2007xt,Engelhardt:2014gca}. 
 It should be possible to use the duality to give a first principle derivation of the ``island'' prescription used in the calculation of the Page curve for an evaporating black hole~\cite{Penington:2019npb,Almheiri:2019psf,Almheiri:2019hni}.

 There have also been exciting developments~\cite{Witten:2021unn,Chandrasekaran:2022eqq,Chandrasekaran:2022cip} in 
 understanding operator algebras in quantum gravity when $\hbar G_N$ corrections are included, which have shed new light onto the nature of the de Sitter and black hole entropies. It is of great interest to understand whether that construction can be applied for general spacetime regions to define entropies for general spacelike surfaces. 

It is also important to understand the role played by couplings in the nature of {emergent} operator algebras. Our boundary discussion has been
concerned with only the large $N$ limit. But the explicit examples we have considered are all in the vacuum state, where the behavior of two-point functions follows from conformal symmetry and is independent of couplings. When considering general states, the effects of couplings become important, and the bulk gravity descriptions are {only available} in the strong coupling limit. It is of immense importance to understand whether various {bulk} geometric concepts persist away from strong couplings, which corresponds to the stringy regime in the bulk.

\vspace{0.2in}   \centerline{\bf{Acknowledgements}} \vspace{0.2in}
We would like to thank Netta Engelhardt, Thomas Faulkner,  Daniel Harlow, Stefan Hollands, Finn Larsen, Nima Lashkari, Javier Magan, Jonathan Sorce, and Edward Witten for discussions. 
This work is supported by the Office of High Energy Physics of U.S. Department of Energy under grant Contract Number  DE-SC0012567 and DE-SC0020360 (MIT contract \# 578218).
SL acknowledges the support of NSF grant No. PHY-2209997.

\appendix

\section{Rindler decomposition of the generalized free field} \label{app:rind}
In this appendix we derive the change of basis from oscillators associated to a Rindler decomposition, $\{a^{(S)}_k,~ a^{(\bar S)}_k\},$ to those of the Minkowski mode expansion, $\{b_p\}.$ We first recall the global and local mode expansions of the generalized free field and then use equivalence of these mode expansions to derive the change of basis. We then briefly mention how the change of basis for Rindler decompositions about an arbitrary point $x_0$ can be derived. Finally, we discuss the equivalence of the entanglement wedge and causal domain algebras, $\sX_S = \sY_{\hat S}$ in this case.

\subsection{Global and local mode expansions}
Recall that the generalized free field can be written in terms of a Minkowski mode expansion
\bega \label{hjeApp}
\sO (x) = \int {d^2 p \ov (2 \pi)^2} \, f_p e^{i p \cdot x}  b_p , \quad 
f_p  = {\sqrt{\pi} \ov \Ga(\De) 2^{\De - 1}}  \theta (-p^2)    \ka^{\De -1},  \quad p \cdot x = - p^0 x^0 + p^1 x^1,
 \\
 \ka = \sqrt{- p^2},
\quad
 b_p^\da = b_{-p} , 
 \quad  [b_p, b_{p'}] = (2 \pi)^2 \ep (p^0) \de^{(2)} (p+p') ,
\end{gather}
on all of $\mathbb{R}^{1,1}.$ 

Within the right (left) Rindler wedge, ${\hat S}: x^1 > |x^0|$ (${\hat {\bar S}}: x^1 < |x^0|$), we have (with $x^{\pm} = x^0 \pm x^1$) a Rindler mode expansion
\be 
\begin{aligned}
	&\sO (x \in {\hat S}) = \int \frac{d^2 k}{(2\pi)^2} u_k^{(S)}(x) a_{k}^{(S)}, \qquad u_k^{(S)}(x) = N_k \le(-x^-\ri)^{-\bar q_+} \le(x^+\ri)^{-q^-} \\
	&\sO (x \in {\hat {\bar S}}) = \int \frac{d^2 k}{(2\pi)^2} u_k^{(\bar S)}(x) a_{k}^{(\bar S)}, \qquad u_k^{(\bar S)}(x) = N_k \le(x^-\ri)^{- q_+} \le(-x^+\ri)^{-\bar q^-} \\
&q_+ = {\De \ov 2} + i k^+, \quad q_- = {\De \ov 2} + i k^- , \quad \bar q_+ = {\De \ov 2} - i k^+, \quad \bar q_- = {\De \ov 2} - i k^- \\
&k^+ = {\om + q \ov 2}, \quad k^- = {\om - q \ov 2}, \quad N_k =  {\sqrt{\sinh \pi |\om| }\ov \sqrt{2 \pi} \Ga (\De)} \le|\Gamma\left(q_+ \right) \Gamma\left(q_- \right) \ri|\\
&(a_k^{(\al)})^\da = a_{-k}^{(\al)} , \qquad    [a_k^{(\al)} , a_{k'}^{(\b)}] = \ep (\om)  (2\pi)^2 \de (k+k') \de_{\al \b} \ ,
\end{aligned}
\label{leftRightRindExp}
\ee
where $\om,~q$ are the quantum numbers associated respectively to the boost and dilatation symmetries of the Rindler wedge.

The mode functions $u_k^{(S, {\bar S})}$ can be analytically continued to the entire Minkowski plane in the lower (upper) half $x^{\pm}$ planes for positive (negative) $\om$ to the (normalized) functions
\be 
\begin{aligned}
	w_k^{(S)}(x) &= {e^{\pi|\om| \ov 2} \ov \sqrt{2\sinh\pi|\om|}} N_k \le(-x^- +i\ep(\om)\ep\ri)^{-\bar q_+} \le(x^+ -i\ep(\om)\ep\ri)^{-q_-} \\
	w_k^{(\bar S)}(x) &= {e^{\pi|\om| \ov 2} \ov \sqrt{2\sinh\pi|\om|}} N_k \le(x^- -i\ep(\om)\ep\ri)^{-q_+} \le(-x^+ +i\ep(\om)\ep\ri)^{-\bar q_-} \ ,
\end{aligned}
\label{boostModeFcns}
\ee
which are supported on purely positive (negative) frequency Minkowski modes for positive (negative) $\om.$ One can associate oscillators $\{c_k^{(\al)}\}$ to these mode functions with
\be 
c_k^{(\al)} \ket{\Om} = 0 , \quad \al = S, {\bar S}, \quad \om > 0 , \quad \le(c_k^{(\al)} \ri)^\da =  c_{-k}^{(\al)} \ ,
\ee
which yields a mode expansion of the generalized free field on the entire Minkowski spacetime in terms of the $\{c_k^{(\al)}\}$ 
\be \label{boostRindModeExp}
	\sO(x) = \sum_{\al = S, {\bar S}} \int \frac{d^2 k}{(2\pi)^2} w_k^{(\al)}(x) c_{k}^{(\al)} = \sum_{\al = S, {\bar S}} \int \frac{d^2 k}{(2\pi)^2} u_k^{(\al)}(x) a_{k}^{(\al)} \ ,
\ee
where the `extended' Rindler mode functions are explicitly given by
\bega \label{contBdryR}
 u^{(S)}_k(x) =  \frac{N_k}{\sinh\pi\omega}
	\begin{cases}
		\sinh\pi\om ~(-x^-)^{-\bar q_+} (x^+)^{-q_-} , &x^- < 0, ~x^+ > 0 \; (R) \cr
		-i\sin\pi q_- ~(x^-)^{-\bar q_+} (x^+)^{-q_-} , &x^-> 0,~ x^+ > 0  \; (F) \cr
		0 , &x^-> 0,~ x^+ < 0 \; (L) \cr
		i\sin\pi {\bar q_+} ~(-x^-)^{-\bar q_+} (-x^+)^{-q_-} , &x^-< 0,~ x^+ < 0 \; (P) \cr
	\end{cases}, \\
 \label{contBdryL}
	u^{(\bar S)}_k(x) =  \frac{N_k}{\sinh\pi\omega}
	\begin{cases}
		0 , &x^- < 0, ~x^+ > 0 \; (R) \cr
		-i\sin q_+ ~(x^-)^{-q_+} (x^+)^{-\bar q_-} , &x^-> 0,~ x^+ > 0 \; (F) \cr
		\sinh\pi\omega ~(x^-)^{- q_+} (-x^+)^{- \bar q_-} , &x^-> 0,~ x^+ < 0 \; (L) \cr
		i \sin\pi {\bar q_-} ~(-x^-)^{- q_+} (-x^+)^{- \bar q_-} , &x^-< 0,~ x^+ < 0 \; (P) \cr
	\end{cases} .
\end{gather} 

The mode expansion in terms of the Rindler oscillators in~\eqref{boostRindModeExp}, $\{a_k^{(\al)}\}$, is obtained via the Bogoliubov transform
\bega\label{c1}
c^{(\al)}_k  =  {1 \ov \sqrt{2 \sinh \pi |\om|}} \left( e^{ {\pi |\om| \ov 2} } a^{(\al)}_k - e^{-{\pi |\om| \ov 2}}  a^{(\bar \al)}_{-k}  \ri) \ .
\end{gather} 
The inverse Bogoliubov transform is
\bega
a^{(\al)}_k  =  {1 \ov \sqrt{2 \sinh \pi |\om|}} \left( e^{ {\pi |\om| \ov 2} } c^{(\al)}_k + e^{-{\pi |\om| \ov 2}}  c^{(\bar \al)}_{-k}  \ri) \ .
\label{a2} 
\end{gather} 

\subsection{Deriving the change of basis} \label{app:rindBaseChange}
The expansions~\eqref{hjeApp} and~\eqref{boostRindModeExp} are for the same operator, $\sO(x),$ thus, equating these two expressions allows us to express the Minkowski and Rindler oscillators in terms of each other. Explicitly, we may write
\be\label{bogo1}
 b_p =\sum_{\al = S, {\bar S}} \int {d^2k \ov (2\pi)^2} A_{kp}^{(\al)} a_k^{(\al)}, \quad  a_k^{(\al)} =  \int {d^2p \ov (2\pi)^2} B_{p k}^{(\al)} b_p  \ .
\ee
Substituting the first definition into~\eqref{hjeApp} and equating with~\eqref{boostRindModeExp}, or alternatively the second definition into~\eqref{boostRindModeExp} and equating with~\eqref{hjeApp}, one obtains
\be\label{bogo2}
u_k^{(\al)} = \int {d^2p \ov (2\pi)^2} A_{kp}^{(\al)} f_p e^{i p \cdot x}  , \quad A_{kp}^{(\al)} = {1 \ov f_p} \int d^2x \, e^{- i p \cdot x} u_k^{(\al)} (x), 
 \quad f_p e^{i p \cdot x} = \int {d^2k \ov (2\pi)^2} B_{p k}^{(\al)} u_k^{(\al)}  \ ,
\ee
and, with $\de_{k k'} = (2 \pi)^2 \de^{(2)} (k-k') = \ha (2 \pi)^2  \de(k^\pm - k'^\pm)$,
\be \label{jeh} 
\int {d^2p \ov (2\pi)^2} A_{kp}^{(\al)} B_{p k'}^{(\b)} = \de_{k k'} \de_{\al \b}, \qquad \sum_\al \int {d^2k \ov (2\pi)^2} B_{p k}^{(\al)}  A_{kp'}^{(\al)} = \de_{p p'}  \ .
\ee
Note that
\be \label{jeh1}
A^{(\bar S)}_{kp} =  A^{(S)}_{-k,-p} = A^{(S)*}_{kp} , \qquad B^{(\bar S)}_{pk} =  B^{(S)}_{-p,-k} = B^{(S)*}_{pk} \ .
\ee
The explicit expressions for $A$ and $B$ can be computed using the middle equation of~\eqref{bogo2} and equations~\eqref{jeh}. One finds
\bega \label{rindToMink}
 A^{(S)}_{kp}  ={\pi \ep (\om) \ov  \sqrt{2 \sinh \pi |\om|}} 
  e^{-i \th (k^+, k^-)} \bca e^{\frac{\pi}{2}\om}  (p^+)^{-ik^+ - \frac{1}{2}} (p^-)^{ik^- - \frac{1}{2}}, \quad & p^\pm > 0  \cr
 - e^{-\frac{\pi}{2}\om}  (-p^+)^{-ik^+ - \frac{1}{2}} (-p^-)^{ik^- - \frac{1}{2}}, \quad & p^\pm < 0
  \eca , \\
  \label{minkToRind}
  B_{pk}^{(S)} =  \frac{\pi }{\sqrt{2\sinh(\pi|\om|)}}  e^{i \th (k^+, k^-)} (p^+)^{ik^+ - \frac{1}{2}} (p^-)^{-ik^- - \frac{1}{2}} 
\bca  e^{\frac{\pi}{2} \omega} & p^\pm > 0 \cr
e^{-\frac{\pi}{2} \omega} & p^\pm < 0 
\eca \ ,
 \end{gather} 
 where 
 \bega
 e^{i \th (k^+, k^-)} = \frac{\left|\Gamma\left(\frac{\Delta}{2} - ik^+\right)\right|}{\Gamma\left(\frac{\Delta}{2} + ik^+\right)}\frac{\left|\Gamma\left(\frac{\Delta}{2} + ik^-\right)\right|}{\Gamma\left(\frac{\Delta}{2} - ik^-\right)} = {|\Ga (q_+) \Ga (q_-)| \ov \Ga (q_+) \Ga (\bar q_-)}, \\
\th (-k^+, -k^-) = - \th (k^+, k^-) , \qquad   \th (-k^-, -k^+) =  \th (k^+, k^-) \ .
\end{gather} 

\subsection{Shifted Rindler wedges} 
One can easily consider a Rindler decomposition about a point $(x^+,x^-) = (x^+_0, x^-_0)$ that is not the origin. The corresponding mode functions are given by  
\be
u^{(\al, x_0)}_k(x) = u^{(\al)}_k(x-x_0), \quad w^{(\al,x_0)}_k(x) = w^{(\al)}_k(x-x_0),
\ee
 where $ u^{(\al)}_k, w^{(\al)}_k$ are mode functions for the Rindler decomposition about the origin, given in~\eqref{contBdryR}--\eqref{contBdryL} and~\eqref{boostModeFcns}. 
From~\eqref{bogo2} we thus find that 
\be 
A_{kp}^{(\al,x_0)} =  {1 \ov f (p)} \int dx \, e^{- i p \cdot x} u_k^{(\al)} (x-x_0)
= e^{- i p \cdot x_0} A_{kp}^{(\al)}, \quad B_{kp}^{(\al)} = e^{i p \cdot x_0} B_{kp}^{(\al)} \ ,
\ee
where $A_{kp}^{(\al)},~B_{kp}^{(\al)}$ are the basis transformation matrices for the Rindler decomposition the origin. They are given explicitly in~\eqref{rindToMink} and~\eqref{minkToRind}.

\subsection{Equivalence of $\sX_S$ and $\sY_{\hat S}$}
For this particular choice of spatial subregion on the $x^0 = 0$ slice, $S: x^1 > 0,$ the entanglement wedge and casual domain subalgebras, $\sX_S$ and $\sY_{\hat S},$ are equal in the vacuum representation. Recall that $\sY_{\hat S}$ is generated by $\{\sO(x), x\in \hat S\}.$ Using~\eqref{leftRightRindExp}, one has 
\be 
	a^{(S)}_k = {1 \ov N_k} \int_{x \in \hat S} d^2x \le(-x^-\ri)^{\bar q_+ - 1} \le(x^+\ri)^{q_- - 1} \sO(x) \ ,
\ee
so, equivalently one can say that $\sY_{\hat S}$ is generated by $\{a^{(S)}_k,~ k \in \mathbb{R}^2\}.$

With the definition of $\sX_S$ given in~\eqref{modT0}--\eqref{im2} one has that $\sX_S$ is generated by $\{\sO(s;x^1),~ s\in \mathbb{R},~ x^1 \in S\}.$ The modular flow for a Rindler region in the vacuum is a boost. Recalling that $\om$ is the quantum number associated to boosts, one has that 
\be 
	\De_0^{is} a^{(S)}_k \De_0^{-is} = e^{-i\om s} a^{(S)}_k \ ,
\ee
which leads to 
\be 
	\sO(s;x^1) = \int {d^2k \ov (2\pi)^2} N_k \le(x^1\ri)^{-\bar q_+} \le(x^1\ri)^{-q_-} e^{-i\om s} a^{(S)}_k = \sO\le(x^- = -x^1 e^{-s},~ x^- = x^1 e^s\ri) \ .
\ee
For any $s\in \mathbb{R},~ x^1 > 0,$ there is exactly one $x_* \in \hat S$ such that $\sO(s;x^1) = \sO(x_*)$. Thus we find $\{\sO(s;x^1),~ s\in \mathbb{R},~ x^1 \in S\} = \{\sO(x), x\in \hat S\},$ so $\sX_S$ and $\sY_{\hat S}$ are generated by the same set of operators and thus are equivalent. This is consistent with the coincidence of the entanglement wedge and causal domain for this situation.

\section{Interval decomposition of the generalized free field} \label{app:interval}
In this appendix we derive the change of basis from interval/complement oscillators $\{a^{(I)}_k,~a^{(O)}_k\}$ to Minkowski oscillators, $\{b_p\}.$ We first discuss the localized mode expansion of the generalized free field and then derive the basis change. We also explicitly demonstrate the superadditivity $\sY_{\hat O} = \sM_{\hat O_L} \lor \sM_{\hat O_R} \subset \sX_{O}$ in the final subsection.

\subsection{Local mode expansion} \label{app:intAnalCont}
We are interested in decomposing the generalized free field in terms of mode expansions associated to the regions $I: |x^1| < a$ and $O = \bar{I}: |x^1| > a$ on the $x^0 = 0$ slice. Consider the conformal transformation (with $b,c >0$)
\be \label{ToDiApp}
	x^- = a~  {y^- + b \ov b - y^-}, \qquad x^+ = a~  {y^+ - c \ov y^+ + c} \ ,
\ee 
which smoothly maps the right Rindler wedge $\hat S$ of the $y-$plane to the diamond region $\hat{I}: |x^{\pm}| < a$ and the region $\{0 < y^- < b,~-c < y^+ < 0\}$ to the (shifted) Rindler region $\hat{O}_L : x^- > a,~ x^+ < -a$. 
The inverse transformation is
\be \label{inverseConfDiamondMinkCoord}
	y^- = b~  {x^- - a \ov x^- + a}, \qquad y^+ = c~  {x^+ + a \ov a - x^+} , 
\ee
and we can compute the conformal factor by the change in the metric
\be \label{confFactorDiamondMinkCoord}
	ds^2 = -dy^+ dy^- = - \Om^{2} dx^+ dx^-, \qquad \Om^2 = {4 a^2 b c \ov (x^- + a)^2 (a- x^+ )^2} \ .
\ee

Recall that $\sO(x)$ is the vacuum representation of a scalar primary of dimension $\De,$ thus, under a conformal transformation we have 
\be \label{confXfmScalarMT}
	\sO(y) \to \sO(x) = \Om^{\De} \sO(y) 
	 , \quad
	\Om (x) = \le({4 a^2 b c \ov (x^- + a)^2 (a- x^+ )^2}\ri)^{1 \ov 2} \ ,
\ee
with $y = y(x).$ Using the Rindler mode expansion~\eqref{leftRightRindExp} in the $y-$plane and the transformation~\eqref{confXfmScalarMT} we can obtain mode expansions for the generalized free field in $\hat{I}$ and $\hat{O}_L.$

One finds (setting $b=c=1$)
\bega\label{oi1App}
\sO (x \in \hat I)  = \int {d^2k \ov (2\pi)^2} u^{(I)}_k(x)~ a^{(I)}_k \ ,  \\
\sO(x \in \hat O_L) 
	= \int {d^2k \ov (2\pi)^2} u^{(O)}_k(x)~ a^{(O)}_k \ ,
	\label{ol1App}\\
u^{(I)}_k   (x) = N_k (2a)^{\De} 
(a - x^-)^{-\bar q_+} (x^- + a)^{- q_+} (a - x^+)^{-\bar q_-} (x^+ + a)^{- q_-} ,
\label{iMTApp} \\	
\label{eMTApp}
	u^{(O)}_k (x) = N_k (2a)^{\De} 
	(x^- - a)^{-q_+} (x^- + a)^{-\bar q_+} (a - x^+)^{- q_-} (-x^+ - a)^{-\bar q_-} \ .
\end{gather} 

The mode functions $u_k^{(I, O)}$ can be analytically continued to the entire Minkowski plane in the lower (upper) half $x^{\pm}$ planes for positive (negative) $\om$ to the (normalized) functions
\be 
\begin{aligned}
	w_k^{(I)}(x) &= {e^{\pi|\om| \ov 2} N_k (2a)^{\De}  \ov \sqrt{2\sinh\pi|\om|}} 
	(a - x^- + i\ep(\om)\ep)^{-\bar q_+} (x^- + a - i\ep(\om)\ep)^{- q_+} (a - x^+ + i\ep(\om)\ep)^{-\bar q_-} (x^+ + a - i\ep(\om)\ep)^{- q_-} \\
	w_k^{(O)}(x) &= {e^{\pi|\om| \ov 2} N_k (2a)^{\De}  \ov \sqrt{2\sinh\pi|\om|}}  
	(x^- - a - i\ep(\om)\ep)^{-q_+} (x^- + a - i\ep(\om)\ep)^{-\bar q_+} (a - x^+ + i\ep(\om)\ep)^{- q_-} (-x^+ - a + i\ep(\om)\ep)^{-\bar q_-} \ ,
\end{aligned}
\label{boostModeFcnsInt}
\ee
which are supported on purely positive (negative) frequency Minkowski modes for positive (negative) $\om.$ One immediate property of these functions is
\be \label{cplxConjWInt}
	\le(w^{(I)}_k(x)\ri)^* = w^{(I)}_{-k}(x), \qquad \le(w^{(O)}_k(x)\ri)^* = w^{(O)}_{-k}(x) \ .
\ee

One can associate oscillators $\{c_k^{(\al)}\}$ to the mode functions~\eqref{boostModeFcnsInt} with
\be 
c_k^{(\al)} \ket{\Om} = 0 , \quad \al = I, O, \quad \om > 0 , \quad \le(c_k^{(\al)} \ri)^\da =  c_{-k}^{(\al)} \ ,
\ee
which yields a mode expansion of the generalized free field on the entire Minkowski spacetime in terms of the $\{c_k^{(\al)}\}$ 
\be \label{boostIntModeExp}
	\sO(x) = \sum_{\al = I, O} \int \frac{d^2 k}{(2\pi)^2} w_k^{(\al)}(x) c_{k}^{(\al)} = \sum_{\al = I, O} \int \frac{d^2 k}{(2\pi)^2} u_k^{(\al)}(x) a_{k}^{(\al)} \ ,
\ee
where the `extended' interval/complement mode functions are explicitly given by
\be \label{bdryIntInterior}
\begin{aligned}
u^{(I)}_k(x) &= {N_k (2a)^{\De}  \ov \sinh\pi\om} \\ 
&\cdot 
\begin{cases}
	-i\sin\pi\De \le(x^- - a\ri)^{-\bar q_+}\le(x^- + a\ri)^{- q_+} \le(x^+ - a\ri)^{-\bar q_-}\le(x^+ + a\ri)^{- q_-}, & x \in F_M \\
	-i\sin\pi q_- \le(x^- - a\ri)^{-\bar q_+}\le(x^- + a\ri)^{- q_+} \le(a - x^+\ri)^{-\bar q_-}\le(x^+ + a\ri)^{- q_-}, & x \in F_L \\
	-i\sin\pi q_+ \le(a - x^-\ri)^{-\bar q_+}\le(x^- + a\ri)^{- q_+} \le(x^+ - a\ri)^{-\bar q_-}\le(x^+ + a\ri)^{- q_-}, & x \in F_R \\
	0, & x \in \hat{O}_L \\
	0, & x \in \hat{O}_R \\
	\sinh\pi\om \le(a - x^-\ri)^{-\bar q_+}\le(x^- + a\ri)^{- q_+} \le(a -x^+\ri)^{-\bar q_-}\le(x^+ + a\ri)^{- q_-}, & x \in \hat{I} \\
	i\sin\pi\bar q_+ \le(a - x^-\ri)^{-\bar q_+}\le(x^- + a\ri)^{- q_+} \le(a -x^+\ri)^{-\bar q_-}\le(-x^+ - a\ri)^{- q_-}, & x \in P_L \\
	i\sin\pi\bar q_- \le(a - x^-\ri)^{-\bar q_+}\le(-x^- - a\ri)^{- q_+} \le(a -x^+\ri)^{-\bar q_-}\le(x^+ + a\ri)^{- q_-}, & x \in P_R \\
	i\sin\pi\De \le(a - x^-\ri)^{-\bar q_+}\le(-x^- - a\ri)^{- q_+} \le(a -x^+\ri)^{-\bar q_-}\le(-x^+ - a\ri)^{- q_-}, & x \in P_M \\
\end{cases} \ ,
\end{aligned}
\ee
\be \label{bdryIntExterior}
\begin{aligned}
u^{(O)}_k(x) &= {N_k (2a)^{\De}   \ov \sinh\pi\om} \\ 
&\cdot
\begin{cases}
	-i\sin\pi\le(\De + i\om\ri) \le(x^- - a\ri)^{-q_+}\le(x^- + a\ri)^{-\bar q_+} \le(x^+ - a\ri)^{-q_-}\le(x^+ + a\ri)^{-\bar q_-}, & x \in F_M \\
	-i\sin\pi q_+ \le(x^- - a\ri)^{-q_+}\le(x^- + a\ri)^{-\bar q_+} \le(a - x^+\ri)^{-q_-}\le(x^+ + a\ri)^{-\bar q_-}, & x \in F_L \\
	-i\sin\pi q_- \le(a - x^-\ri)^{-q_+}\le(x^- + a\ri)^{-\bar q_+} \le(x^+ - a\ri)^{-q_-}\le(x^+ + a\ri)^{-\bar q_-}, & x \in F_R \\
	\sinh\pi\om \le(x^- - a\ri)^{-q_+}\le(x^- + a\ri)^{-\bar q_+} \le(a - x^+\ri)^{-q_-}\le(-x^+ - a\ri)^{-\bar q_-}, & x \in \hat{O}_L \\
	\sinh\pi\om \le(a - x^-\ri)^{-q_+}\le(-x^- - a\ri)^{-\bar q_+} \le(x^+ - a\ri)^{-q_-}\le(x^+ + a\ri)^{-\bar q_-}, & x \in \hat{O}_R \\
	0, & x \in \hat{I} \\
	i\sin\pi\bar q_- \le(a - x^-\ri)^{-q_+}\le(x^- + a\ri)^{-\bar q_+} \le(a -x^+\ri)^{-q_-}\le(-x^+ - a\ri)^{-\bar q_-}, & x \in P_L \\
	i\sin\pi\bar q_+ \le(a - x^-\ri)^{-q_+}\le(-x^- - a\ri)^{-\bar q_+} \le(a -x^+\ri)^{-q_-}\le(x^+ + a\ri)^{-\bar q_-}, & x \in P_R \\
	i\sin\pi\le(\De - i\om\ri) \le(a - x^-\ri)^{-q_+}\le(-x^- - a\ri)^{-\bar q_+} \le(a -x^+\ri)^{-q_-}\le(-x^+ - a\ri)^{-\bar q_-}, & x \in P_M \\
\end{cases} \ .
\end{aligned}
\ee
As in the Rindler case, the expansion of $\sO(x)$ in terms of $\{a^{(I)}_k,~a^{(O)}_k\}$ is obtained from that in terms of $\{c^{(I)}_k,~c^{(O)}_k\}$ via the Bogoliubov transform~\eqref{c1} (with $\al = I,O$).

\subsection{Derivation of the basis change} \label{app:intBaseChange}
Equating the mode expansions~\eqref{hjeApp} and~\eqref{boostIntModeExp} one can relate the interval/complement oscillators $\{a^{(I)}_k,~a^{(O)}_k\}$ to the Minkowski oscillators $\{b_p\}.$ Defining the change of basis `matrices' $P$ and $Q$ by 
\be  \label{pMatrix}
b_p = \int {d^2k \ov (2 \pi)^2}\, \le(Q^{(I)}_{kp}  c^{(I)}_k + Q^{(O)}_{kp} c^{(O)}_k \ri)
=  \int {d^2k \ov (2 \pi)^2}\, \le(P^{(I)}_{kp}  a^{(I)}_k + P^{(O)}_{kp} a^{(O)}_k \ri) \ ,
\ee
we have 
\be \label{qMatIntegral}
 Q^{(I)}_{kp} = f_p^{-1}  \int d^2 x \, e^{-i p \cdot x}  w^{(I)}_k (x), \qquad 
 Q^{(O)}_{kp} =f_p^{-1}  \int d^2 x \, e^{-i p \cdot x}  w^{(O)}_k (x)  \ .
\ee
Since $w^{(I, O)}_k (x)$ is analytic in lower (upper) $x^\pm$-plane for $\om > 0$ ($\om < 0$), we have 
\bega 
 Q^{(I)}_{kp} = \th (\om) \th (p^+) \th (p^-) A (k, p) +  \th (-\om) \th (-p^+) \th (-p^-) B  (k, p)\\
 Q^{(O)}_{kp} = \th (\om) \th (p^+) \th (p^-) \tilde A  (k, p)+  \th (-\om) \th (-p^+) \th (-p^-) \tilde B (k, p) \ .
 \end{gather} 
 Note that, from~\eqref{qMatIntegral} and~\eqref{cplxConjWInt}, we have
 \be 
 (Q^{(I)}_{kp})^* =  Q^{(I)}_{(-k)(-p)}, \quad (Q^{(O)}_{kp})^* =  Q^{(O)}_{(-k)(-p)}, \quad B (-k, - p)= A^* (k, p), \quad \tilde B  (-k, - p) = \tilde A^* (k, p)   \ .
 \ee
From the Bogoliubov transform~\eqref{c1}, we have 
\bega
P^{(I)}_{kp}  = {e^{ {\pi |\om| \ov 2} } \ov \sqrt{2 \sinh \pi |\om|}} Q^{(I)}_{kp} - {e^{ - {\pi |\om| \ov 2} } \ov \sqrt{2 \sinh \pi |\om|}} 
Q^{(O)}_{(-k)p} , \\
P^{(O)}_{kp}  = {e^{ {\pi |\om| \ov 2} } \ov \sqrt{2 \sinh \pi |\om|}} Q^{(O)}_{kp} - {e^{ - {\pi |\om| \ov 2} } \ov \sqrt{2 \sinh \pi |\om|}} 
Q^{(I)}_{(-k)p} \ .
 \end{gather} 
Using the explicit form of the mode functions~\eqref{boostModeFcnsInt} we have
\bega 
A(k,p) = {1 \ov 2 f_p} {e^{\pi|\om|\ov 2} \ov \sqrt{2\sinh\pi|\om|}} N_k (2a)^{\De} 
 I(p^+; \bar q_+, q_+) I(p^-; \bar q_-, q_-)\\
\tilde A(k,p) = {1 \ov 2 f_p} {e^{\pi|\om|\ov 2} \ov \sqrt{2\sinh\pi|\om|}}  N_k (2a)^{\De} 
\tilde I  (p^+; \bar q_+, q_+) (\tilde I  (p^-; \bar q_-, q_-) )^*
\end{gather} 
where 
\bega
I (p^+; \bar q_+, q_+)  = \int dx \, e^{i p^+ x} \, (a + i \ep - x)^{-\bar q_+} (x + a- i \ep)^{- q_+} ,  \\
\tilde I  (p^+; \bar q_+, q_+) =  \int dx \, e^{i p^+ x} \,(x + a- i \ep)^{-\bar q_+}  (x - a - i \ep)^{-q_+}  \ .  
\end{gather} 
The above integrals can be evaluated to give  
\bega
I (p^+; \bar q_+, q_+)   =\th (p^+)  {2\pi e^{-\pi k^+} \ov \Ga(\De)} e^{-ip^+ a} (p^+)^{\De - 1}  {}_1F_1\le(\bar q_+ ; \De; 2iap^+\ri)
\\
\tilde I  (p^+; \bar q_+, q_+) = \th (p^+)   e^{-ip^+ a} (p^+)^{\De - 1} {2\pi e^{i\pi \De \ov 2} \ov \Ga(\De)} {}_1F_1\le( q_+ ; \De; 2iap^+\ri)
\end{gather} 
which gives 
\bega
A(k,p) =C  \ka^{\nu}  e^{- {\pi \ov 2} \om} e^{- i p^0 a} 
   {}_1F_1\le(\bar q_+ ; \De; 2iap^+\ri)
    {}_1F_1\le(\bar q_- ; \De; 2iap^-\ri) , \\
    \tilde A(k,p) =C^* \ka^{\nu}  e^{\pi \om\ov 2}  e^{-ip^1 a}
 {}_1F_1\le( q_+ ; \De; 2iap^+\ri)
    {}_1F_1\le( \bar q_- ; \De; -2iap^-\ri) , \\
    C =  { \pi   \ov   2^{\De -1} ( \Ga(\De) )^2 }  
  \le|\Gamma\left(q_+ \right) \Gamma\left(q_- \right) \ri|
 (2a)^{\De} 
 \ . 
\end{gather} 
With the identity 
\be \label{hsn}
 {}_1F_1(\bar q_- ; \De ; -2iap^-) = e^{-2iap^-} {}_1F_1(q_- ; \De ; 2iap^-)
 \ee
we find
\be 
\tilde A (k, p) = A (-k, p)  \ .
\ee

Using the expression for $P$ in terms of $Q$ and $Q$ in terms of $A, \tilde{A}, B, \tilde{B}$ one obtains the explicit expression for the change of basis from interval/complement oscillators to Minkowski oscillators
\be \label{explicitPMatrix}
\begin{aligned}
	P^{(I)}_{kp} &= {\ep(\om)\ep(p^0)\theta(p^+ p^-) \ov \sqrt{2\sinh\pi|\om|}} C \kappa^\nu e^{-ip^0 a} {}_1F_1\le({\bar q_+}; \De ; 2ia p^+ \ri){}_1F_1\le({\bar q_-}; \De ; 2ia p^- \ri) \\
	P^{(O)}_{kp} &= {\ep(\om)\ep(p^0)\theta(p^+ p^-) e^{\pi \ep(p^0)\om} \ov \sqrt{2\sinh\pi|\om|}} C^* \kappa^\nu e^{-ip^0 a} {}_1F_1\le({ q_+}; \De ; 2ia p^+ \ri){}_1F_1\le({ q_-}; \De ; 2ia p^- \ri) \ .
\end{aligned}
\ee

The inverse transformation from Minkowski to interval/complement oscillators such that
\be
	a^{(\al)}_k = \int {d^2p \ov (2\pi)^2} R^{(\al)}_{pk} b_p
\ee
is
\be 
\begin{aligned}
	R^{(I)}_{pk} &= {\theta(p^+ p^-) \ov \sqrt{2\sinh\pi|\om|}} C^* \kappa^\nu e^{-ip^0 a} {}_1F_1\le({\bar q_+}; \De ; 2ia p^+ \ri){}_1F_1\le({\bar q_-}; \De ; 2ia p^- \ri) \\
	R^{(O)}_{pk} &= {\theta(p^+ p^-) e^{\pi \ep(p^0)\om} \ov \sqrt{2\sinh\pi|\om|}} C \kappa^\nu e^{-ip^0 a} {}_1F_1\le({ q_+}; \De ; 2ia p^+ \ri){}_1F_1\le({ q_-}; \De ; 2ia p^- \ri) \ .
\end{aligned}
\ee

\subsection{Relations between $\sX_{I,O}$ and $\sY_{\hat{I},\hat{O}}$} \label{app:nonInvO}
For the subregion $\hat{I},$ which is conformally equivalent to a Rindler wedge, the modular flow with respect to the vacuum is local. We will see that this leads to the equivalence of $\sX_I$ and $\sY_{\hat I}.$ Recall that $\sY_{\hat I}$ is defined as the algebra generated by $\{\sO(x),~ x \in \hat I\}.$
From the definition of $\sX_I$ given in~\eqref{modT0}--\eqref{im2} one has that $\sX_I$ is generated by $\{\sO(s;x^1),~ s\in \mathbb{R},~ x^1 \in I\}.$ Since $\om$ was the modular frequency for the Rindler wedge (i.e. quantum number for the boost symmetry) before the conformal transformation it still has this interpretation after the conformal transformation. In particular, we have 
\be 
	\De_{0(I)}^{is} a^{(I)}_k \De_{0(I)}^{-is} = e^{-i\om s} a^{(I)}_k \ ,
\ee
which leads to 
\be 
\begin{aligned}
	\sO(s;x^1) &= \int {d^2k \ov (2\pi)^2} N_k (2a)^{\De} \le(\Om_s(-x^1,x^1)\ri)^{\De} \le(a - x^-_s\ri)^{-\bar q_+}\le(x^-_s + a\ri)^{- q_+} \le(a -x^+_s\ri)^{-\bar q_-}\le(x^+_s + a\ri)^{- q_-}  a^{(I)}_k \\
	&= \le(\Om_s(-x^1,x^1)\ri)^{\De} \sO\le(x^-_s, x^+_s\ri) \ .
\end{aligned}
\ee
In the expression above, $x^-_s$ is evaluated at $x^- = -x^1,$ i.e. $x^-_s \to x^-_s(-x^1)$ and $x^+_s$ is evaluated at $x^+ = x^1,$ i.e. $x^+_s \to x^+_s(x^1)$. We see that, as expected, the modular flow acts as a conformal transformation with
\be 
	x_s^-(x^-) = -a~ {a(1-e^s) - (1+e^s)x^- \ov a(1+e^s) - (1-e^s)x^-}, \qquad x_s^+(x^+) = a~ {a(1-e^{-s}) + (1+e^{-s})x^+ \ov a(1+e^{-s}) + (1-e^{-s})x^+} \ ,
\ee
and conformal factor
\be 
	ds^2 = -dx^-_s dx^+_s = - \Om_s^2 dx^- dx^+,~ \le(\Om_s(x^-,x^+)\ri)^{2} = \le({(2a)^2 \ov (a(1+e^s) - (1-e^s)x^-)(a(1+e^{-s}) + (1-e^{-s})x^+)}\ri)^2 \ . 	
\ee
This is in exact agreement with the computation of modular flow for an interval in~\cite{Casini:2011kv}. We see that for each $s \in \mathbb{R},~ x^1 \in I$ there is exactly one $x_* \in \hat{I}$ such that $\sO(s;x^1) = \Om_s^{\De} \sO(x_*).$ We then have $\{\sO(s;x^1),~ s\in \mathbb{R},~ x^1 \in I\} = \{\sO(x),~ x \in \hat I\},$ so $\sX_I$ and $\sY_{\hat I}$ are generated by the same operators and thus coincide.

Now we turn to the complementary region $O.$ The algebra $\sY_{\hat O}$ is generated by $\{\sO(x),~ x \in \hat O\},$ while $\sX_O$ is, via~\eqref{modT0}--\eqref{im2}, generated by $\{\sO(s;x^1),~ s \in \mathbb{R}, x^1 \in O\}.$ Before the conformal transformation, the oscillators $\{a^{(O)}_k\}$ were associated to the left Rindler wedge and $\om$ was the modular frequency, thus $\om$ still has this interpretation after the conformal transformation though it is not associated to any spacetime symmetry of the region $\hat O.$ We therefore have 
\be 
	\De_{0(O)}^{is} a^{(O)}_k \De_{0(O)}^{-is} = e^{-i\om s} a^{(O)}_k \ ,
\ee
which leads to 
\be 
\begin{aligned}
	\sO(s;x^1) &= \int {d^2k \ov (2\pi)^2} N_k (2a)^{\De} \le|x^- - a\ri|^{-q_+}\le|x^- + a\ri|^{-\bar q_+} \le|a -x^+\ri|^{- q_-}\le|-x^+ - a\ri|^{-\bar q_-} e^{-i\om s} a^{(O)}_k  \ .
\end{aligned}
\ee
This expansion of $\sO(s;x^1)$ can be inverted to solve for $\{a^{(O)}_k\}$ as
\bega \label{i1uApp}
	a^{(O)}_k = \int_{x^1 \in O} dx^1 ds ~ f_k(x^1, s) \sO(s; x^1) , \\
	f_k(x^1, s) = {1 \ov N_k} \le({(x^1)^2 - a^2 \ov 2a}\ri)^{\De -1} \le({x^1-a \ov x^1 + a}\ri)^{-iq} e^{i\om s}  \ .
\end{gather} 
We therefore have that $\{\sO(s;x^1),~ s \in \mathbb{R}, x^1 \in O\}$ and $\{a^{(O)}_k,~ k \in \mathbb{R}^2\}$ generate the same algebra. Thus $\sX_O$ can be viewed as being generated by the $\{a^{(O)}_k\}.$

We will now prove that $\{\sO(x),~ x \in \hat O\}$ does not generate the same algebra as $\{a^{(O)}_k\}.$ From~\eqref{boostIntModeExp}--\eqref{bdryIntExterior} it is clear that all $\sO(x \in \hat O)$ can be expressed in terms of the $a^{(O)}_k,$ so clearly $\sY_{\hat O} \subseteq \sX_O.$ However, we will see that the $\{a^{(O)}_k\}$ cannot be expressed in terms of the $\{\sO(x \in \hat O)\},$ so we have a proper inclusion $\sY_{\hat O} \subset \sX_O.$ 

In order for it be possible to express the $\{a^{(O)}_k\}$ in terms of the $\{\sO(x \in \hat O)\}$ we would need to find a set of functions $\{f_k(x), k \in \mathbb{R}^2\}$ supported for $x \in \hat O$ such that
\be \label{cantWriteAOkApp}
	a^{(O)}_k = \int_{x \in \hat O} d^2x~ f_k(x) \sO(x) , \quad \forall k\  .
\ee

Consider the mode function $u^{(O)}_k(x \in \hat{O} = \hat{O}_L \cup \hat{O}_R).$ We may view $u^{(O)}_k(x \in \hat{O})$ as a matrix taking us from the basis of vectors $\{a^{(O)}_k, k \in \mathbb{R}^2\}$ to the set of vectors $\{\sO(x), x \in \hat{O}\}.$ An expression such as~\eqref{cantWriteAOkApp} would require $f_k(x)$ to be the left inverse of $u^{(O)}_k(x \in \hat{O}).$ Namely,~\eqref{cantWriteAOkApp} requires the set of functions $\{f_k(x), k \in \mathbb{R}^2\}$ to satisfy
\be 
	\int_{x \in \hat{O}} d^2x f_k(x) u^{(O)}_{k'}(x) = (2\pi)^2 \delta^{(2)}(k - k') \ .
\ee
Viewed as a matrix, $u^{(O)}_k(x \in \hat{O}),$ only has a left inverse (i.e. the $f_k(x)$ only exist) if the nullspace of $u^{(O)}_k(x \in \hat{O})$ is trivial. We will see that this is not the case and hence such $f_k(x)$ do not exist.\\

The most general vector in the domain of $u^{(O)}_k(x \in \hat{O})$ is an arbitrary function of the two momenta, $\hat{g}(k^+,k^-).$ For convenience, we will remove some non-zero factors from $\hat{g}(k^+,k^-)$ and write 
\be \label{hatGFromG}
	\hat{g}(k^+, k^-) = {
	1 \ov N_k (2a)^{\De}} g(k^+,k^-) \ .
\ee
We have a complete basis for functions on $\mathbb{R}^2$ given by the Fourier modes. i.e. for any function $g(k^+,k^-)$ we may write
\be \label{fourierDecompG}
	 g(k^+, k^-) = \int d^2y~ e^{ik^+y^- + ik^-y^+} \tilde{g}(y^-, y^+) \ ,
\ee
where $y^-$ is the conjugate variable to $k^+$ and $y^+$ the conjugate variable to $k^-.$\\

Let us now consider $u^{(O)}_k(x \in \hat{O})$ acting on the most general vector $\hat{g}(k^+,k^-).$ The result is, of course, some function of $x$ on $\hat{O}.$ Naming the result of the multiplication of $\hat{g}(k^+,k^-)$ by this matrix $r_g(x),$ we have 
(for $\hat{O}_L: x^- > a, x^+ < -a$)
\be \label{resultInOL}
\begin{aligned}
	r_g(x \in \hat{O}_L) &= \int {d^2k \ov (2\pi)^2} \tilde u^{(O)}_k(x \in O_L)~ \hat{g}(k^+,k^-) \\
	&= \int d^2y \int {d^2k \ov (2\pi)^2} (x^- - a)^{-q_+} (x^- + a)^{-\bar q_+} (a - x^+)^{-q_-} (-x^+ - a)^{-\bar q_-} e^{ik^+ y^- + ik^- y^+} \tilde{g}(y^-, y^+) \\
	&= 2\le(((x^-)^2 -a^2)((x^+)^2 -a^2)\ri)^{-{\De \ov 2}} \\
	&\cdot \int d^2y~ \tilde{g}(y^-, y^+) \delta\le(y^- - \log\le(x^- - a \ov x^- + a\ri)\ri) \delta\le(y^+ - \log\le(a - x^+ \ov -x^+ - a\ri)\ri) \\ 
	&= 2\le(((x^-)^2 - a^2)((x^+)^2 -a^2)\ri)^{-{\De \ov 2}} \tilde{g}\le(\log\le(x^- - a \ov x^- + a\ri), \log\le(a - x^+ \ov -x^+ - a\ri)\ri) \ .
\end{aligned}
\ee
and for $\hat{O}_R: x^- < -a, x^+ > a$
\be \label{resultInOR}
\begin{aligned}
	r_g(x \in \hat{O}_R) &= \int {d^2k \ov (2\pi)^2} \tilde u^{(O)}_k(x \in O_R)~ \hat{g}(k^+,k^-) \\
	&= \int d^2y \int {d^2k \ov (2\pi)^2} (a - x^-)^{-q_+} (-x^- - a)^{-\bar q_+} (x^+ - a)^{-q_-} (x^+ + a)^{-\bar q_-} e^{ik^+ y^- + ik^- y^+} \tilde{g}(y^-, y^+) \\
	&= 2\le(((x^-)^2 -a^2)((x^+)^2 -a^2)\ri)^{-{\De \ov 2}} \\
	&\cdot \int d^2y~ \tilde{g}(y^-, y^+) \delta\le(y^- - \log\le(a - x^- \ov -x^- - a\ri)\ri) \delta\le(y^+ - \log\le(x^+ - a \ov x^+ + a\ri)\ri) \\ 
	&= 2\le(((x^-)^2 - a^2)((x^+)^2 -a^2)\ri)^{-{\De \ov 2}} \tilde{g}\le(\log\le(a - x^- \ov -x^- - a\ri), \log\le(x^+ - a \ov x^+ + a\ri)\ri) \ .
\end{aligned}
\ee

Equations~\eqref{resultInOL}--\eqref{resultInOR} completely describe the function obtained by the multiplication of $u^{(O)}_k(x \in \hat{O})$ on the vector $\hat{g}(k).$ Notice in~\eqref{resultInOL} that the first and second arguments of $\tilde{g}$ on the final line range over $(-\infty, 0)$ and $(0,\infty)$ respectively while $x$ ranges over all of $\hat{O}_L.$ Similarly, in~\eqref{resultInOR} the first and second arguments of $\tilde{g}$ on the final line range over $(0, \infty)$ and $(-\infty, 0)$ respectively while $x$ ranges over all of $\hat{O}_R.$\\

Thus, any function $\hat{g}(k)$ such that its (rescaled) Fourier transform $g(y^-,y^+)$ has support only for $y^- y^+ > 0$ gives $r_g(x \in \hat{O}) = 0.$ Such functions of $k$ need not be identically zero, thus we have found a family of non-trivial functions of $k$ that are annihilated by $u^{(O)}_k(x \in \hat{O}).$ Thus the null-space of $u^{(O)}_k(x \in \hat{O})$ is non-trivial so it is {\bf not left-invertible}, showing that $a^{(O)}_k$ cannot be written in terms of position-space operators supported only in $\hat{O}.$ Thus, $\{a^{(O)}_k\}$ and $\{\sO(x \in \hat{O})\}$ do not generate the same algebra and we have $\sX_O \supset \sY_{\hat O}.$

\subsubsection{Lack of invertibility and superadditivity}
The lack of invertibility of the mode expansion is crucial for $\sX_O$ to describe the entanglement wedge, rather than the causal wedge of the union of the two Rindler regions $\hat{O}_L$ and $\hat{O}_R$. One can see immediately that invertibility of the mode expansion precludes the possibility for $\sX_{O}$ to describe the entanglement wedge. Suppose that it was possible to find a set of $f_k(x)$ such that~\eqref{cantWriteAOkApp} was satisfied. We could then split each $f_k(x)$ into a piece supported only on $\hat{O}_L$ and a piece supported only on $\hat{O}_R,$ i.e. we could write $f_k(x) = f^L_k(x)+f^R_k(x).$ Using the Rindler mode expansions on $\hat{O}_L$ and $\hat{O}_R$ separately,~\eqref{cantWriteAOkApp} could be written as
\be \label{additiveProof}
	a^{(O)}_k = \int {d^2k' \ov (2\pi)^2} M^{k'}_{k} a^{(S,(0,a))}_{k'} + \int {d^2k' \ov (2\pi)^2} N^{k'}_{k} a^{({\bar S},(0,-a))}_{k'} \ ,
\ee 
for some functions $M^{k'}_k,~N^{k'}_k.$ Since $\sX_{O}$ is generated by $\{ a^{(O)}_k \}$,~\eqref{additiveProof} would imply that all elements of $\sX_{O}$ can be written in terms of $\{ a^{(S,(0,a))}_k,~ a^{({\bar S},(0,-a))}_k \},$ the generators of the algebras $\sY_{\hat{O}_R} = \sX_{O_R}$ and $\sY_{\hat{O}_L} = \sX_{O_L},$ respectively. Namely,~\eqref{additiveProof} would imply additivity: $\sX_{O} = \sX_{O_L} \vee \sX_{O_R}.$ However, from AdS-Rindler reconstruction and additivity of the bulk free field theory we know that $\sX_{O_L} \vee \sX_{O_R}$ only describes the causal wedge of $O_L \cup O_R.$ Thus, for $\sX_{O}$ to describe the entanglement wedge, additivity of the boundary GFF theory must be violated and thus there cannot exist any expression like~\eqref{cantWriteAOkApp}.

\section{RT surface without entropy: Rindler} \label{app:rindRT}
We consider a bulk scalar field $\phi$, which has the expansion 
\bega 
\phi (X) = \int {d^2 p \ov (2 \pi)^2} v_p (X) b_p  = \int {d^2k \ov (2\pi)^2} \le(\tilde v^{(S)}_k(X) a^{(S)}_k + \tilde v^{(\bar S)}_k(X) a^{(\bar S)}_k\ri) \ ,\\
v_p (X) = \theta (-p^2)  \sqrt{\pi} z^{d \ov 2} J_{\De -1} (pz)  e^{i p \cdot x } , \qquad p = \sqrt{- p^2}  , 
\end{gather}
as in~\eqref{phex}. The expansion in terms of Rindler oscillators is obtained by using the basis change from Rindler to Minkowski oscillators as in~\eqref{bogo1}. The expansion coefficients are then
\bega
\tilde v^{(S)}_k(X) =  \int {d^2 p \ov (2 \pi)^2} v_p (X) A_{kp}^{(S)}, \qquad \tilde v^{(\bar S)}_k(X) =  \int {d^2 p \ov (2 \pi)^2} v_p (X) A_{kp}^{(\bar S)} \ .
\end{gather} 
We would like to show that 
\be \label{yen1}
\tilde v^{(S)}_k(X)  = 0, \quad \forall z \; \text{and} \; x^\pm \in \hat{\bar S} \ .
\ee
From~\eqref{jeh1} we have 
\be 
\tilde v^{(\bar S)}_{-k} (z, x^\mu) =\tilde v^{(S)}_{k} (z, -x^\mu) 
\ee
thus, if~\eqref{yen1} is true it implies that $\tilde v^{(\bar S)}_{k}(X) =0$ for $x \in \hat{S}$. 

More explicitly, from~\eqref{rindToMink}, we have
\be 
\begin{aligned} 
\tilde v^{(S)}_k(X) &=  \int {d^2 p \ov (2 \pi)^2} \theta (-p^2)  \sqrt{\pi} z^{d \ov 2} J_\nu (pz)  e^{i p \cdot x } A_{kp}^{(S)} \\
 &= \sqrt{\pi} z^{d \ov 2} {\pi \ep (\om) \ov  \sqrt{2 \sinh \pi |\om|}} 
  e^{-i \th (k^+, k^-)} \le(e^{\pi\om \ov 2}I_+ - e^{-{\pi\om \ov 2}} I_- \ri) 
\end{aligned} 
\ee
where 
\bega 
I_+ (z, x^\mu) = \int {d^2 p \ov (2 \pi)^2} \theta (-p^2)  \th (p^0)  J_\nu (pz)  e^{i p \cdot x }  (p^+)^{-ik^+ - \frac{1}{2}} (p^-)^{ik^- - \frac{1}{2}} ,  \\
I_- (z, x^\mu) = \int {d^2 p \ov (2 \pi)^2} \theta (-p^2)  \th (-p^0)  J_\nu (pz)  e^{i p \cdot x }  (-p^+)^{-ik^+ - \frac{1}{2}} (-p^-)^{ik^- - \frac{1}{2}} = I_+ (z, -x^\mu)  \ .
\end{gather}

We can change integration variables in $I_+$ as follows
\be 
\begin{aligned}  
I_+ &= {1 \ov 2 \pi^2} \int_0^\infty dp^+ dp^- \, J_\nu (2 \sqrt{p^+ p^-} z) e^{- i p^+ x^- - i p^- x^+}  (p^+)^{-ik^+ - \frac{1}{2}} (p^-)^{ik^- - \frac{1}{2}} \\
&=  {1 \ov \pi^2} \int_0^\infty {p dp dr \ov r} \, J_\nu (2 p z) p^{- i k -1} r^{-{i \om }} e^{- i p (r x^- + {1 \ov r} x^+)}
\end{aligned}
\ee
where $p = \sqrt{p^+ p^-}$ and $r = \sqrt{p^+\ov p^-}$. If $x \in \hat{\bar S}$, i.e. $x^+ < 0, x^- > 0$, we can rotate the $r$-contour to the negative imaginary axis , i.e. 
\bega 
I_+ (z, x^{\mu}) = {1 \ov \pi^2} \int_0^\infty dp \, J_\nu (2 p z) p^{- i k}  \int_0^{-i \infty}  {dr \ov r} \,  r^{-{i \om }} e^{- i p (r x^- + {1 \ov r} x^+)} \\
=  {e^{- {\pi \om \ov 2}} \ov \pi^2} \int_0^\infty dp \, J_\nu (2 p z) p^{- i k}  
 \int_0^{\infty}  {dr \ov r} \,  r^{-{i \om }} e^{- p (r x^- - {1 \ov r} x^+)} 
 \equiv e^{- {\pi \om \ov 2}}  I 
 \ .
\end{gather} 
Similarly we can rotate the $r-$contour to the positive imaginary axis for $I_+ (z, -x^{\mu})$,  
\be 
I_+ (z, -x^{\mu}) = {e^{ {\pi \om \ov 2}} \ov \pi^2} \int_0^\infty dp \, J_\nu (2 p z) p^{- i k}  
 \int_0^{\infty}  {dr \ov r} \,  r^{-{i \om }} e^{- p (r x^- - {1 \ov r} x^+)}  = e^{\pi\om \ov 2} I\ .
\ee
We thus conclude that 
\be 
e^{\pi\om \ov 2}I_+ - e^{-{\pi\om \ov 2}} I_-  = 0, \quad x \in \hat{\bar S} \quad \Rightarrow \quad \tilde v^{(S)}_k(X) =0 \ .
\ee
If $x \in \hat{S}$ we have instead 
\be 
I_+ =  e^{{\pi\om \ov 2}} I, \quad I_- = e^{-{\pi\om \ov 2}} I,
\ee
which gives 
\be
\tilde v^{(S)}_k(X)  =  {z^{d \ov 2} \pi^{3 \ov 2} \ep (\om) \ov  \sqrt{2 \sinh \pi |\om|}} 
  e^{-i \th (k^+, k^-)} 2 \sinh \pi \om \, I
  =  \pi^{3 \ov 2}  z^{d \ov 2}  \sqrt{2 \sinh \pi |\om|}
  e^{-i \th (k^+, k^-)}  I \ .
  \ee
Similarly, if $x^+,x^- >0$ or $x^+, x^- < 0,$ one has $v^{(S)}_k(X) \neq 0$ and $v^{(\bar S)}_k(X) \neq 0.$ Thus, only for $x \in \hat{\bar S}$ does $v^{(S)}_k(X)$ vanish identically, demonstrating that $\fc_{\hat{\bar S}} = \{X=(z,x^{\mu}),~ x \in \hat{\bar S}\}$ is the algebraic entanglement wedge, $\fb_{\bar S},$ for the half-space $\bar S$.

The region $\fb_{\bar S} = \fc_{\hat{\bar S}}$ found as the algebraic entanglement wedge for the half-space ${\bar S}: x^1 < 0$ (on the $x^0=0$ slice) exactly coincides with the entanglement wedge that one expects from the RT prescription for the boundary spatial subregion $\bar S$. In particular, $\fb_{\bar S}$ is the bulk domain of dependence of the homology hypersurface $E_{\bar S}: \{(z,x^{\mu}),~x^0=0, x^1 < 0\}$ whose boundary is $\pt E_{\bar S} = \ga_{\bar S} \cup {\bar S}$ with $\ga_{\bar S} = \{(z,x^{\mu}),~ x^0=0, x^1=0\}$ where $\bar S$ is viewed as a curve on the conformal boundary ($z=0$) of Poincar\'e AdS$_3.$ Defining the RT surface as the (bulk part of the) boundary of the bulk spatial subregion whose domain of dependence is the algebraic entanglement wedge, we find the RT surface to be given by $\ga_{\bar S}.$ This exactly agrees with the RT surface one obtains using the usual extremal surface prescription.

\section{RT surface without entropy: interval and its complement} \label{app:int}
In this appendix we show that the algebra $\sX_I = \sY_{\hat I}$ of a diamond-shaped boundary region is dual to the expected bulk region given by the domain of dependence of the homology hypersurface between a time-slice of the boundary diamond and the RT surface. We also show that the algebra, $\sX_{O},$ of the union, $O = O_L \cup O_R,$ of two finitely separated half-spaces, $O_L : x^1 < -a$ and $O_R: x^1 > a,$ describes the bulk region expected from the RT prescription. In particular, we will explicitly see that the entanglement wedges are `superadditive,' i.e. $\fb_O \supset \fb_{O_L} \vee \fb_{O_R}.$\\

From our discussion of appendix~\ref{app:nonInvO} we see that the algebra $\sX_I$ is generated by oscillators $\{a^{(I)}_k\}$ and the algebra $\sX_O$ is generated by oscillators $\{a^{(O)}_k\}.$ Thus, any operator that can be written in terms of only $\{a^{(I)}_k\} ~(\{a^{(O)}_k\})$ clearly lies in $\sX_{I} ~(\sX_{O}).$\\

We consider a bulk scalar field $\phi$, which has the expansion 
\bega 
\phi (X) = \int {d^2 p \ov (2 \pi)^2} v_p (X) b_p  = \int {d^2k \ov (2\pi)^2} \le(\tilde v^{(I)}_k(X) a^{(I)}_k + \tilde v^{(O)}_k(X) a^{(O)}_k\ri) \ ,\\
v_p (X) = \theta (-p^2)  \sqrt{\pi} z^{d \ov 2} J_{\De -1} (pz)  e^{i p \cdot x } , \qquad p = \sqrt{- p^2}  , 
\end{gather}
as in~\eqref{phex}. The expansion in terms of interval/complement oscillators is obtained by using the basis change from interval/complement to Minkowski oscillators as in~\eqref{pMatrix}. The expansion coefficients are then
\bega \label{expCoeffInt}
\tilde v^{(I)}_k(X) =  \int {d^2 p \ov (2 \pi)^2} v_p (X) P_{kp}^{(I)}, \qquad \tilde v^{(O)}_k(X) =  \int {d^2 p \ov (2 \pi)^2} v_p (X) P_{kp}^{(O)} \ .
\end{gather} 

We may then derive an explicit expression for a bulk operator at any point $X$ on the Poincar\'e patch in terms of the interval/complement basis. Using~\eqref{explicitPMatrix}, we find the coefficient of the interval oscillator to be 
\be \label{splitGenPCkernel}
\begin{aligned}
	\tilde v^{(I)}_k(X) &= {N_k z (2a)^{\De} 
	\pi^2 \ov 2^{\De - 2} \Ga(\De) \sinh\pi\om} \le(I_+(k) - I_-(k)\ri), \qquad I_{\pm}(k) = \int_0^{\infty} {dp^+~dp^- \ov (2\pi)^2}~ j_k(\pm p^+, \pm p^-), \\
	j_k(p^+,p^-) &= (4 p^+ p^-)^{\De - 1 \ov 2} {}_1F_1\le(\bar q_+ ; \De ; 2ia p^+\ri) {}_1F_1\le(\bar q_- ; \De ; 2ia p^-\ri) e^{-i(x^- +a)p^+ -i(x^+ +a)p^-} J_{\De -1}(2 z \sqrt{p^+ p^-}) \ .
\end{aligned}
\ee
From the form of $P^{(O)}_{kp}$ in~\eqref{explicitPMatrix} we find that $\tilde v^{(O)}_k(X)$ can easily be expressed in terms of the same quantities as $\tilde v^{(I)}_k(X).$ Explicitly we have
\be  \label{suppExteriorOscGenPCKer}
	\tilde v^{(O)}_k(X) = {N_k z (2a)^{\De} 
	\pi^2 \ov 2^{\De - 2} \Ga(\De) \sinh\pi\om} \le(e^{\pi\om} I_+(-k) - e^{-\pi\om} I_-(-k)\ri) \ .
\ee

An explicit calculation of $I_{\pm}(k)$ (see appendix~\ref{app:calcIpm} for details) gives
\be \label{expIPlus}
\begin{aligned}
	I_+(k)	&= {2^{\nu}  \ov (2\pi)^2} z^{\nu} (i(x^-+a-i\ep))^{-q_+} (i(x^--a-i\ep))^{-\bar q_+} \Ga(\De) \le[i\le({(x^++a)(x^-+a-i\ep)-z^2 \ov x^- + a-i\ep}\ri)\ri]^{-i\om} \\
	&\cdot \le[i\le({(x^+ - a)(x^-+a-i\ep)-z^2 \ov x^- + a-i\ep}\ri)\ri]^{-\bar q_-} \le[i\le({(x^+ + a)(x^- - a-i\ep)-z^2 \ov x^- - a-i\ep}\ri)\ri]^{-\bar q_+} \\
	&\cdot {}_2F_1\le(\bar q_+, \bar q_- ; \De ; {4a^2z^2 \ov ((x^+-a)(x^- + a -i\ep) - z^2)((x^+ +a)(x^- - a - i\ep) - z^2)}\ri) \ ,
\end{aligned}
\ee
with $\ep > 0$ needed to ensure convergence of the integrals. In all factors except the hypergeometric function we can safely take $\ep$ to zero. The argument of the hypergeometric function to first order in $\ep$ is
\be 
\begin{aligned}
	&{4a^2z^2 \ov ((x^+-a)(x^- + a) - z^2)((x^+ + a)(x^- - a) - z^2)} - {8 a^2 z^2 ((a^2 - (x^+)^2)x^- + z^2 x^+) \ov (a^4 - a^2((x^-)^2+(x^+)^2-2z^2)+(z^2-x^-x^+)^2)^2}~i\ep + O(\ep^2) \\
	&= {4a^2z^2 \ov ((x-a)^2 +z^2 -t^2)((x+a)^2 +z^2 - t^2)} - ((a^2 - (x^+)^2)x^- + z^2 x^+)i\ep \ ,
\end{aligned}
\ee
where we absorb positive definite factors into $\ep.$ When the real part of the expression above is in $(1,\infty),$ we are on the branch cut of the hypergeometric function and thus must use the $i\ep$ prescription to understand from which direction the branch cut is approached.\\

A similar calculation (see appendix~\ref{app:calcIpm} for details) yields
\be \label{expIMinus}
\begin{aligned}
	I_-(k)	&= {2^{\nu}  \ov (2\pi)^2} z^{\nu} (-i(x^-+a+i\ep))^{-q_+} (-i(x^--a+i\ep))^{-\bar q_+} \Ga(\De) \le[-i\le({(x^++a)(x^-+a+i\ep)-z^2 \ov x^- + a+i\ep}\ri)\ri]^{-i\om} \\
	&\cdot \le[-i\le({(x^+ - a)(x^-+a+i\ep)-z^2 \ov x^- + a+i\ep}\ri)\ri]^{-\bar q_-} \le[-i\le({(x^+ + a)(x^- - a+i\ep)-z^2 \ov x^- - a+i\ep}\ri)\ri]^{-\bar q_+} \\
	&\cdot {}_2F_1\le(\bar q_+, \bar q_- ; \De ; {4a^2z^2 \ov ((x^+-a)(x^- + a +i\ep) - z^2)((x^+ +a)(x^- - a + i\ep) - z^2)}\ri) \ .
\end{aligned}
\ee
One finds that, due to the complex powers and branch cuts,~\eqref{expIPlus} and~\eqref{expIMinus} take on distinct functional forms depending on the location of the bulk point $X = (t, x, z).$ In particular, we find that the Poincar\'e patch is split into twelve distinct regions. They are defined as follows
\be \label{twelveRegs}
\begin{aligned}
	\fr_1 &= \{(x^0,x^1,z) ~|~ (x^1)^2 + z^2 > (a + |x^0|)^2 \} \\
	\fr_2 &= \{(x^0,x^1,z) ~|~ (x^1)^2 + z^2 < (a - |x^0|)^2,~ |x^0| < a \} \\
	\fr_3 &= \{(x^0,x^1,z) ~|~ (x^0)^2 > (\sqrt{(x^1)^2+z^2} + a)^2,~ x^0 > a\} \\
	\fr_4 &= \{(x^0,x^1,z) ~|~ (x^0)^2 > (x^1+a)^2 + z^2,~ (x^0)^2 < (x^1-a)^2 + z^2,~ x^0 > 0\} \\
	\fr_5 &= \{(x^0,x^1,z) ~|~ (x^0)^2 < (x^1+a)^2 + z^2,~ (x^0)^2 > (x^1-a)^2 + z^2,~ x^0 > 0\} \\
	\fr_6 &= \{(x^0,x^1,z) ~|~ (x^0)^2 > (\sqrt{(x^1)^2+z^2} + a)^2,~ x^0 < -a\} \\
	\fr_7 &= \{(x^0,x^1,z) ~|~ (x^0)^2 > (x^1+a)^2 + z^2,~ (x^0)^2 < (x^1-a)^2 + z^2,~ x^0 < 0\} \\
	\fr_8 &= \{(x^0,x^1,z) ~|~ (x^0)^2 < (x^1+a)^2 + z^2,~ (x^0)^2 > (x^1-a)^2 + z^2,~ x^0 < 0\} \\
	\fr_9 &= \{(x^0,x^1,z) ~|~ (x^0)^2 < (x^1+a)^2 + z^2,~ (x^0)^2 < (x^1-a)^2 + z^2,~ (x^0)^2 > (\sqrt{(x^1)^2+z^2} - a)^2,~ x^0 > 0\} \\
	\fr_{10} &= \{(x^0,x^1,z) ~|~ (x^0)^2 < (x^1+a)^2 + z^2,~ (x^0)^2 < (x^1-a)^2 + z^2,~ (x^0)^2 > (\sqrt{(x^1)^2+z^2} - a)^2,~ x^0 < 0\} \\
	\fr_{11} &= \{(x^0,x^1,z) ~|~ (x^0)^2 > (x^1+a)^2 + z^2,~ (x^0)^2 > (x^1-a)^2 + z^2,~ (x^0)^2 < (\sqrt{(x^1)^2+z^2} + a)^2,~ x^0 < -a\} \\
	\fr_{12} &= \{(x^0,x^1,z) ~|~ (x^0)^2 > (x^1+a)^2 + z^2,~ (x^0)^2 > (x^1-a)^2 + z^2,~ (x^0)^2 < (\sqrt{(x^1)^2+z^2} + a)^2,~ x^0 > a\} \ .
\end{aligned}
\ee
The region $\fr_1$ is exactly the entanglement wedge of $O$ that one expects from the RT prescription. Similarly, $\fr_2$ is the expected entanglement wedge of $I.$\\

Defining
\be 
\begin{aligned}
	L_1 &\equiv (x^+ + a)(x^- - a)-z^2 \\
	L_2 &\equiv (x^+ - a)(x^- + a)-z^2 \\
	L_3 &\equiv (x^+ + a)(x^- + a)-z^2 \ ,
\end{aligned}
\ee
the expressions for $\tilde v^{(I)}_k (X)$ and $\tilde v^{(O)}_k (X)$ can be written in a (somewhat) compact manner.\\

The explicit results are
\be \label{explVInt} 
\begin{aligned}
\tilde v^{(I)}_k (X) &= N_k  (2a)^{\De} 
z^{\De} \\
&\cdot
\begin{cases}
	0 ,&X \in \fr_1 \\
	\le(-L_1\ri)^{-\bar q_+} \le(-L_2\ri)^{-\bar q_-} L_3^{-i\om} {}_2F_1\le(\bar q_+, \bar q_- ; \De ; {4 a^2 z^2 \ov L_1 L_2}\ri) ,&X \in \fr_2 \\
	-i{\sin\pi\De \ov \sinh\pi\om} L_1^{-\bar q_+} L_2^{-\bar q_-} L_3^{-i\om} {}_2F_1\le(\bar q_+, \bar q_- ; \De ; {4 a^2 z^2 \ov L_1 L_2}\ri) ,&X \in \fr_3 \\
	-i{\sin\pi q_- \ov \sinh\pi\om} L_1^{-\bar q_+} \le(-L_2\ri)^{-\bar q_-} L_3^{-i\om} {}_2F_1\le(\bar q_+, \bar q_- ; \De ; {4 a^2 z^2 \ov L_1 L_2}\ri) ,&X \in \fr_4 \\
	-i{\sin\pi q_+ \ov \sinh\pi\om} \le(-L_1\ri)^{-\bar q_+} L_2^{-\bar q_-} L_3^{-i\om} {}_2F_1\le(\bar q_+, \bar q_- ; \De ; {4 a^2 z^2 \ov L_1 L_2}\ri) ,&X \in \fr_5 \\
	i{\sin\pi\De \ov \sinh\pi\om} L_1^{-\bar q_+} L_2^{-\bar q_-} L_3^{-i\om} {}_2F_1\le(\bar q_+, \bar q_- ; \De ; {4 a^2 z^2 \ov L_1 L_2}\ri) ,&X \in \fr_6 \\
	i{\sin\pi \bar q_+ \ov \sinh\pi\om} L_1^{-\bar q_+} \le(-L_2\ri)^{-\bar q_-} \le(-L_3\ri)^{-i\om} {}_2F_1\le(\bar q_+, \bar q_- ; \De ; {4 a^2 z^2 \ov L_1 L_2}\ri) ,&X \in \fr_7 \\
	i{\sin\pi \bar q_- \ov \sinh\pi\om} \le(-L_1\ri)^{-\bar q_+} L_2^{-\bar q_-} \le(-L_3\ri)^{-i\om} {}_2F_1\le(\bar q_+, \bar q_- ; \De ; {4 a^2 z^2 \ov L_1 L_2}\ri) ,&X \in \fr_8 \\
	{\Ga(\De) \Ga(i\om) \ov \Ga(q_+)\Ga(q_-)} \le(-L_1\ri)^{-\bar q_+} \le(-L_2\ri)^{-\bar q_-} L_3^{-i\om} {}_2F_1\le(\bar q_+, \bar q_- ; 1 - i\om ; 1 - {4 a^2 z^2 \ov L_1 L_2}\ri) ,&X \in \fr_9 \\
	{\Ga(\De) \Ga(-i\om) \ov \Ga(\bar q_+)\Ga(\bar q_-)} \le(-L_1\ri)^{-\bar q_+} \le(-L_2\ri)^{-\bar q_-} \le(-L_3\ri)^{-i\om} \le({4 a^2 z^2 \ov L_1 L_2} - 1\ri)^{i\om} {}_2F_1\le(q_+, q_- ; 1 + i\om ; 1 - {4 a^2 z^2 \ov L_1 L_2}\ri) ,&X \in \fr_{10} \\
	{i \ov \sinh\pi\om} L_1^{-\bar q_+} L_2^{-\bar q_-} \le(-L_3\ri)^{-i\om} \le( \sin\pi(\De - i\om) {\Ga(\De) \Ga(i\om) \ov \Ga(q_+)\Ga(q_-)} {}_2F_1\le(\bar q_+, \bar q_- ; 1 - i\om ; 1 - {4 a^2 z^2 \ov L_1 L_2}\ri) \ri. &{} \\
	\le. + \sin\pi\De {\Ga(\De) \Ga(-i\om) \ov \Ga(\bar q_+)\Ga(\bar q_-)} \le({4 a^2 z^2 \ov L_1 L_2} - 1\ri)^{i\om} {}_2F_1\le(q_+, q_- ; 1 + i\om ; 1 - {4 a^2 z^2 \ov L_1 L_2}\ri) \ri) ,&X \in \fr_{11} \\
	{-i \ov \sinh\pi\om} L_1^{-\bar q_+} L_2^{-\bar q_-} L_3^{-i\om} \le( \sin\pi\De {\Ga(\De) \Ga(i\om) \ov \Ga(q_+)\Ga(q_-)} {}_2F_1\le(\bar q_+, \bar q_- ; 1 - i\om ; 1 - {4 a^2 z^2 \ov L_1 L_2}\ri) \ri. &{} \\
	\le. + \sin\pi(\De + i\om) {\Ga(\De) \Ga(-i\om) \ov \Ga(\bar q_+)\Ga(\bar q_-)} \le({4 a^2 z^2 \ov L_1 L_2} - 1\ri)^{i\om} {}_2F_1\le(q_+, q_- ; 1 + i\om ; 1 - {4 a^2 z^2 \ov L_1 L_2}\ri) \ri) ,&X \in \fr_{12} \\
\end{cases}
\end{aligned}
\ee

\be \label{explVExt} 
\begin{aligned}
\tilde v^{(O)}_k (X) &= N_k  (2a)^{\De} 
z^{\De} \\
&\cdot
\begin{cases}
	\le(-L_1\ri)^{- q_+} \le(-L_2\ri)^{- q_-} \le(-L_3\ri)^{i\om} {}_2F_1\le( q_+,  q_- ; \De ; {4 a^2 z^2 \ov L_1 L_2}\ri) ,&X \in \fr_1 \\
	0 ,&X \in \fr_2 \\
	-i{\sin\pi(\De + i\om) \ov \sinh\pi\om} L_1^{- q_+} L_2^{- q_-} L_3^{i\om} {}_2F_1\le( q_+,  q_- ; \De ; {4 a^2 z^2 \ov L_1 L_2}\ri) ,&X \in \fr_3 \\
	-i{\sin\pi q_+ \ov \sinh\pi\om} L_1^{- q_+} \le(-L_2\ri)^{- q_-} L_3^{i\om} {}_2F_1\le( q_+,  q_- ; \De ; {4 a^2 z^2 \ov L_1 L_2}\ri) ,&X \in \fr_4 \\
	-i{\sin\pi q_- \ov \sinh\pi\om} \le(-L_1\ri)^{- q_+} L_2^{- q_-} L_3^{i\om} {}_2F_1\le( q_+,  q_- ; \De ; {4 a^2 z^2 \ov L_1 L_2}\ri) ,&X \in \fr_5 \\
	i{\sin\pi(\De -i\om) \ov \sinh\pi\om} L_1^{- q_+} L_2^{- q_-} L_3^{i\om} {}_2F_1\le( q_+,  q_- ; \De ; {4 a^2 z^2 \ov L_1 L_2}\ri) ,&X \in \fr_6 \\
	i{\sin\pi \bar q_- \ov \sinh\pi\om} L_1^{- q_+} \le(-L_2\ri)^{- q_-} \le(-L_3\ri)^{i\om} {}_2F_1\le( q_+,  q_- ; \De ; {4 a^2 z^2 \ov L_1 L_2}\ri) ,&X \in \fr_7 \\
	i{\sin\pi \bar q_+ \ov \sinh\pi\om} \le(-L_1\ri)^{- q_+} L_2^{- q_-} \le(-L_3\ri)^{i\om} {}_2F_1\le( q_+,  q_- ; \De ; {4 a^2 z^2 \ov L_1 L_2}\ri) ,&X \in \fr_8 \\
	{\Ga(\De) \Ga(i\om) \ov \Ga(q_+)\Ga(q_-)} \le(-L_1\ri)^{- q_+} \le(-L_2\ri)^{- q_-} L_3^{i\om} \le({4 a^2 z^2 \ov L_1 L_2} - 1\ri)^{-i\om} {}_2F_1\le(\bar q_+, \bar q_- ; 1 - i\om ; 1 - {4 a^2 z^2 \ov L_1 L_2}\ri) ,&X \in \fr_9 \\
	{\Ga(\De) \Ga(-i\om) \ov \Ga(\bar q_+)\Ga(\bar q_-)} \le(-L_1\ri)^{- q_+} \le(-L_2\ri)^{- q_-} \le(-L_3\ri)^{i\om} {}_2F_1\le( q_+,  q_- ; 1 + i\om ; 1 - {4 a^2 z^2 \ov L_1 L_2}\ri) ,&X \in \fr_{10} \\
	{i \ov \sinh\pi\om} L_1^{- q_+} L_2^{- q_-} \le(-L_3\ri)^{i\om} \le( \sin\pi\De {\Ga(\De) \Ga(-i\om) \ov \Ga(\bar q_+)\Ga(\bar q_-)} {}_2F_1\le( q_+, q_- ; 1 + i\om ; 1 - {4 a^2 z^2 \ov L_1 L_2}\ri) \ri. \\
	\le. + \sin\pi(\De -i\om) {\Ga(\De) \Ga(i\om) \ov \Ga( q_+)\Ga( q_-)} \le({4 a^2 z^2 \ov L_1 L_2} - 1\ri)^{-i\om} {}_2F_1\le(\bar q_+, \bar q_- ; 1 - i\om ; 1 - {4 a^2 z^2 \ov L_1 L_2}\ri) \ri) ,&X \in \fr_{11} \\
	{-i \ov \sinh\pi\om} L_1^{- q_+} L_2^{- q_-} \le(-L_3\ri)^{i\om} \le( \sin\pi(\De + i\om) {\Ga(\De) \Ga(-i\om) \ov \Ga(\bar q_+)\Ga(\bar q_-)} {}_2F_1\le( q_+, q_- ; 1 + i\om ; 1 - {4 a^2 z^2 \ov L_1 L_2}\ri) \ri. \\
	\le. + \sin\pi\De {\Ga(\De) \Ga(i\om) \ov \Ga( q_+)\Ga( q_-)} \le({4 a^2 z^2 \ov L_1 L_2} - 1\ri)^{-i\om} {}_2F_1\le(\bar q_+, \bar q_- ; 1 - i\om ; 1 - {4 a^2 z^2 \ov L_1 L_2}\ri) \ri) ,&X \in \fr_{12} \ .
\end{cases}
\end{aligned}
\ee

Notice that we have $\tilde v^{(O)}_k (X) = 0$ when $X \in \fr_2.$ Thus local bulk field operators in $\fr_2$ are supported only on $\{a^{(I)}_k\}$ oscillators and thus are elements of $\sX_I$. Thus the algebraic entanglement wedge of $I$ is $\fb_I = \fr_2.$ This region is the bulk domain of dependence of the hypersurface $E_I = \{(z,x^{\mu}),~ x^0=0, 0 < z < \sqrt{a^2 - (x^1)^2}\}$ whose boundary is $\pt E_I = \ga_I \cup I.$ With the RT surface defined as the (bulk part of the) boundary of this hypersurface whose domain of dependence is $\fb_I,$ we find the RT surface to be $\ga_I = \{(z,x^{\mu}),~ x^0=0, z= \sqrt{a^2 - (x^1)^2}, |x^1| < a\}.$ This is exactly as expected from the RT prescription.\\

When $X \in \fr_1,$ we have $\tilde v^{(I)}_k = 0.$  This shows that local bulk field operators located in $\fr_1$ are supported solely on the $a^{(O)}_k$ oscillators and thus are elements of $\sX_{O}.$ The algebraic entanglement wedge of $O$ is $\fb_O = \fr_1.$ Looking at the definition of $\fr_1$ from~\eqref{twelveRegs} we see that it is the bulk domain of dependence of the hypersurface $E_O = \{(z,x^{\mu}) ~|~ x^0=0, (x^1)^2 + z^2 > a^2 \}.$ Defining the RT surface as the (bulk part of the) boundary, $\pt E_O = \ga_O \cup O,$ of this hypersurface whose domain of dependence is $\fb_O,$ we find the RT surface to be\footnote{Note that $\ga_O = \ga_I$ in this case since $\sX_O$ and $\sX_I$ are commutants.} $\ga_O = \{(z,x^{\mu}),~ x^0=0, z= \sqrt{a^2 - (x^1)^2}, |x^1| < a\}.$ Again, this is exactly as expected from the RT prescription.

An important point is that we never used the RT prescription. Instead we simply asked which bulk operators can be described by $\sX_{O}$ and we find that the region $\fr_1$ emerges as a result of the calculation whose only inputs are global reconstruction and the basis change from Minkowski oscillators to interval/complement oscillators.

\subsection{Calculation of $I_{\pm}(k)$} \label{app:calcIpm}
We now explicitly calculate the integrals $I_{\pm}(k),$ defined in~\eqref{splitGenPCkernel}, which are relevant to the computation of the $a^{(I)}_k/a^{(O)}_k$ mode expansion of local bulk operators. For convenience, we copy~\eqref{splitGenPCkernel} below
\be \label{splitGenPCkernelApp}
\begin{aligned}
	I_{\pm}(k) &= \int_0^{\infty} {dp^+~dp^- \ov (2\pi)^2}~ j_k(\pm p^+, \pm p^-), \\
	j_k(p^+,p^-) &= (4 p^+ p^-)^{\De - 1 \ov 2} {}_1F_1\le(\bar q_+ ; \De ; 2ia p^+\ri) {}_1F_1\le(\bar q_- ; \De ; 2ia p^-\ri) e^{-i(x^- +a)p^+ -i(x^+ +a)p^-} J_{\nu}(2 z \sqrt{p^+ p^-}) \ ,
\end{aligned}
\ee
where we define $\nu = \De -1$ to simplify the expressions below.\\

Noting that
\be 
	J_{\nu}(2\sqrt{p^+p^-}z) = {(2\sqrt{p^+p^-}z)^{\nu} \ov (i2\sqrt{p^+p^-}z)^{\nu}} I_{\nu}(i2\sqrt{p^+p^-}z) = e^{-{i\pi\nu \ov 2}} I_{\nu}(2i\sqrt{p^+p^-}z) \ ,
\ee
we can apply the following formula~\cite{PBMvol4}
\be \label{PBMvol4:33562}
	\int_0^{\infty} dx~ e^{-px} x^{B-1 \ov 2} I_{B-1}(\sigma \sqrt{x}) {}_1F_1\le(A;B; \om x\ri) = \le(\sigma \ov 2\ri)^{B-1} p^{A-B} (p-\om)^{-A} e^{\sigma^2 \ov 4p} {}_1F_1\le(A;B; {\sigma^2 \om \ov 4p (p-\om)}\ri) 
\ee
to~\eqref{splitGenPCkernelApp}.~\eqref{PBMvol4:33562} applies for
\be 
	\Re(B) > 0,~ \Re(p) > 0,~ \Re(p-\om) > 0 \ .
\ee 
We will perform the $p^+$ integral of~\eqref{splitGenPCkernelApp} using~\eqref{PBMvol4:33562}. We can then perform the $p^-$ integral using the formula~\cite{PBMvol4}
\be \label{PBMvol4:33572}
\begin{aligned}
	\int_0^{\infty} dx~ &e^{-px} x^{B-1} {}_1F_1\le(A' ;B; \sigma x\ri) {}_1F_1\le(A;B; \om x\ri) \\
	&= \Ga(B) p^{A+A'-B} (p-\sigma)^{-A'} (p-\om)^{-A} {}_2F_1\le(A,A'; B; {\sigma \om \ov (p-\sigma)(p-\om)}\ri) \ ,
\end{aligned}
\ee
which applies for
\be 
	\Re(B) > 0,~ \Re(p) > 0,~ \Re(p-\sigma) > 0,~ \Re(p-\omega)>0,~ \Re(p-\sigma-\omega) > 0 \ .
\ee
 
In the computation of $I_+,$ in order to apply~\eqref{PBMvol4:33562} we need to regulate by sending $x^- \to x^- - i\ep,$ for infinitesimal $\ep > 0.$ In particular, recalling that we have $\nu = \De - 1,$ the parameters in~\eqref{PBMvol4:33562} for this application are
\be \label{IPlusIntParamsPPlusInt}
\begin{aligned}
	p &= i(x^- + a) \to_{x^- \to x^- - i\ep} i(x^- + a) + \ep \\
	A &= \bar q_+,~~ B = \De \\
	\sigma &= 2iz \sqrt{p^-},~~ \om = 2ia \ .
\end{aligned}
\ee
With these parameters and using~\eqref{PBMvol4:33562} we obtain
\be 
\begin{aligned}
	I_+(k) &= {2^{\nu} e^{-{i\pi\nu \ov 2}} \ov (2\pi)^2} \int_0^{\infty} dp^- (p^-)^{\nu \ov 2} e^{-i(x^++a)p^-} {}_1F_1\le(\bar q_- ; \De ; 2iap^- \ri) \\
	&\cdot \le(iz\sqrt{p^-}\ri)^{\nu} (i(x^-+a-i\ep))^{-q_+} (i(x^--a-i\ep))^{-\bar q_+} e^{{i z^2 \ov x^-+a-i\ep} p^-} {}_1F_1\le(\bar q_+ ; \De ; {2iaz^2 \ov (x^-+a-i\ep)(x^--a-i\ep)} p^- \ri) \\
	&= {2^{\nu} e^{-{i\pi\nu \ov 2}} \ov (2\pi)^2} (iz)^{\nu} (i(x^-+a-i\ep))^{-q_+} (i(x^--a-i\ep))^{-\bar q_+} \\
	&\cdot \int_0^{\infty} dp^- (p^-)^{\nu} e^{-i(x^++a)p^-} e^{{i z^2 \ov x^-+a-i\ep} p^-} {}_1F_1\le(\bar q_- ; \De ; 2iap^- \ri){}_1F_1\le(\bar q_+ ; \De ; {2iaz^2 \ov (x^-+a-i\ep)(x^--a-i\ep)} p^- \ri) \ .
\end{aligned}
\ee
With the $p^+$ integral performed, we may now apply~\eqref{PBMvol4:33572} to the $p^-$ integral. From the final line above, we see that the parameters are
\be \label{IPlusIntParamsPMinusInt}
\begin{aligned}
	p &= i(x^+ + a) - {iz^2 \ov x^- + a - i\ep} \\
	A' &= \bar q_-,~~ B = \De ,~~ A = \bar q_+ \\
	\sigma &= 2ia,~~ \om = {2iaz^2 \ov (x^-+a-i\ep)(x^--a-i\ep)} \ .
\end{aligned}
\ee
With these parameter values we have
\be 
	\begin{aligned}
		\Re(B) &= \De > 0,~ \Re(p) = \Re(p-\sigma) = {z^2 \ov (x^-+a)^2 + \ep^2} \ep > 0 \\
		\Re(p-\om) &= \Re(p-\sigma-\om) = {z^2 \ov (x^- - a)^2 + \ep^2} \ep > 0 \ ,
	\end{aligned}
\ee
so~\eqref{PBMvol4:33572} applies without any need for further regulation.\\

Using~\eqref{PBMvol4:33572} we then obtain
\be 
\begin{aligned}
	I_+(k)	&= {2^{\nu}  \ov (2\pi)^2} z^{\nu} (i(x^-+a-i\ep))^{-q_+} (i(x^--a-i\ep))^{-\bar q_+} \Ga(\De) \le[i\le({(x^++a)(x^-+a-i\ep)-z^2 \ov x^- + a-i\ep}\ri)\ri]^{-i\om} \\
	&\cdot \le[i\le({(x^+ - a)(x^-+a-i\ep)-z^2 \ov x^- + a-i\ep}\ri)\ri]^{-\bar q_-} \le[i\le({(x^+ + a)(x^- - a-i\ep)-z^2 \ov x^- - a-i\ep}\ri)\ri]^{-\bar q_+} \\
	&\cdot {}_2F_1\le(\bar q_+, \bar q_- ; \De ; {4a^2z^2 \ov ((x^+-a)(x^- + a -i\ep) - z^2)((x^+ +a)(x^- - a - i\ep) - z^2)}\ri) \ ,
\end{aligned}
\ee
as quoted in~\eqref{expIPlus}.\\

Similarly, in the computation of $I_-,$ in order to apply~\eqref{PBMvol4:33562} to the $p^+$ integral, we now need to regulate by sending $x^- \to x^- + i\ep,$ for infinitesimal $\ep > 0.$ For this application the parameters in~\eqref{PBMvol4:33562} are
\be \label{IMinusIntParamsPPlusInt}
\begin{aligned}
	p &= -i(x^- + a) \to_{x^- \to x^- + i\ep} -i(x^- + a) + \ep \\
	A &= \bar q_+,~~ B = \De \\
	\sigma &= 2iz \sqrt{p^-},~~ \om = -2ia \ .
\end{aligned}
\ee
With these parameters and using~\eqref{PBMvol4:33562} we obtain
\be 
\begin{aligned}
	&I_-(k) = {2^{\nu} e^{-{i\pi\nu \ov 2}} \ov (2\pi)^2} \int_0^{\infty} dp^- (p^-)^{\nu \ov 2} e^{i(x^++a)p^-} {}_1F_1\le(\bar q_- ; \De ; -2iap^- \ri) \\
	&\cdot \le(iz\sqrt{p^-}\ri)^{\nu} (-i(x^-+a+i\ep))^{-q_+} (-i(x^--a+i\ep))^{-\bar q_+} e^{-{i z^2 \ov x^-+a+i\ep} p^-} {}_1F_1\le(\bar q_+ ; \De ; {-2iaz^2 \ov (x^-+a+i\ep)(x^--a+i\ep)} p^- \ri) \\
	&= {2^{\nu} e^{-{i\pi\nu \ov 2}} \ov (2\pi)^2} (iz)^{\nu} (-i(x^-+a+i\ep))^{-q_+} (-i(x^--a+i\ep))^{-\bar q_+} \\
	&\cdot \int_0^{\infty} dp^- (p^-)^{\nu} e^{i(x^++a)p^-} e^{-{i z^2 \ov x^-+a+i\ep} p^-} {}_1F_1\le(\bar q_- ; \De ; -2iap^- \ri){}_1F_1\le(\bar q_+ ; \De ; {-2iaz^2 \ov (x^-+a+i\ep)(x^--a+i\ep)} p^- \ri) \ .
\end{aligned}
\ee

With the $p^+$ integral performed, we may now apply~\eqref{PBMvol4:33572} to the $p^-$ integral. The parameters are
\be \label{IMinusIntParamsPMinusInt}
\begin{aligned}
	p &= -i(x^+ + a) + {iz^2 \ov x^- + a + i\ep} \\
	A' &= \bar q_-,~~ B = \De ,~~ A = \bar q_+ \\
	\sigma &= -2ia,~~ \om = {-2iaz^2 \ov (x^-+a+i\ep)(x^--a+i\ep)} \ .
\end{aligned}
\ee
With these parameter values we have
\be 
	\begin{aligned}
		\Re(B) &= \De > 0,~ \Re(p) = \Re(p-\sigma) = {z^2 \ov (x^-+a)^2 + \ep^2} \ep > 0 \\
		\Re(p-\om) &= \Re(p-\sigma-\om) = {z^2 \ov (x^- - a)^2 + \ep^2} \ep > 0 \ ,
	\end{aligned}
\ee
so~\eqref{PBMvol4:33572} applies without any further regulation.\\

Using~\eqref{PBMvol4:33572} we then obtain
\be 
\begin{aligned}
	I_-(k)	&= {2^{\nu}  \ov (2\pi)^2} z^{\nu} (-i(x^-+a+i\ep))^{-q_+} (-i(x^--a+i\ep))^{-\bar q_+} \Ga(\De) \le[-i\le({(x^++a)(x^-+a+i\ep)-z^2 \ov x^- + a+i\ep}\ri)\ri]^{-i\om} \\
	&\cdot \le[-i\le({(x^+ - a)(x^-+a+i\ep)-z^2 \ov x^- + a+i\ep}\ri)\ri]^{-\bar q_-} \le[-i\le({(x^+ + a)(x^- - a+i\ep)-z^2 \ov x^- - a+i\ep}\ri)\ri]^{-\bar q_+} \\
	&\cdot {}_2F_1\le(\bar q_+, \bar q_- ; \De ; {4a^2z^2 \ov ((x^+-a)(x^- + a +i\ep) - z^2)((x^+ +a)(x^- - a + i\ep) - z^2)}\ri) \ ,
\end{aligned}
\ee
as quoted in~\eqref{expIMinus}.\\

\section{Showing the additivity and intersection anomalies} \label{app:addAnom}
In this appendix we use the results of appendix~\ref{app:int} to demonstrate the second relation of~\eqref{uha1} and relation~\eqref{uha2}. The first relation of~\eqref{uha1} is discussed in appendix~\ref{app:bdrySuperadd}.

\subsection{Showing superadditivity}
The superadditivity phenomenon we studied in section~\ref{sec:union} is the relation
\be \label{addAnomApp}
	\sX_{R_+} \vee \sX_{R_-} \subset \sX_{R_U} \ .
\ee
From the definitions of $\sX_{R_{\pm}},~\sX_{R_U}$ it is clear that $\sX_{R_+} \vee \sX_{R_-} \subseteq \sX_{R_U}.$ It therefore remains to show that this is a proper inclusion. Consider the operator $\phi(x^0=0, x^1 = 0, z)$ with $\sqrt{a^2-b^2} < z < a+b.$ From~\eqref{explVExt} (with half-width $a \to a+b$) we see that, since $z < a+b,$ $\phi(0,0,z)$ can be interpreted as a boundary operator with support only on $\{a^{R_U}_k\}$ (and no support on $\{a^{\bar{R_U}}_k\}$). Thus, $\phi(0,0,z) \in \sX_{R_U}.$

We will now show that $\phi(0,0,z)$ is not in $\sX_{R_+} \vee \sX_{R_-},$ establishing the proper inclusion. In fact, $\phi(0,0,z) \in \le(\sX_{R_-} \cup \sX_{R_+}\ri)'.$ To see this, consider~\eqref{explVInt}--\eqref{explVExt} translated to $R_{\pm}$ which is centered at $t=0,~x=\pm b$ rather than the origin. The overlap of $\phi(0,0,z)$ with the corresponding oscillators is then given by~\eqref{explVInt}--\eqref{explVExt} with $x \to x \mp d$. Since $z > \sqrt{a^2-b^2}$ we see that $(0,0,z)$ lies in the $\fr_1$ region with respect to both $R_+$ and $R_-.$ Thus, $\phi(0,0,z)$ has no overlap with $\{a^{(R_{\pm})}_k\}$ and lies in the commutant of both $\sX_{R_+}$ and $\sX_{R_-}.$ Thus, $\phi(0,0,z)$ is a non-zero (non-central) operator in $\le(\sX_{R_+} \cup \sX_{R_-}\ri)'$ and thus it does not lie in $\le(\sX_{R_+} \cup \sX_{R_-}\ri)'' = \sX_{R_+} \vee \sX_{R_-}.$ It does however lie in $\sX_{R_U}$ establishing the proper inclusion~\eqref{addAnomApp}.

\subsection{Showing the intersection anomaly}
The intersection anomaly~\eqref{uha2} of section~\ref{sec:union} reads
\be \label{intAnomApp}
	\sX_{R_I} \subset \sX_{R_-} \land \sX_{R_+} \ .
\ee
Clearly $\sX_{R_{\pm}} \supset \sX_{R_I}$ so the inclusion
\be
	\sX_{R_I} \subseteq \sX_{R_-} \land \sX_{R_+} 
\ee
is immediate. It then remains to show that there are operators in $\sX_{R_-} \land \sX_{R_+}$ that are not in $\sX_{R_I}.$ Consider the operator $\phi(x^0=0,x^1=0,z)$ with $a-b < z < \sqrt{a^2-b^2}.$ 

This operator has localized mode expansions in $R_{\pm}$ described by~\eqref{explVInt}--\eqref{explVExt} with $x \to x \mp b.$ Since $z < \sqrt{a^2-b^2},$ we see that $(0,0,z)$ is in $\fr_2$ with respect to both $R_+$ and $R_-$ and thus $\phi(0,0,z)$ has no overlap with $\{a^{\bar R_{\pm}}_k\}.$ Thus $\phi(0,0,z)$ can be expanded in terms of either the $\{a^{(R_-)}_k\}$ or $\{a^{(R_+)}_k\}$ and is therefore part of both $\sX_{R_-}$ and $\sX_{R_+}.$ i.e. $\phi(0,0,z) \in \sX_{R_-} \land \sX_{R_+}.$

We may also express $\phi(0,0,z)$ in terms of $\{a^{R_I}_k,~a^{\bar R_I}_k\}.$ The explicit description is given by~\eqref{explVInt}--\eqref{explVExt} with half-width $a \to a - b.$ Since $z > a-b,$ $(0,0,z)$ is in $\fr_1$ with respect to the $R_I$ decomposition. Thus $\phi(0,0,z)$ can be expressed in terms of the $\{a^{\bar R_I}_k\}$ only and cannot be expressed in terms of the $\{a^{ R_I}_k\}.$ Thus, $\phi(0,0,z) \notin \sX_{R_I}.$ This, along with the discussion of the previous paragraph, then establishes that $\phi(0,0,z)$ is an operator in $\sX_{R_-} \land \sX_{R_+}$ but not in $\sX_{R_I}$ which shows that we have the proper inclusion~\eqref{intAnomApp}.

\section{Boundary Geometric Superadditivity} \label{app:bdrySuperadd}
We consider a Minkowski boundary in the vacuum state. For the overlapping spatial subregions $R_-: x^1 \in (-b-a,-b+a)$ and $R_+: x^1 \in (b-a,b+a)$ with $a > b > 0$ we have boundary causal completions
\be 
	{\hat R}_{\pm} = \{(x^0,x^1) ~|~ |x^0 + (x^1 \mp b)| < a,~ |x^0 - (x^1 \mp b)| < a \} \ .
\ee
These regions overlap in a small diamond of half-width $a-b$ centered at the origin. The union, $R_U = R_+ \cup R_-,$ has causal completion
\be 
	{\hat R}_{U} = \{(x^0,x^1) ~|~ |x^0 + x^1| < a+b,~ |x^0 - x^1| < a+b \} \ ,
\ee
and the intersection $R_I = R_+ \cap R_-,$ has causal completion
\be 
	{\hat R}_{I} = \{(x^0,x^1) ~|~ |x^0 + x^1| < a-b,~ |x^0 - x^1| < a-b \} \ .
\ee
We are interested in understanding the relations among $\sX_{R_-} \lor \sX_{R_+},~ \sX_{R_-} \land \sX_{R_+}, \sX_{R_U},~ \sX_{R_I}.$ In particular, we will see that $\sX_{R_-} \lor \sX_{R_+} = \sM_{\hat{R_-}} \lor \sM_{\hat{R_+}} \supset \sM_{\hat{R_-} \cup \hat{R_+}},$ i.e. the addition of the algebras in $\hat{R_-}$ and $\hat{R_+}$ leads to a larger algebra then expected from the boundary spacetime union of regions.

\subsection{Bulk dual of $\sX_{R_-} \lor \sX_{R_+}$}
We first construct the bulk dual of the algebra $\sX_{R_-} \lor \sX_{R_+} = \sM_{\hat{R}_+} \lor \sM_{\hat{R}_-}$. Bulk additivity implies that this algebra is dual to the causal completion of the union of causal domains of the two boundary diamonds, $\le(\fc_{\hat{R}_+} \cup \fc_{\hat{R}_-}\ri)''$. We construct this causal completion directly in the bulk using the bulk causal structure.

The causal domains of the boundary diamonds can be explicitly described on Poincar\'e AdS$_3$ by
\be 
	\fc_{{\hat R}_{\pm}} = \{(x^0,x^1,z) ~|~ (x^1 \mp b)^2 + z^2 < (a - |x^0|)^2,~ |x^0| < a\} \ .
\ee
The union $\fc_{{\hat R}_{-}} \cup \fc_{{\hat R}_{+}}$ is not a causally complete bulk subregion. However, its causal complement is causally complete and thus the domain of dependence of a subregion $\bar H$ of a bulk Cauchy slice (in this case, the $t=0$ slice). The spatial subregion $\bar H$ is described by 
\be 
	{\bar H} = \{(x^0,x^1,z) ~|~ x^0=0, (x^1-b)^2 + z^2 > a^2\} \cap \{(x^0,x^1,z) ~|~ x^0=0, (x^1+b)^2 + z^2 > a^2\} \ .
\ee
The causal complement is then $\le(\fc_{{\hat R}_{-}} \cup \fc_{{\hat R}_{+}}\ri)' = {\bar H}''.$ We are interested in the causal completion of $\fc_{{\hat R}_{-}} \cup \fc_{{\hat R}_{+}}$ which we can clearly see is ${\bar H}'.$ 

Since ${\bar H}'$ is causally complete in the bulk, it is the domain of dependence of the subregion $H,$ the complement of ${\bar H}$ on the $x^0=0$ bulk Cauchy slice. Thus we have established that $\le(\fc_{{\hat R}_{-}} \cup \fc_{{\hat R}_{+}}\ri)'' = H''.$ If $H$ had a smooth boundary, the boundary of $H''$ would be generated by orthogonal null geodesics sent from the boundary of $H.$ However, in this case the boundary of $H$ has a kink at the point $(x^0 = 0, x^1 = 0, z = \sqrt{a^2 - b^2}),$ and thus there is not a well-defined normal at that point. For this reason we construct $H''$ as ${\bar H}'$ rather than from null geodesics.

Clearly $H''$ has both $x^0-$ reflection and $x^1-$reflection symmetry about the $x^0=0$ and $x^1=0$ axes, respectively. For this reason we only need to consider the part of $H''$ with $x^0>0,~x^1>0$ and the rest can be obtained by reflections. We first exclude from $H''$ all points connected by pure $x^0-$translation to ${\bar H}$. Notice that any point with $0 < x^1 < a+b, z > \sqrt{a^2-(x^1-b)^2}$ is clearly timelike separated (by a constant $(x^1,z)$ curve) from ${\bar H}$ and thus doesn't lie in $H''.$ Thus there are no points with $z > a$ in $H''.$ For fixed $z < a,$ there can also not be any points with $x^1-b > \sqrt{a^2 - z^2}.$ Similarly, when $\sqrt{a^2-b^2} < z < a,$ there cannot be any points with $0 < x^1 < b-\sqrt{a^2-z^2}.$

Thus, from considerations of pure $x^0-$translations alone, we conclude that points in $H''$ (with $x^1 >0$) must satisfy $z < a$ and furthermore, for $\sqrt{a^2-b^2} < z$ they must have $|x^1-b| > \sqrt{a^2-z^2}$, while for $0 < z < \sqrt{a^2-b^2}$ we must have $x^1-b > \sqrt{a^2-z^2}.$

Since Poincar\'e AdS is conformally flat (i.e. null curves are the same as in Minkowski space), a point $(x^0,x^1,z)$ is only in $H'' = {\bar H}'$ if for all $(\bar{x^1}, {\bar z}) \in {\bar H}$ we have
\be \label{spaceToHBar}
	(x^0)^2 < (x^1-\bar{x^1})^2 + (z-\bar{z})^2 \ .
\ee
Notice that the `closest' points to $H''$ in ${\bar H}$ are precisely those which lie on its boundary with $H.$ For $\bar{x^1}>0,$ the boundary is the curve $\bar{z} = \sqrt{a^2 - (\bar{x^1}-b)^2}.$ Thus, for fixed $(x^1,z)$ that could potentially be associated to a point in $H'',$ the minimal value of the RHS of~\eqref{spaceToHBar} over all $(\bar{x^1},\bar{z}) \in \bar{H}$ will be obtained at a point on the boundary curve for which the RHS of ~\eqref{spaceToHBar} reads
\be 
	F(\bar{x^1} ; x^1 ,z) = (x^1-\bar{x^1})^2 + \le(z-\sqrt{a^2 - (\bar{x^1}-b)^2}\ri)^2 \ .
\ee
For fixed (allowed) $(x^1,z)$, $F$ is minimized over all $\bar{x^1} \in (0, b+a)$ at $\bar{x^1}_*$ with 
\be 
\bar{x^1}_* = 
\begin{cases}
	b + {a(x^1-b) \ov \sqrt{(x^1-b)^2 +z^2}}, & (\sqrt{a^2-b^2} < z <a) \lor \le(z < \sqrt{a^2 -b^2} \land x^1 > b \le(1 - \sqrt{z^2 \ov a^2 -b^2}\ri) \ri) \\
	0, & z < \sqrt{a^2 -b^2} \land 0 < x^1 < b \le(1 - \sqrt{z^2 \ov a^2 -b^2} \ri) .
\end{cases}
\ee

We therefore obtain that the points $(x^0,x^1,z) \in H''$ with $x^1 >0$ must satisfy $|x^0| < t_{\max}(x^1,z)$ where
\be 
t_{\max}(x^1,z) = 
\begin{cases}
	\sqrt{(x^1)^2 + (z - \sqrt{a^2-b^2})^2}, & z < \sqrt{a^2 -b^2} \land 0 < x^1 < b \le(1 - \sqrt{z^2 \ov a^2 -b^2} \ri)\\
	a - \sqrt{(x^1-b)^2 + z^2}, & (\sqrt{a^2-b^2} < z < a) \lor \le(z < \sqrt{a^2 -b^2} \land x^1 > b \le(1 - \sqrt{z^2 \ov a^2 -b^2}\ri) \ri) ,\\
	
\end{cases}
\ee
and recall that we also must have $|x^1-d| < \sqrt{a^2 -z^2}.$ 

Using the $x^0-$ and $x^1-$reflection symmetries we obtain the complete characterization of the causal completion of the union of the two causal domains of the boundary diamonds, $H''.$ It is the set of bulk points $(t,x^1,z)$ such that $z < a$ and $|x^0| < t_{\max}(x^1,z)$ with
\be
t_{\max}(x^1,z < \sqrt{a^2 - b^2}) = 
\begin{cases}
	a - \sqrt{(x^1+b)^2+z^2}, & x^1 \in \le(-b-\sqrt{a^2-z^2},-b \cdot \le(\sqrt{a^2-b^2} - z \ov \sqrt{a^2-b^2}\ri)\ri) \\
	\sqrt{(x^1)^2 + (\sqrt{a^2-b^2} - z)^2}, & x^1 \in \le(-b \cdot \le(\sqrt{a^2-b^2} - z \ov \sqrt{a^2-b^2}\ri), b \cdot \le(\sqrt{a^2-b^2} - z \ov \sqrt{a^2-b^2}\ri)\ri) \\
	a - \sqrt{(x^1-b)^2+z^2}, & x^1 \in \le(b \cdot \le(\sqrt{a^2-b^2} - z \ov \sqrt{a^2-b^2}\ri), b+\sqrt{a^2-z^2}\ri) \\
\end{cases}
\ee
\be
t_{\max}(x^1, \sqrt{a^2 - b^2} < z < a ) = 
\begin{cases}
	a - \sqrt{(x^1+b)^2+z^2}, & x^1 \in \le(-b-\sqrt{a^2-z^2},-b+\sqrt{a^2-z^2}\ri) \\
	0, & x^1 \in \le(-b+\sqrt{a^2-z^2}, b-\sqrt{a^2-z^2}\ri) \\
	a - \sqrt{(x^1-b)^2+z^2}, & x^1 \in \le(b-\sqrt{a^2-z^2}, b+\sqrt{a^2-z^2}\ri) \\
\end{cases}
\ee

\begin{figure}
        \centering
        \begin{subfigure}[b]{0.45\textwidth}
            \centering
            \includegraphics[width=\textwidth]{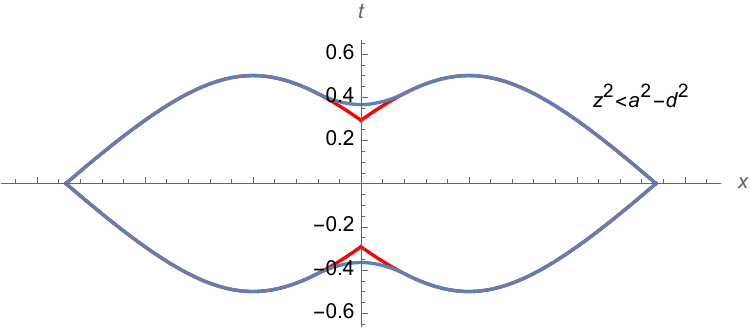}
            \caption[]%
            {{\small }}    
        \end{subfigure}
        \hfill
        \begin{subfigure}[b]{0.45\textwidth}   
            \centering 
            \includegraphics[width=\textwidth]{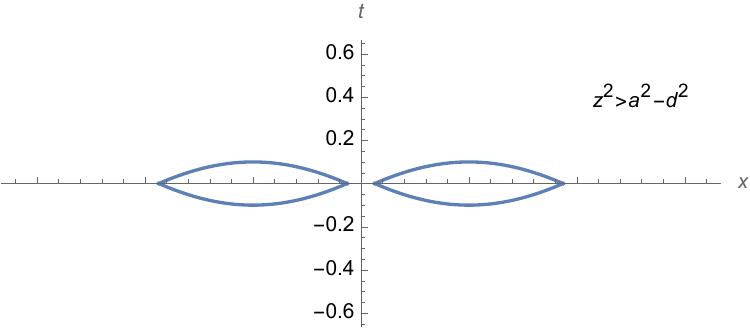}
            \caption[]%
            {{\small }}    
        \end{subfigure}
        \caption[  ]
        {\justifying\small Characterizing $H'',$ for $a =1,~ b=0.5$. (a) ($z = 0.5$) $z < \sqrt{a^2-b^2}$ cross-section of $H''.$ The points between the blue and red curves are in $H''$ but not in the union of causal domains. (b) ($z = 0.9$) $z > \sqrt{a^2-b^2}$ cross-section of $H''$. } 
        \label{fig:causalCompCWUnion}
\end{figure}

\begin{figure}[h]
\begin{centering}
\includegraphics[width=4.5cm]{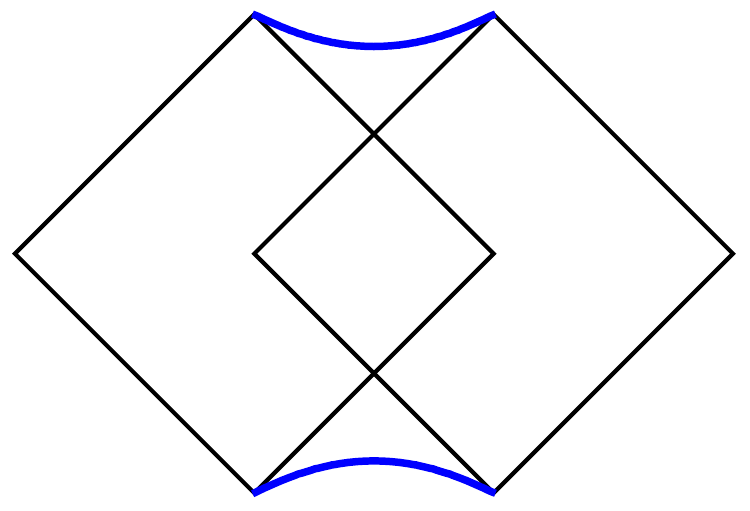}
\end{centering}
\caption[]{\justifying\small The near-boundary points of the bulk causal completion of the union of causal domains of two overlapping boundary diamonds. The points between the black and blue curves lie in $R_e$ but not in $\hat{R}_- \cup \hat{R}_+.$
}
\label{fig:twoDiamSuperAdd}
\end{figure}

Two constant $z$ cross-sections of $H''$ are shown in figure~\ref{fig:causalCompCWUnion}. The intersection of $H''$ with the boundary is shown in figure~\ref{fig:twoDiamSuperAdd}. We will call this boundary subregion $(\hat{R}_+ \cup \hat{R}_-)_{\text{ext}}$ and sometimes use the shorthand $R_e$. Notice that $(\hat{R}_+ \cup \hat{R}_-)_{\text{ext}} \supset \hat{R}_+ \cup \hat{R}_-,$ i.e. the causal completion of the union of the two causal domains contains additional near-boundary points. Thus, bulk additivity suggests that $\sM_{\hat{R}_+ \cup \hat{R}_-} \subset \le(\sM_{\hat{R}_+ \cup \hat{R}_-}\ri)'',$ i.e. $\sM_{\hat{R}_+ \cup \hat{R}_-}$ is not a von Neumann algebra. One may then ask, is $\le(\sM_{\hat{R}_+ \cup \hat{R}_-}\ri)''$ the same as $\le(\sM_{(\hat{R}_+ \cup \hat{R}_-)_{\text{ext}}}\ri)''$? In the next subsection, we give a bulk argument that the answer is yes by identifying $\le(\sM_{(\hat{R}_+ \cup \hat{R}_-)_{\text{ext}}}\ri)''$ with the algebra of bulk operators in its causal domain, which we will see is exactly equal to $H''.$ Moreover, $(\hat{R}_+ \cup \hat{R}_-)_{\text{ext}}$ is the largest boundary subregion whose causal domain is equal to $H''.$

One final interesting point is that the intersection of $H''$ with the boundary is not a causally complete boundary subregion. In particular, we have $(\hat{R}_+ \cup \hat{R}_-)_{\text{ext}} \subset \widehat{(R_+ \cup R_-)}.$ Thus bulk causal completion is {\bf not equivalent} to boundary causal completion. This was already clear from the time-band algebras but the signature in this case is more interesting. In this case, bulk causal completion gives {\it some} extra near-boundary points that are not in the original boundary subregion of interest; however, it does not give all the extra near-boundary points of the boundary causal completion of the original boundary subregion.

\subsection{Causal wedge of $(\hat{R}_+ \cup \hat{R}_-)_{\text{ext}}$}
We saw in the previous subsection that $\le(\fc_{{\hat R}_{-}} \cup \fc_{{\hat R}_{+}}\ri)''$ contains extra near-boundary points that are not in $\hat{R}_+ \cup \hat{R}_-.$ In fact the intersection of $\le(\fc_{{\hat R}_{-}} \cup \fc_{{\hat R}_{+}}\ri)''$ with the boundary is the larger region $(\hat{R}_+ \cup \hat{R}_-)_{\text{ext}}$ shown in figure~\ref{fig:twoDiamSuperAdd}. We can explicitly describe this subregion as the region of the $(x^0,x^1)$ plane with $|x^0| < t_{\max}(x^1),$ where
\be
t_{\max}(x^1) = 
\begin{cases}
	a + (x^1+b), & x^1 \in \le(-b-a,-b \ri) \\
	\sqrt{(x^1)^2 + a^2-b^2}, & x^1 \in \le(-b , b \ri) \\
	a - (x^1-b), & x^1 \in \le(b , b+a\ri) \\
\end{cases}
.
\ee

We wish to understand the causal wedge of this boundary subregion $\fa_{(\hat{R}_+ \cup \hat{R}_-)_{\text{ext}}} = J^+\le((\hat{R}_+ \cup \hat{R}_-)_{\text{ext}}\ri) \cap J^-\le((\hat{R}_+ \cup \hat{R}_-)_{\text{ext}}\ri).$ By $x^0-$reflection symmetry about $x^0=0$ we need only study its bulk causal past as its bulk causal future is simply the $x^0-$reflection of that bulk subregion. In particular, we will study the bulk points $(x^0,x^1,z)$ in the boundary of the causal past with $x^0 > 0.$ For simplicity we will suppress the bulk subregion name and write $J^-$ rather than $J^-\le((\hat{R}_+ \cup \hat{R}_-)_{\text{ext}}\ri)$ and $R_e$ rather than $(\hat{R}_+ \cup \hat{R}_-)_{\text{ext}}$.

A point $(x^0,x^1,z)$ is only in $J^-$ if $x^0 < a ~(=\max_{x^1}(t_{\max}(x^1)))$ and 
\be \label{timelikeRe}
	(x^0 -t')^2 > (x^1-x')^2 + z^2 \ ,
\ee
for some $(t',x') \in R_e.$ The boundary of $J^-$ is given by finding the maximum value of $x^0$ for a fixed $(x^1,z)$ so that~\eqref{timelikeRe} can still be satisfied. Clearly the maximal value of $x^0$ will be obtained by taking $t'$ as large as possible, i.e. $t' = t_{\max}(x').$ To be in the past of $R_e$ it is necessary that $x^0 < t',$ thus,~\eqref{timelikeRe} can be stated in terms of $x^0$ as
\be \label{inPastRe}
	x^0 < t_{\max}(x') - \sqrt{(x^1-x')^2 + z^2} \ .
\ee

The maximum of the RHS is obtained at $x'^*$ where, for $z < \sqrt{a^2-b^2}$
\be \label{intX0Star}
x'^* = 
\begin{cases}
	-b, & x^1 \in \le(-\infty,-b\le(1 - \sqrt{z^2 \ov a^2 - b^2}\ri)\ri) \\
	x^1 {\sqrt{a^2-b^2} \ov \sqrt{a^2-b^2} - z}, & x^1 \in \le(-b\le(1 - \sqrt{z^2 \ov a^2 - b^2}\ri), b\le(1 - \sqrt{z^2 \ov a^2 - b^2}\ri)\ri) \\
	b, & x^1 \in \le(b\le(1 - \sqrt{z^2 \ov a^2 - b^2}\ri), \infty\ri) \\
\end{cases}
,
\ee
and for $z > \sqrt{a^2 - b^2}$
\be 
x'^* = 
\begin{cases}
	-b, & x^1 \in \le(-\infty,0\ri) \\
	b, & x^1 \in \le(0, \infty\ri) \\
\end{cases}
.
\ee

Plugging $x'^*$ into~\eqref{inPastRe} we obtain the maximal value of $x^0$ for any fixed $(x^1,z),$ so that $(x^0,x^1,z)$ is in the causal past of $R_e.$ For the intermediate values of $x^1$ in~\eqref{intX0Star} we have 
\be 
\begin{aligned}
	&t_{\max}(x'^*) - \sqrt{(x^1-x'^*)^2 + z^2} \\
	&= \sqrt{\le(x^1 {\sqrt{a^2-b^2} \ov \sqrt{a^2-b^2} - z}\ri)^2 + a^2 -b^2} - \sqrt{\le(x^1-x^1 {\sqrt{a^2-b^2} \ov \sqrt{a^2-b^2} - z}\ri)^2+z^2} \\
	&= \sqrt{(\sqrt{a^2-b^2} - z)^2 + (x^1)^2} \ .
\end{aligned}
\ee
For the other values of $x^1$ the evaluation of $t_{\max}(x'^*) - \sqrt{(x^1-x'^*)^2 + z^2}$ is trivial, using $t_{\max}(\pm b) = a.$ We therefore obtain the maximum value of $x^0$ for any fixed $(x^1,z)$ to be in the past of $R_e$
\be
t_{\max}^{\text{(past)}}(x^1,z < \sqrt{a^2 - b^2}) = 
\begin{cases}
	a - \sqrt{(x^1+b)^2+z^2}, & x^1 \in \le(-\infty,-b\le(1 - \sqrt{z^2 \ov a^2 - b^2}\ri)\ri) \\
	\sqrt{(x^1)^2 + (\sqrt{a^2-b^2} - z)^2}, & x^1 \in \le(-b\le(1 - \sqrt{z^2 \ov a^2 - b^2}\ri), b\le(1 - \sqrt{z^2 \ov a^2 - b^2}\ri)\ri) \\
	a - \sqrt{(x^1-b)^2+z^2}, & x^1 \in \le(b\le(1 - \sqrt{z^2 \ov a^2 - b^2}\ri), \infty\ri) \\
\end{cases}
\ee
\be
t_{\max}^{\text{(past)}}(x^1, z > \sqrt{a^2 - b^2} ) = 
\begin{cases}
	a - \sqrt{(x^1+b)^2+z^2}, & x^1 \in \le(-\infty,0\ri) \\
	a - \sqrt{(x^1-b)^2+z^2}, & x^1 \in \le(0, \infty\ri) \\
\end{cases}
.
\ee

However, for $(x^0,x^1,z)$ to be in the causal wedge of $R_e$ it must also be in the bulk causal future of $R_e,$ not only in its bulk causal past. This places a minimum on the possible value of $x^0$ for fixed $(x^1,z).$ We thus only have $(x^0,x^1,z)$ in the causal wedge of $R_e$ if
\be 
	t^{\text{(future)}}_{\min}(x^1,z) < x^0 < t_{\max}^{\text{(past)}}(x^1,z) \ .
\ee
By time reflection symmetry
\be 
	t^{\text{(future)}}_{\min}(x^1,z) = -t_{\max}^{\text{(past)}}(x^1,z) \ .
\ee

Clearly, no points with spatial coordinates $(x^1,z)$ can be in the causal wedge of $R_e$ when $t_{\max}^{\text{(past)}}(x^1,z) < 0.$ This prevents any points with $z > a,$ or with $z < a$ but $x^1 < -b -\sqrt{a^2 -z^2}$ or $x^1 > b + \sqrt{a^2-z^2}$ from lying in the causal wedge of $R_e.$ The final result is then that the causal wedge of $R_e$ is the set of points $(x^0,x^1,z)$ with $|x^0| < t_{\max}(x^1,z)$ where
\be
t_{\max}(x^1,z < \sqrt{a^2 - b^2}) = 
\begin{cases}
	a - \sqrt{(x^1+b)^2+z^2}, & x^1 \in \le(-b-\sqrt{a^2-z^2},-b\le(1 - \sqrt{z^2 \ov a^2 - b^2}\ri)\ri) \\
	\sqrt{(x^1)^2 + (\sqrt{a^2-b^2} - z)^2}, & x^1 \in \le(-b\le(1 - \sqrt{z^2 \ov a^2 - b^2}\ri), b\le(1 - \sqrt{z^2 \ov a^2 - b^2}\ri)\ri) \\
	a - \sqrt{(x^1-b)^2+z^2}, & x^1 \in \le(b\le(1 - \sqrt{z^2 \ov a^2 - b^2}\ri), b+\sqrt{a^2-z^2}\ri) \\
\end{cases}
\ee
\be
t_{\max}(x^1, \sqrt{a^2 - b^2} < z < a) = 
\begin{cases}
	a - \sqrt{(x^1+b)^2+z^2}, & x^1 \in \le(-b-\sqrt{a^2-z^2},-b+\sqrt{a^2-z^2}\ri) \\
	a - \sqrt{(x^1-b)^2+z^2}, & x^1 \in \le(b-\sqrt{a^2-z^2}, b+\sqrt{a^2-z^2}\ri) \\
\end{cases}
.
\ee

This then shows that $\fa_{R_e} = \le(\fc_{\hat{R}_+} \cup \fc_{\hat{R}_-}\ri)'',$ i.e. the causal wedge of the extended boundary subregion $R_e$ is equal to the bulk causal completion of the union of causal wedges for the two boundary diamonds. In particular, this establishes that $\fa_{R_e}$ is itself causally complete (i.e. $\fa_{R_e} = \fc_{R_e} \equiv (\fa_{R_e})''$) and thus expectations from the bulk field theory imply that it should have an associated von Neumann algebra which should be identified with $\sM_{R_e} = \le(\sM_{R_e}\ri)''.$ Identifying $\sM_{\hat{R}_+} \lor \sM_{\hat{R}_-}$ (which is by construction a vN algebra) with the bulk causal completion of the union of causal wedges and the equivalence of this bulk subregion with the causal wedge of $R_e$ then suggests
\be \label{superaddExample}
	\sM_{\hat{R}_+} \lor \sM_{\hat{R}_-} = \sM_{R_e} \ .
\ee
Since $R_e$ properly contains $\hat{R}_+ \cup \hat{R}_-,$ this is a phenomenon of {\bf superadditivity}, i.e. the addition of algebras associated to two spacetime subregions is equivalent to the algebra of a region that is larger than the geometric union of the original spacetime subregions.

\subsection{A boundary only approach to superadditivity} \label{app:hbs}

Here we discuss a boundary approach to finding the double commutant. In particular, we would like to show that the operator $\sO(x)$ lies in $\sM_{\hat{R}_-} \vee \sM_{\hat{R}_+}$ whenever $x \in R_e.$ We first characterize the commutant $\le(\sM_{\hat{R}_-} \cup \sM_{\hat{R}_+}\ri)'.$ 

Consider an operator $\sO_g$ supported solely on oscillators $a^{({\bar R_-})}_k,$ i.e. in the commutant of $\sM_{\hat{R_-}}.$ Such an operator has a mode expansion
\be 
	\sO_g = \int {d^2 k \ov (2\pi)^2} g(k) a^{({\bar R_-})}_k \ .
\ee
Using the basis change to Minkowski oscillators, we can equivalently write
\be \label{specialMinkMode}
	\sO_g = \int {dp^+ dp^- \ov (2\pi)^2} \theta(p^+ p^-) K_p(g) b_p \ ,
\ee
where
\be \label{vanishR1Form}
	K_p(g) = \int {d^2 k \ov (2\pi)^2} R^{(\bar R_-)}_{pk} g(k) \ .
\ee
If we wish $\sO_g$ to also lie in the commutant of $\sM_{\hat{R}_+},$ then we must demand that its overlap with $a^{R_+}_k$ vanishes. Using the basis change from $R_+$ interval oscillators to Minkowski oscillators, this condition is
\be 
	\int {dp^+ dp^- \ov (2\pi)^2} \theta(p^+p^-) K_p(g) P^{(R_+)}_{k'p} = 0,~ \forall k' \ .
\ee 
This is a stringent condition on $K_p(g)$ (and therefore $g$).

For a spatial interval of width $2a$ centered at at a point $x_0,$ the basis change from interval to Minkowski oscillators can be explicitly computed to be
\be 
\begin{aligned}
	P^{(R_{x_0})}_{kp} &= e^{ip^+ x_0^- + ip^- x_0^+} {\ep(\om)\ep(p^0)\theta(p^+ p^-) \ov \sqrt{2\sinh\pi|\om|}} C \kappa^\nu e^{-ip^0 a} {}_1F_1\le({\bar q_+}; \De ; 2ia p^+ \ri){}_1F_1\le({\bar q_-}; \De ; 2ia p^- \ri) \\
	P^{(\bar R_{x_0})}_{kp} &= e^{ip^+ x_0^- + ip^- x_0^+} {\ep(\om)\ep(p^0)\theta(p^+ p^-) e^{\pi \ep(p^0)\om} \ov \sqrt{2\sinh\pi|\om|}} C^* \kappa^\nu e^{-ip^0 a} {}_1F_1\le({ q_+}; \De ; 2ia p^+ \ri){}_1F_1\le({ q_-}; \De ; 2ia p^- \ri) \ .
\end{aligned}
\ee

The inverse transformation from Minkowski to interval oscillators is
\be 
\begin{aligned}
	R^{(R_{x_0})}_{pk} &= e^{-ip^+ x_0^- - ip^- x_0^+} {\theta(p^+ p^-) \ov \sqrt{2\sinh\pi|\om|}} C^* \kappa^\nu e^{-ip^0 a} {}_1F_1\le({\bar q_+}; \De ; 2ia p^+ \ri){}_1F_1\le({\bar q_-}; \De ; 2ia p^- \ri) \\
	R^{(\bar R_{x_0})}_{pk} &= e^{-ip^+ x_0^- - ip^- x_0^+} {\theta(p^+ p^-) e^{\pi \ep(p^0)\om} \ov \sqrt{2\sinh\pi|\om|}} C \kappa^\nu e^{-ip^0 a} {}_1F_1\le({ q_+}; \De ; 2ia p^+ \ri){}_1F_1\le({ q_-}; \De ; 2ia p^- \ri) \ .
\end{aligned}
\ee

The explicit formula for $K_p(g)$ is then 
\be 
\begin{aligned}
	K_p(g) &= \theta(p^+ p^-){\pi \kappa^\nu \ov 2^\nu} (2a)^\De e^{-ip^+ (a+b) - ip^- (a-b)} \\
	&\cdot \int {d^2k \ov (2\pi)^2}  { e^{\pi \ep(p^0)\om} \ov \sqrt{2\sinh\pi|\om|}} {|\Ga(q_+)\Ga(q_-)|\ov \Ga(\De)^2} b^{ik^+}c^{-ik^-}  {}_1F_1\le({ q_+}; \De ; 2ia p^+ \ri){}_1F_1\le({ q_-}; \De ; 2ia p^- \ri) g(k) \ .
\end{aligned}
\ee

The condition for $\sO_g$ to also lie in the commutant of $\sM_{R_+}$ is explicitly
\be 
\begin{aligned}
	0 &= {\pi^2 (2a)^{2\De} \ep(\om') b^{ik'^+} c^{-ik'^-} \ov \sqrt{2\sinh\pi|\om'|}} {|\Ga(q'_+)\Ga(q'_-)| \ov \Ga(\De)^4} \\
	&\cdot \int {dp^+dp^- \ov (2\pi)^2} (p^+p^-)^{\De -1} e^{-2ip^+ (a+b) - 2ip^- (a-b)} \ep(p^0)\theta(p^+ p^-)  {}_1F_1\le({\bar q'_+}; \De ; 2ia p^+ \ri){}_1F_1\le({\bar q'_-}; \De ; 2ia p^- \ri)  \\
	&\cdot \int {d^2k \ov (2\pi)^2}  { e^{\pi \ep(p^0)\om} \ov \sqrt{2\sinh\pi|\om|}} |\Ga(q_+)\Ga(q_-)| b^{ik^+}c^{-ik^-}  {}_1F_1\le({ q_+}; \De ; 2ia p^+ \ri){}_1F_1\le({ q_-}; \De ; 2ia p^- \ri) g(k) \ .
\end{aligned}
\ee

The $p^+,~p^-$ integrals can be done explicitly leaving a $k-$integral condition on $g(k).$ The condition for $\sO_g$ to lie in $\le(\sM_{\hat{R}_-} \cup \sM_{\hat{R}_+}\ri)'$ is then
\be \label{constrKpFromRPlus}
\begin{aligned}
	0 &= \int {d^2k \ov (2\pi)^2} g(k) {|\Ga(q_+)\Ga(q_-)| \ov \sqrt{2\sinh\pi|\om|}} b^{ik^+}c^{-ik^-} \le({a+b \ov b}\ri)^{i(k^+ - k'^+)} \le({a-b \ov b}\ri)^{i(k^- - k'^-)} \\
	&\cdot \le(e^{\pi(k^++k'^-)} F\le(q_+, {\bar q'_+}; \De ; {a^2 \ov b^2} + i\ep\ri) F\le(q_-, {\bar q'_-}; \De ; {a^2 \ov b^2} - i\ep\ri) \ri. \\
	&\le. - e^{-\pi(k^++k'^-)} F\le(q_+, {\bar q'_+}; \De ; {a^2 \ov b^2} - i\ep\ri) F\le(q_-, {\bar q'_-}; \De ; {a^2 \ov b^2} + i\ep\ri) \ri) \ .
\end{aligned}
\ee

Computing the commutator of $\sO_g$ with $\sO(x)$ one obtains
\be 
	[\sO(x), \sO_g] = \int {dp^+dp^- \ov (2\pi)^2} u_p(x) K_{-p}(g) \ep(p^0) \ .
\ee
This can be explicitly written as
\be 
\begin{aligned}
	&\int {dp^+dp^- \ov (2\pi)^2} {\sqrt{\pi} \ov \Ga(\De)} e^{-ip^+ x^- -ip^- x^+} \theta(p^+ p^-)\pi (p^+p^-)^{\De -1} (2a)^\De e^{+ip^+ (a+b) + ip^- (a-b)} \\
	&\cdot \int {d^2k \ov (2\pi)^2}  { e^{-\pi \ep(p^0)\om} \ov \sqrt{2\sinh\pi|\om|}} {|\Ga(q_+)\Ga(q_-)|\ov \Ga(\De)^2} b^{ik^+}c^{-ik^-}  {}_1F_1\le({ q_+}; \De ; -2ia p^+ \ri){}_1F_1\le({ q_-}; \De ; -2ia p^- \ri) g(k) \ .
\end{aligned}
\ee

Again the $p^+,~ p^-$ integrals can be explicitly performed to yield
\be \label{genCommOutsideRMinus}
\begin{aligned}
	&{\pi \sqrt{\pi} (2a)^\De \ov \Ga(\De)(2\pi)^2} \int {d^2k \ov (2\pi)^2} \le(e^{\pi(k^++k^-)} e^{{i\pi \ov 2} \ep(x^- -b -a) {\bar q_+}} e^{{i\pi \ov 2} \ep(x^- -b +a) { q_+}} e^{{i\pi \ov 2} \ep(x^+ +b -a) {\bar q_-}} e^{{i\pi \ov 2} \ep(x^+ +b +a) { q_+}} \ri. \\
	&\le. - e^{-\pi(k^++k^-)} e^{-{i\pi \ov 2} \ep(x^- -b -a) {\bar q_+}} e^{-{i\pi \ov 2} \ep(x^- -b +a) { q_+}} e^{-{i\pi \ov 2} \ep(x^+ +b -a) {\bar q_-}} e^{-{i\pi \ov 2} \ep(x^+ +b +a) { q_+}}\ri) \\
	&{|\Ga(q_+)\Ga(q_-)| \ov \sqrt{2\sinh\pi|\om|}} b^{ik^+}c^{-ik^-}  |x^- - b -a|^{-\bar q_+} |x^- - b +a|^{- q_+} |x^+ + b -a|^{-\bar q_-} |x^+ + b +a|^{- q_-} g(k) \ .
\end{aligned}
\ee

The case we are most interested in is when $(x^+,x^-)$ is in the ``$F_R$'' region with respect to $R_-$ as this is a region that contains points in $R_e$ that are not in $\hat{R}_- \cup \hat{R}_+.$ In this case, $-a < x^- - b < a$ and $x^+ + b > a > -a.$ The commutator expression then simplifies to 
\be \label{commWithOpOutsideRMinus}
\begin{aligned}
	&{\pi \sqrt{\pi} (2a)^\De \ov \Ga(\De)(2\pi)^2} \int {d^2k \ov (2\pi)^2} 2i\sin\pi{\bar q_-} \\
	&{|\Ga(q_+)\Ga(q_-)| \ov \sqrt{2\sinh\pi|\om|}} b^{ik^+}c^{-ik^-}  |x^- - b -a|^{-\bar q_+} |x^- - b +a|^{- q_+} |x^+ + b -a|^{-\bar q_-} |x^+ + b +a|^{- q_-} g(k) \ .
\end{aligned}
\ee

We expect that~\eqref{commWithOpOutsideRMinus} must vanish whenever we have $x^+x^- < a^2 - b^2, x^+ > a - b,~ x^- > a - b$ and $g$ satisfies~\eqref{constrKpFromRPlus} but we have not been able to demonstrate this explicitly.

\subsubsection{Double commutant is not the causal completion}
One can easily show, using the global mode expansion, that the bulk operator $\phi(t=0,x=0,z=\sqrt{a^2-b^2 + \ep})$ fails to commute with a boundary operator $\sO(t',x')$ only when $(t'-x')(t'+x') > a^2 -b^2 + \ep,$ for $\ep > 0.$ This is exactly what we expect from bulk causality. We have previously shown that this bulk operator, when written in terms of boundary GFF modes, is supported only on $a^{(\bar R_-)}_k$ modes in the $\{a^{(R_-)}_k,~a^{(\bar R_-)}_k\}$ basis while it only has support on $a^{(\bar R_+)}_k$ modes when written in the $\{a^{(R_+)}_k,~a^{(\bar R_+)}_k\}$ basis. Thus, such an operator clearly lies in $\le(\sM_{\hat{R}_-} \cup \sM_{\hat{R}_+}\ri)'.$ It does not commute with boundary operators with $(t'-x')(t'+x') > a^2 -b^2 + \ep,$ thus, such boundary operators cannot be in $\le(\sM_{\hat{R}_-} \cup \sM_{\hat{R}_+}\ri)''.$ 

In particular, for $x'=0$, the operators with $t' > \sqrt{a^2-b^2}$ are not in the double commutant. This is interesting since this includes operators with $\sqrt{a^2-b^2} < t' < a + b$ which are in the causal completion of the boundary subregion.

\section{Proof of geodesic convexity} \label{app:geodConvex}
We wish to show that the complement, $D_w = \overline{C_w},$ of the bulk dual of a time-reflection symmetric time-band on a bulk Cauchy slice is geodesically convex. This means that for any two points $p_1,~p_2 \in D_w,$ there is a unique minimizing geodesic, $\ga_{12},$ on the $t=0$ bulk slice that connects them and this geodesic lies entirely in $D_w.$ Let $(x_1,z_1),~(x_2,z_2)$ be the coordinates of $p_1,~p_2$, respectively. Without loss of generality we take $x_2 > x_1.$ Since the induced geometry on the $t=0$ bulk slice is simply that of the Poincar\'e half-plane, there is a unique geodesic connecting these two points given by 
\be \label{geod12}
	\ga_{12} = \{(x-x_c)^2+z^2 = w_c^2 ~|~ x_1 < x < x_2,~ z>0\},
\ee
with
\be \label{geodParam}
	x_c = {x_1^2 + z_1^2 - (x_2^2+z_2^2) \ov 2(x_1-x_2)}, \qquad w_c = {\sqrt{\le((x_1-x_2)^2 +(z_1-z_2)^2\ri)\le((x_1-x_2)^2 +(z_1+z_2)^2\ri)}\ov 2|x_1-x_2|} \ .
\ee
An explicit computation shows that, for $\bar{x} \in (x_1,x_2),$ the point $(\bar{x},\bar{z})$ only lies on $\ga_{12}$ if
\be \label{zSqGeod}
	\bar{z}^2 = z_1^2 \le({x_2 - \bar{x} \ov x_2 - x_1}\ri) + z_2^2 \le({\bar{x} - x_1 \ov x_2 - x_1}\ri) + (\bar{x}-x_1)(x_2 - \bar{x}) \ .
\ee
Since $p_1,~p_2$ are in $D_w$ we have $z_{i} > z_w(x_{i})$ or more explicitly
\be \label{zLargeEndpt}
	z_{i}^2 > \max_{x_0 \in I} \le(w(x_0)^2 - (x_i - x_0)^2\ri) \ ,
\ee
for $i=1,2.$ Since $x_1 < \bar{x} < x_2$ we can use~\eqref{zLargeEndpt} in~\eqref{zSqGeod} to conclude that for $(\bar{x},\bar{z}) \in \ga_{12},$ we have
\be \label{zSqGeodIneq}
	\bar{z}^2 > \le(\max_{x_0 \in I} \le(w(x_0)^2 - (x_1 - x_0)^2\ri)\ri) \le({x_2 - \bar{x} \ov x_2 - x_1}\ri) + \le(\max_{x_0' \in I} \le(w(x_0')^2 - (x_2 - x_0')^2\ri)\ri) \le({\bar{x} - x_1 \ov x_2 - x_1}\ri) + (\bar{x}-x_1)(x_2 - \bar{x}) \ .
\ee
The geodesic $\ga_{12}$ remains in $D_w$ so long as for each $\bar{x}\in (x_1,x_2)$ we have $\bar{z} > z_w(\bar{x}).$ At the coordinate $x=\bar{x},$ the curve~\eqref{t0TBCurve} has\footnote{Here we assume $\bar{x}$ is such that $\max_{x_0 \in I} \le(w(x_0)^2 - (\bar{x}-x_0)^2\ri) > 0,$ otherwise we have $z_w(\bar{x}) = 0$ by definition and clearly since $\ga_{12}$ is a geodesic in the Poincar\'e half-plane we always have $\bar{z}>0,$ so at such points $\ga_{12}$ is obviously in $D_w.$}
\be \label{curveAtXbar}
	z_w(\bar{x})^2 = \max_{x_0 \in I} \le(w(x_0)^2 - (\bar{x}-x_0)^2\ri) = w(x_0^*)^2 - (\bar{x}-x_0^*)^2 \ ,
\ee
where $x_0^*$ is the value of $x_0 \in I$ where the maximum is obtained. In~\eqref{zSqGeodIneq}, each of the maxima is taken over $I$ and the result multiplies a positive number. Thus, the right-hand-side of~\eqref{zSqGeodIneq} must be larger than the same expression evaluated at $x_0 \to x_0^*$ and $x_0' \to x_0^*.$ In particular we have
\be \label{zSqGeodIneqStar}
	\bar{z}^2 > \le(w(x_0^*)^2 - (x_1 - x_0^*)^2\ri) \le({x_2 - \bar{x} \ov x_2 - x_1}\ri) + \le(w(x_0^*)^2 - (x_2 - x_0^*)^2\ri) \le({\bar{x} - x_1 \ov x_2 - x_1}\ri) + (\bar{x}-x_1)(x_2 - \bar{x}) \ .
\ee
The right-hand-side of~\eqref{zSqGeodIneqStar} can be explicitly computed to be $w(x_0^*)-(\bar{x}-x_0^*)^2 = z_w(\bar{x}).$ So,~\eqref{zSqGeodIneqStar} implies that for any $(\bar{x},\bar{z}) \in \ga_{12}$ we have $\bar{z}^2 > z_w(\bar{x})^2$ and thus $(\bar{x},\bar{z}) \in D_w,$ completing the proof that 
\be 
	\ga_{12}\subset D_w \ ,
\ee
and thus showing that $D_w$ is geodesically convex.


\begin{thebibliography}{99}


\bibitem{shortPaper}
S.~Leutheusser and H.~Liu,
[arXiv:2110.05497 [hep-th]].

\bibitem{longPaper}
S.~Leutheusser and H.~Liu,
[arXiv:2112.12156 [hep-th]].

\bibitem{Witten:2021jzq}
E.~Witten,
[arXiv:2112.11614 [hep-th]].

\bibitem{Witten:2021unn}
E.~Witten,
JHEP \textbf{10}, 008 (2022)
doi:10.1007/JHEP10(2022)008
[arXiv:2112.12828 [hep-th]].

\bibitem{Schlenker:2022dyo}
J.~M.~Schlenker and E.~Witten,
JHEP \textbf{07}, 143 (2022)
doi:10.1007/JHEP07(2022)143
[arXiv:2202.01372 [hep-th]].

\bibitem{Chandrasekaran:2022cip}
V.~Chandrasekaran, R.~Longo, G.~Penington and E.~Witten,
[arXiv:2206.10780 [hep-th]].

\bibitem{Chandrasekaran:2022eqq}
V.~Chandrasekaran, G.~Penington and E.~Witten,
[arXiv:2209.10454 [hep-th]].

\bibitem{Faulkner:2022ada}
T.~Faulkner and M.~Li,
[arXiv:2211.12439 [hep-th]].

\bibitem{Gao:2021tzr}
P.~Gao and L.~Lamprou,
JHEP \textbf{06}, 143 (2022)
doi:10.1007/JHEP06(2022)143
[arXiv:2111.14010 [hep-th]].

\bibitem{Chandrasekaran:2022qmq}
V.~Chandrasekaran and A.~Levine,
JHEP \textbf{06}, 039 (2022)
doi:10.1007/JHEP06(2022)039
[arXiv:2203.05058 [hep-th]].

\bibitem{Dabholkar:2022mxo}
A.~Dabholkar,
[arXiv:2207.03624 [hep-th]].

\bibitem{Sugishita:2022ldv}
S.~Sugishita and S.~Terashima,
JHEP \textbf{11}, 041 (2022)
doi:10.1007/JHEP11(2022)041
[arXiv:2207.06455 [hep-th]].

\bibitem{Gomez:2022eui}
C.~Gomez,
[arXiv:2207.06704 [hep-th]].

\bibitem{Bahiru:2022mwh}
E.~D.~Bahiru,
[arXiv:2208.04258 [hep-th]].

\bibitem{Verlinde:2022xkw}
H.~Verlinde,
[arXiv:2210.08306 [hep-th]].

\bibitem{deBoer:2022zps}
J.~de Boer, D.~L.~Jafferis and L.~Lamprou,
[arXiv:2211.16512 [hep-th]].

\bibitem{Seo:2022pqj}
M.~S.~Seo,
[arXiv:2212.05637 [hep-th]].

\bibitem{Donnelly:2022kfs}
W.~Donnelly, L.~Freidel, S.~F.~Moosavian and A.~J.~Speranza,
[arXiv:2212.09120 [hep-th]].

\bibitem{Bzowski:2022kgf}
A.~Bzowski,
[arXiv:2212.10652 [hep-th]].

\bibitem{VanRaamsdonk:2009ar}
M.~Van Raamsdonk,
[arXiv:0907.2939 [hep-th]].

\bibitem{Czech:2012bh}
B.~Czech, J.~L.~Karczmarek, F.~Nogueira and M.~Van Raamsdonk,
Class. Quant. Grav. \textbf{29}, 155009 (2012)
doi:10.1088/0264-9381/29/15/155009
[arXiv:1204.1330 [hep-th]].

\bibitem{Czech:2012be}
B.~Czech, J.~L.~Karczmarek, F.~Nogueira and M.~Van Raamsdonk,
Class. Quant. Grav. \textbf{29}, 235025 (2012)
doi:10.1088/0264-9381/29/23/235025
[arXiv:1206.1323 [hep-th]].

\bibitem{Wall:2012uf}
A.~C.~Wall,
Class. Quant. Grav. \textbf{31}, no.22, 225007 (2014)
doi:10.1088/0264-9381/31/22/225007
[arXiv:1211.3494 [hep-th]].

\bibitem{Lewkowycz:2013nqa}
A.~Lewkowycz and J.~Maldacena,
JHEP \textbf{08}, 090 (2013)
doi:10.1007/JHEP08(2013)090
[arXiv:1304.4926 [hep-th]].

\bibitem{Faulkner:2013ana}
T.~Faulkner, A.~Lewkowycz and J.~Maldacena,
JHEP \textbf{11}, 074 (2013)
doi:10.1007/JHEP11(2013)074
[arXiv:1307.2892 [hep-th]].

\bibitem{Headrick:2014cta}
M.~Headrick, V.~E.~Hubeny, A.~Lawrence and M.~Rangamani,
JHEP \textbf{12}, 162 (2014)
doi:10.1007/JHEP12(2014)162
[arXiv:1408.6300 [hep-th]].

\bibitem{Almheiri:2014lwa}
A.~Almheiri, X.~Dong and D.~Harlow,
JHEP \textbf{04}, 163 (2015)
doi:10.1007/JHEP04(2015)163
[arXiv:1411.7041 [hep-th]].

\bibitem{Jafferis:2014lza}
D.~L.~Jafferis and S.~J.~Suh,
JHEP \textbf{09}, 068 (2016)
doi:10.1007/JHEP09(2016)068
[arXiv:1412.8465 [hep-th]].

\bibitem{Pastawski:2015qua}
F.~Pastawski, B.~Yoshida, D.~Harlow and J.~Preskill,
JHEP \textbf{06}, 149 (2015)
doi:10.1007/JHEP06(2015)149
[arXiv:1503.06237 [hep-th]].

\bibitem{Jafferis:2015del}
D.~L.~Jafferis, A.~Lewkowycz, J.~Maldacena and S.~J.~Suh,
JHEP \textbf{06}, 004 (2016)
doi:10.1007/JHEP06(2016)004
[arXiv:1512.06431 [hep-th]].

\bibitem{Hayden:2016cfa}
P.~Hayden, S.~Nezami, X.~L.~Qi, N.~Thomas, M.~Walter and Z.~Yang,
JHEP \textbf{11}, 009 (2016)
doi:10.1007/JHEP11(2016)009
[arXiv:1601.01694 [hep-th]].

\bibitem{Dong:2016eik}
X.~Dong, D.~Harlow and A.~C.~Wall,
Phys. Rev. Lett. \textbf{117}, no.2, 021601 (2016)
doi:10.1103/PhysRevLett.117.021601
[arXiv:1601.05416 [hep-th]].

\bibitem{Harlow:2016vwg}
D.~Harlow,
Commun. Math. Phys. \textbf{354}, no.3, 865-912 (2017)
doi:10.1007/s00220-017-2904-z
[arXiv:1607.03901 [hep-th]].

\bibitem{Faulkner:2017vdd}
T.~Faulkner and A.~Lewkowycz,
JHEP \textbf{07}, 151 (2017)
doi:10.1007/JHEP07(2017)151
[arXiv:1704.05464 [hep-th]].

\bibitem{Cotler:2017erl}
J.~Cotler, P.~Hayden, G.~Penington, G.~Salton, B.~Swingle and M.~Walter,
Phys. Rev. X \textbf{9}, no.3, 031011 (2019)
doi:10.1103/PhysRevX.9.031011
[arXiv:1704.05839 [hep-th]].

\bibitem{Ryu:2006bv}
S.~Ryu and T.~Takayanagi,
Phys. Rev. Lett. \textbf{96}, 181602 (2006)
doi:10.1103/PhysRevLett.96.181602
[arXiv:hep-th/0603001 [hep-th]].

\bibitem{Hubeny:2007xt}
V.~E.~Hubeny, M.~Rangamani and T.~Takayanagi,
JHEP \textbf{07}, 062 (2007)
doi:10.1088/1126-6708/2007/07/062
[arXiv:0705.0016 [hep-th]].


\bibitem{Casini:2019kex}
H.~Casini, M.~Huerta, J.~M.~Mag\'an and D.~Pontello,
JHEP \textbf{02}, 014 (2020)
doi:10.1007/JHEP02(2020)014
[arXiv:1905.10487 [hep-th]].

\bibitem{Benedetti:2022aiw}
V.~Benedetti, H.~Casini and P.~J.~Martinez,
[arXiv:2210.00013 [hep-th]].


\bibitem{Mintun:2015qda}
E.~Mintun, J.~Polchinski and V.~Rosenhaus,
Phys. Rev. Lett. \textbf{115}, no.15, 151601 (2015)
doi:10.1103/PhysRevLett.115.151601
[arXiv:1501.06577 [hep-th]].

\bibitem{Freivogel:2016zsb}
B.~Freivogel, R.~Jefferson and L.~Kabir,
JHEP \textbf{04}, 119 (2016)
doi:10.1007/JHEP04(2016)119
[arXiv:1602.04811 [hep-th]].

\bibitem{Casini:2020rgj}
H.~Casini, M.~Huerta, J.~M.~Magan and D.~Pontello,
JHEP \textbf{04}, 277 (2021)
doi:10.1007/JHEP04(2021)277
[arXiv:2008.11748 [hep-th]].

\bibitem{Papadodimas:2012aq}
K.~Papadodimas and S.~Raju,
JHEP \textbf{10}, 212 (2013)
doi:10.1007/JHEP10(2013)212
[arXiv:1211.6767 [hep-th]].

\bibitem{Papadodimas:2013jku}
K.~Papadodimas and S.~Raju,
Phys. Rev. D \textbf{89}, no.8, 086010 (2014)
doi:10.1103/PhysRevD.89.086010
[arXiv:1310.6335 [hep-th]].


\bibitem{gelfandNaimark}
I.~Gelfand and M.~Naimark,
Matematiceshij sbornik \textbf{54} (1943) 197-217.

\bibitem{segal}
I.E.~Segal,
Bulletin of the American Mathematical Society \textbf{53} (1947) 73-88.

\bibitem{Duetsch:2002hc}
M.~Duetsch and K.~H.~Rehren,
Annales Henri Poincare \textbf{4}, 613-635 (2003)
doi:10.1007/s00023-003-0141-9
[arXiv:math-ph/0209035 [math-ph]].

\bibitem{El-Showk:2011yvt}
S.~El-Showk and K.~Papadodimas,
JHEP \textbf{10}, 106 (2012)
doi:10.1007/JHEP10(2012)106
[arXiv:1101.4163 [hep-th]].

\bibitem{Banks:1998dd}
T.~Banks, M.~R.~Douglas, G.~T.~Horowitz and E.~J.~Martinec,
[arXiv:hep-th/9808016 [hep-th]].

\bibitem{Bena:1999jv}
I.~Bena,
Phys. Rev. D \textbf{62}, 066007 (2000)
doi:10.1103/PhysRevD.62.066007
[arXiv:hep-th/9905186 [hep-th]].

\bibitem{Hamilton:2006az}
A.~Hamilton, D.~N.~Kabat, G.~Lifschytz and D.~A.~Lowe,
Phys. Rev. D \textbf{74}, 066009 (2006)
doi:10.1103/PhysRevD.74.066009
[arXiv:hep-th/0606141 [hep-th]].

\bibitem{Araki:1964}
H.~Araki,
Journal of Mathematical Physics \textbf{5} no.1, (1964) 1-13.

\bibitem{Hubeny:2012wa}
V.~E.~Hubeny and M.~Rangamani,
JHEP \textbf{06}, 114 (2012)
doi:10.1007/JHEP06(2012)114
[arXiv:1204.1698 [hep-th]].

\bibitem{Hubeny:2013gba}
V.~E.~Hubeny, M.~Rangamani and E.~Tonni,
JHEP \textbf{10}, 059 (2013)
doi:10.1007/JHEP10(2013)059
[arXiv:1306.4324 [hep-th]].

\bibitem{ArakiTT}
H.~Araki,
Helv. Phys. Acta \textbf{36} (1963) 132-9.

\bibitem{addPaper}
S.~Leutheusser and H.~Liu,
[arXiv:2411.04183 [hep-th]].

\bibitem{Kang:2018xqy}
M.~J.~Kang and D.~K.~Kolchmeyer,
[arXiv:1811.05482 [hep-th]].

\bibitem{Faulkner:2020hzi}
T.~Faulkner,
[arXiv:2008.04810 [hep-th]].

\bibitem{Kelly:2016edc}
W.~R.~Kelly,
JHEP \textbf{03}, 153 (2017)
doi:10.1007/JHEP03(2017)153
[arXiv:1610.00669 [hep-th]].

\bibitem{Brown:2019rox}
A.~R.~Brown, H.~Gharibyan, G.~Penington and L.~Susskind,
JHEP \textbf{08}, 121 (2020)
doi:10.1007/JHEP08(2020)121
[arXiv:1912.00228 [hep-th]].

\bibitem{Engelhardt:2021mue}
N.~Engelhardt, G.~Penington and A.~Shahbazi-Moghaddam,
Class. Quant. Grav. \textbf{38}, no.23, 234001 (2021)
doi:10.1088/1361-6382/ac2de5
[arXiv:2102.07774 [hep-th]].

\bibitem{Engelhardt:2021qjs}
N.~Engelhardt, G.~Penington and A.~Shahbazi-Moghaddam,
Class. Quant. Grav. \textbf{39}, no.9, 094002 (2022)
doi:10.1088/1361-6382/ac3e75
[arXiv:2105.09316 [hep-th]].

\bibitem{Banerjee:2016mhh}
S.~Banerjee, J.~W.~Bryan, K.~Papadodimas and S.~Raju,
JHEP \textbf{05}, 004 (2016)
doi:10.1007/JHEP05(2016)004
[arXiv:1603.02812 [hep-th]].

\bibitem{Bahiru:2022oas}
E.~Bahiru, A.~Belin, K.~Papadodimas, G.~Sarosi and N.~Vardian,
[arXiv:2209.06845 [hep-th]].


\bibitem{Balasubramanian:2013lsa}
V.~Balasubramanian, B.~D.~Chowdhury, B.~Czech, J.~de Boer and M.~P.~Heller,
Phys. Rev. D \textbf{89}, no.8, 086004 (2014)
doi:10.1103/PhysRevD.89.086004
[arXiv:1310.4204 [hep-th]].

\bibitem{Headrick:2014eia}
M.~Headrick, R.~C.~Myers and J.~Wien,
JHEP \textbf{10}, 149 (2014)
doi:10.1007/JHEP10(2014)149
[arXiv:1408.4770 [hep-th]].

\bibitem{Bousso:2022hlz}
R.~Bousso and G.~Penington,
[arXiv:2208.04993 [hep-th]].

\bibitem{Cruz:1994ir}
N.~Cruz, C.~Martinez and L.~Pena,
Class. Quant. Grav. \textbf{11}, 2731-2740 (1994)
doi:10.1088/0264-9381/11/11/014
[arXiv:gr-qc/9401025 [gr-qc]].

\bibitem{Lin:2022zxd}
H.~W.~Lin, J.~Maldacena, L.~Rozenberg and J.~Shan,
[arXiv:2207.00408 [hep-th]].

\bibitem{Lin:2022rzw}
H.~W.~Lin, J.~Maldacena, L.~Rozenberg and J.~Shan,
[arXiv:2207.00407 [hep-th]].

\bibitem{Lin:2022rbf}
H.~W.~Lin,
JHEP \textbf{11}, 060 (2022)
doi:10.1007/JHEP11(2022)060
[arXiv:2208.07032 [hep-th]].

\bibitem{Engelhardt:2014gca}
N.~Engelhardt and A.~C.~Wall,
JHEP \textbf{01}, 073 (2015)
doi:10.1007/JHEP01(2015)073
[arXiv:1408.3203 [hep-th]].

\bibitem{Penington:2019npb}
G.~Penington,
JHEP \textbf{09}, 002 (2020)
doi:10.1007/JHEP09(2020)002
[arXiv:1905.08255 [hep-th]].

\bibitem{Almheiri:2019psf}
A.~Almheiri, N.~Engelhardt, D.~Marolf and H.~Maxfield,
JHEP \textbf{12}, 063 (2019)
doi:10.1007/JHEP12(2019)063
[arXiv:1905.08762 [hep-th]].

\bibitem{Almheiri:2019hni}
A.~Almheiri, R.~Mahajan, J.~Maldacena and Y.~Zhao,
JHEP \textbf{03}, 149 (2020)
doi:10.1007/JHEP03(2020)149
[arXiv:1908.10996 [hep-th]].

\bibitem{Casini:2011kv}
H.~Casini, M.~Huerta and R.~C.~Myers,
JHEP \textbf{05}, 036 (2011)
doi:10.1007/JHEP05(2011)036
[arXiv:1102.0440 [hep-th]].

\bibitem{PBMvol4}
A.~P.~Prudnikov, I.~A.~Brychkov and O.~I.~Marichev,
{\it Integrals and series: direct laplace transforms (Vol. 4),}
Routledge (1992)



\end{thebibliography}
\end{document}